\documentclass[fleqn,usenatbib]{mnras}

\usepackage{newtxtext,newtxmath}
\usepackage[dvipsnames]{xcolor}
\usepackage[T1]{fontenc}

\DeclareRobustCommand{\VAN}[3]{#2}
\let\VANthebibliography\thebibliography
\def\thebibliography{\DeclareRobustCommand{\VAN}[3]{##3}\VANthebibliography}

\usepackage{graphicx}	% Including figure files
\usepackage{amsmath}	% Advanced maths commands
\usepackage{bbm}
\usepackage{gensymb}
\usepackage{subcaption}
\usepackage{booktabs}
\usepackage{multirow}
\usepackage{siunitx}
\usepackage{mathtools}
\usepackage{hyperref}
\usepackage{mathtools}
\title[$f(R)$ emulation using conditional GANs]{Emulation of $f(R)$ modified gravity from \texorpdfstring{$\Lambda$}{Lambda}CDM using conditional GANs}

\author[Gondhalekar et al.]{
Yash Gondhalekar,$^{1}$\thanks{E-mail: yashgondhalekar567@gmail.com}
Sownak Bose,$^{2}$ \thanks{E-mail: sownak.bose@durham.ac.uk}
Baojiu Li,$^{2}$
Carolina Cuesta-Lazaro$^{3,4,5}$
\\
$^{1}$Department of CSIS, BITS Pilani K.K Birla Goa Campus, Goa, 403726, Goa, India\\
$^{2}$Institute for Computational Cosmology, Department of Physics, Durham University, South Road, Durham DH1 3LE, UK\\
$^{3}$The NSF AI Institute for Artificial Intelligence and Fundamental Interactions Massachusetts Institute of Technology, Cambridge, MA 02139, USA\\
$^{4}$Department of Physics, Massachusetts Institute of Technology, Cambridge, MA 02139, USA\\
$^{5}$Harvard-Smithsonian Center for Astrophysics, 60 Garden Street, Cambridge, MA 02138, USA
}

\date{Accepted XXX. Received YYY; in original form ZZZ}

\pubyear{2024}

\begin{document}
\label{firstpage}
\pagerange{\pageref{firstpage}--\pageref{lastpage}}
%TC:endignore
\maketitle

\begin{abstract}

A major aim of cosmological surveys is to test deviations from the standard $\Lambda$CDM model, but the full scientific value of these surveys will only be realised through efficient simulation methods that keep up with the increasing volume and precision of observational data. $N$-body simulations of modified gravity (MG) theories are computationally expensive since highly non-linear equations must be solved. This represents a significant bottleneck in the path to reach the data volume and resolution attained by equivalent $\Lambda$CDM simulations. We develop a field-level neural network-based emulator that generates density and velocity divergence fields under the $f(R)$ gravity MG model from the corresponding $\Lambda$CDM simulated fields. Using attention mechanisms and a complementary frequency-based loss function, our model is able to learn this intricate mapping. We use the idea of latent space extrapolation to generalise our emulator to $f(R)$ models with differing field strengths. The predictions of our emulator agree with the $f(R)$ simulations to within 5\% for matter density and to within 10\% for velocity divergence power spectra up to $k \sim 2\,h$ $\mathrm{Mpc}^{-1}$. But for a few select cases, higher-order statistics are reproduced with $\lesssim$10\% agreement. Latent extrapolation allows our emulator to generalise to different parameterisations of the $f(R)$ model without explicitly training on those variants. Given a $\Lambda$CDM simulation, the GPU-based emulator can reproduce the equivalent $f(R)$ realisation $\sim$600 times faster than full $N$-body simulations. This lays the foundations for a valuable tool for realistic yet rapid mock field generation and robust cosmological analyses.

\end{abstract}

\begin{keywords} 
cosmology: large-scale structure of Universe -- methods: statistical -- methods: numerical
\end{keywords}

%%%%%%%%%%%%%%%%%%%%%%%%%%%%%%%%%%%%%%%%%%%%%%%%%%

%%%%%%%%%%%%%%%%% BODY OF PAPER %%%%%%%%%%%%%%%%%%
\section{Introduction}\label{sec:intro}

The $\Lambda$CDM ($\Lambda$ Cold Dark Matter) cosmological model fits several observational data adequately well \citep[e.g.,][]{Spergel2003,Percival2007,Hinshaw2013,Alam2017,PlanckCollaboration2020,Aiola2020,Abbott2022}. $\Lambda$CDM is thus remarkably predictive, but also mathematically simple, consisting of only six free parameters. However, despite its tremendous success, it is known that the justification for assuming $\Lambda$CDM as a `standard' or `concordance' model may be fragile. The model cannot `explain' the concepts of dark matter, dark energy (which resembles $\Lambda$, the cosmological constant, in $\Lambda$CDM), and inflation, leaving their true largely elusive. It also introduces puzzling questions such as the `fine-tuning' and the `coincidence' problems \citep{Weinberg1989,Carroll2001}. Precision observations over the years have also indicated tensions with the $\Lambda$CDM model \citep[e.g.,][and references therein]{Abdalla2022}. % Precise observations over the years has led to increasingly accurate constraints on the parameters describing $\Lambda$CDM

These limitations have led to the exploration of several alternative models, ranging from simple extensions to $\Lambda$CDM to theories that alter the underlying physics itself \citep[e.g.,][for a review]{Bull2016}. Of these, Modified Gravity (MG) models \citep{Clifton2012} were developed primarily to explain the accelerated expansion of the Universe \citep{Riess1998} since they reproduce the accelerated expansion without dark energy \citep[e.g.,][]{Koyama2016}. MG theories suggest that general relativity (GR), the physical theory underpinning $\Lambda$CDM, may be inaccurate at large cosmological scales and thus modify gravity at these scales. It is unclear whether MG theories are better or worse descriptors of the Universe than $\Lambda$CDM \emph{a priori}, and the only way to find preferred MG models is through comparison with observations. Beyond this, studying MG has several merits: it may improve our understanding of the pertaining issues with $\Lambda$CDM, develop new physics that may provide a comprehensive theory of gravity, and develop robust tests of GR \citep[e.g.,][]{Jain2010}, which is timely since cosmology has entered the `precision' era. In fact, a key science target of ongoing and upcoming surveys such as Legacy Survey of Space and Time at the Vera Rubin Observatory \citep[LSST;][]{Ishak2019} and the {\it Euclid} mission \citep{Laureijs2011} is to test deviations from GR on cosmological scales.

GR has passed stringent observational tests on small scales such as the Solar System \citep[see, e.g.,][for a review]{Will2014}, so any modifications to gravity in an MG model must vanish on these scales. It is convenient to assume modifications to gravity as a long-range fifth force, most commonly mediated by a new scalar degree of freedom. This new fifth force is then well constrained on small scales, since it must be compatible with local tests of gravity. MG theories obtain this constraint using a built-in `screening' mechanism that suppresses the fifth force in the local Universe \citep[e.g.,][]{Joyce2015}.

This screening mechanism introduces additional non-linearities in the field equations due to the scale-dependent interaction between the matter evolution and the fifth force introduced by the scalar degree of freedom. In particular, screening breaks down the superposition principle from GR, due to which the standard linear Poisson equation of GR must be solved together with the non-linear scalar field equation. The latter requires new numerical techniques, making $N$-body simulations of MG theories challenging in terms of numerical convergence and computational expense. Several $N$-body codes for MG models have been developed in the past, e.g., {\sc ECOSMOG} \citep{Li2012}, {\sc MG-GADGET} \citep{Puchwein2013}, {\sc ISIS} \citep{Llinares2014} \citep[e.g.,][for a review of different full $N$-body codes for MG]{Winther2015CompareProject}. As a rough estimate, an MG simulation may require 5--20 times more time than the corresponding $\Lambda$CDM simulation \citep[e.g.,][]{Winther2015,Winther2015CompareProject}, depending on the resolution and the MG model. However, to effectively explore the parameter space of MG models and confront them with observations, it is necessary to alleviate this computational bottleneck \citep[e.g.,][]{Alam2021} (see also Sect.~\ref{subsec:need-fast-MG-emulators}).

To solve this challenge, several studies have developed new recipes to speed up MG simulations while minimally compromising accuracy; we list a few approaches. \citet{Barreira2015} used an approximate approach that involved truncating the number of iterations used to solve the scalar field equation where screening is most active (based on the recognition that scalar field equations need not be solved with high accuracy where they are screened); \citet{Winther2015} devised and solved a linearised scalar field equation along with an analytical estimate of the screening. \citet{Bose2017} used a redefinition of the variables that enter the scalar field equations to make them analytically solvable and thus more efficient than the Newton-Raphson iterations used in traditional relaxation algorithms. Other approaches include work by \citet{Mead2015}, who developed an algorithm based on the cosmological rescaling algorithm of \citet{Angulo2010} to rescale the outputs of $\Lambda$CDM to an MG-equivalent, taking only $\sim$1 min to rescale $512^3$ dark matter particles; \citet{Valogiannis2017} who designed a fast approximate $N$-body method, based on the COmoving Lagrangian Acceleration ({\sc COLA}) approach adapted for MG theories, that evolves linear and non-linear scales differently; this was followed by a similar study \citep{Winther2017} that presented the {\sc MG-PICOLA} code. Recently, \cite{Ruan2022} and \cite{HernandezAguayo2022} introduced the {\sc MG-GLAM} $N$-body code with computationally efficient numerical implementations. The authors demonstrate a significant acceleration in the runtime of MG simulations, requiring only $\sim$2-5 times the time for an equivalent $\Lambda$CDM simulation.

Recently, there has been a significant interest in constructing emulators--functions that approximate the predictions of cosmological simulations--to reproduce the matter power spectrum enhancement in MG compared to $\Lambda$CDM, \citep[e.g.,][]{Cataneo2019,Winther2019,Ramachandra2021,Bose2023,Fiorini2023,Gupta2023,HarnoisDeraps2023,Mauland2023,Casares2024,Bai2024}. This is because the power spectrum is simple to model and produces analytically tractable changes compared to $\Lambda$CDM. Such approaches generally agree with full $N$-body simulations exceptionally well at large scales and with $\gtrsim$3-5\% errors for smaller scales ($k \gtrsim 1-7 \,h$ $\mathrm{Mpc}^{-1}$ depending on the approach), but at extremely low computational costs. The power spectrum emulator trained on the {\sc FORGE} simulation suite by \citet{Arnold2022} was able to achieve $\lesssim$2.5\% errors up to very small scales ($k \sim 10 \,h$ $\mathrm{Mpc}^{-1}$) across a wide range of cosmological parameters and the strength of $f(R)$ modified gravity.

It is widely believed that data simulation approaches capable of leveraging information from observational datasets beyond what traditional summary statistics, such as the power spectrum, can are needed to realise the full potential of cosmological observations. One promising approach to achieve this goal is through fast, field-level mock generation using deep learning models that model the entire non-linear evolution of matter. These field-level emulators are generally empirical, learning physics from data without strong theoretical guidance, and they do not optimise for predefined summary statistics such as the power spectrum. Several examples of such fast emulators can be found in the literature under the umbrella of deep generative models \citep[e.g.,][]{Rodriguez2018,Mustafa2019,Perraudin2019,He2019,Feder2020,Ullmo2021,Jamieson2023}, and they have shown promise. Sect.~\ref{subsec:need-fast-MG-emulators} discusses the importance of quick field-level emulators in the context of MG. % A noticeable trend in several of the works cited above has been the use of machine learning to emulate the power spectrum.

% In particular, emulators that learn physics, make accurate yet quick predictions, and are minimally affected by unphysical factors may prove extremely valuable

% This paper focuses on a pure machine learning-based cosmological emulation whose applications have surged in the last few years owing to their ability to model complex functions and provide major speed-up compared to full $N$-body or even approximate simulations. Such emulators are generally empirical and do not optimise over pre-defined summary statistics like the power spectrum. Several examples of such fast emulators can be found in the literature under the umbrella of deep generative models \citep[e.g.,][]{Rodriguez2018,List2019,Mustafa2019,Perraudin2019,He2019,Feder2020,Ullmo2021,Jamieson2023}. In particular, emulators that learn physics, make accurate yet quick predictions\footnote{Although the general trend in such approaches in the past has been that speed and accuracy conflict with each other, and so the best approach depends on what tradeoff is sought.}, and are minimally affected by unphysical factors may prove extremely valuable.

In this work, we use Generative Adversarial Networks (GANs), a specific class of generative models, to demonstrate fast emulation of matter density and velocity divergence in modified gravity. Although the applications of GANs in astronomy have been well explored, their use for field-level emulation in MG has received little attention. Our focus is specifically on $f(R)$ gravity \citep[e.g.,][]{Sotiriou2010,DeFelice2010}, which is the most widely studied MG model in the literature. $f(R)$ gravity also introduces a mass-dependent modification of the abundance of halos compared to $\Lambda$CDM \citep[e.g.,][]{Li-Hu-2011,Li2012-Halo,Zivick2015,Cataneo2016}, and there have been attempts to quickly reproduce the MG halo distribution using dark matter density fields in $\Lambda$CDM \citep[e.g.,][]{Farieta2024}; however, a detailed analysis of these attempts is beyond the scope of this study.

Since the primary motivation to run MG simulations is to compare them with predictions from a reference, $\Lambda$CDM simulation, it is reasonable to assume that both simulations need to be run to perform cosmological tests. Combining this with the fact that $\Lambda$CDM simulations are generally much faster to run, we develop a GAN that is conditioned on the output of a $\Lambda$CDM simulation (our $\Lambda$CDM simulations are $\sim$10 times faster than the corresponding $f(R)$ simulations). We first run $\Lambda$CDM simulations and then use our emulator to generate samples in $f(R)$ gravity. The architecture of our emulator is based on BicycleGAN, which is a GAN designed for multimodal image-to-image translation tasks to achieve realism and diversity of generated images simultaneously. We have made a few modifications to this architecture to suit our problem. Predictions are made at a fixed redshift ($z = 0$) and cosmological parameters; however, extensions of our study to varying redshift and cosmological parameters would be straightforward using Latin hypercube simulation datasets. We have performed a quantitative evaluation of our emulator in terms of statistical consistency with the MG simulation outputs, runtime, and energy efficiency. Additionally, we attempt to interpret our neural network-based emulator by identifying correlations between the spatial variations of the scalar field in the MG model and the regions in the matter field to which our emulator devotes its attention. This, consequently, provides a {\it physical interpretation} to an otherwise seemingly opaque learning process in neural network-based emulators. 

The $\Lambda$CDM to $f(R)$ mapping is not straightforward from a machine learning point of view since, as we will see later, the differences produced by MG compared to $\Lambda$CDM are much smaller than several previous applications where GANs were used for cosmological field emulation. Such image-to-image mapping problems, where an input image is transformed to an output image using neural networks, have been extensively applied in cosmology \citep[e.g.,][]{List2019,KodiRamanah2019,KodiRamanah2020,Li2021,Shirasaki2021,Bernardini2022}. Concurrent to this work, \citet{Saadeh2024} conducted a related study to emulate $f(R)$ gravity starting from $\Lambda$CDM using a \texttt{map2map} neural network in which the strength of modified gravity was encoded as a style parameter to generate varied outputs.

The paper is organised as follows. Sect.~\ref{sec:fr-emulation} starts with a review of the $f(R)$ modified gravity theory and motivates the need for fast modified gravity emulators. A brief background on GANs and the specific architecture used in this study is provided. Sect.~\ref{sec:method} presents the methodology, followed by the model evaluation in Sect.~\ref{sec:results}. Sect.~\ref{sec:discussion} summarises our findings and potential for future work, followed by concluding remarks in Sect.~\ref{sec:conclusion}.

\section{\texorpdfstring{$f(R)$}{f(R)} modified gravity emulation using GAN\texorpdfstring{\lowercase{s}}{s}}\label{sec:fr-emulation}
\subsection{\texorpdfstring{$f(R)$}{f(R)} modified gravity theory}\label{subsec:fr-theory}
The $f(R)$ theory of gravity extends GR by replacing the Ricci scalar, $R$, in the Einstein--Hilbert action to a generic, algebraic function of $R$:
\begin{equation}\label{eqn:EH-modified-action}
    \mathcal{S} = \frac{1}{2\kappa^2} \int \mathrm{d}^4x\sqrt{-g} [R + f(R)] + \int \mathrm{d}^4x \mathcal{L}_m,
\end{equation}
where $\kappa^2 \equiv 8\pi G$, $G$ is Newton's constant, $g$ is the determinant of the metric tensor, $g_{\mu\nu}$, $\mathcal{L}_m$ is the Lagrangian density for matter fields (including cold dark matter, baryons, neutrinos, photons).

The modified Einstein equation is obtained by taking a variation of the action, Eqn.~\ref{eqn:EH-modified-action}, with respect to $g_{\mu\nu}$ to yield:
\begin{equation}\label{eqn:modified-einstein-equation}
    G_{\mu\nu} = 8\pi GT^m_{\mu\nu} + X_{\mu\nu},
\end{equation}
where $T^m_{\mu\nu}$ is energy-momentum tensor for matter, $G_{\mu\nu} \equiv R_{\mu\nu} - \frac{1}{2}g_{\mu\nu}R$ is the Einstein tensor, $R_{\mu\nu}$ is the Ricci tensor. The term $X_{\mu\nu}$ is the modification to GR and is defined as:
\begin{equation}
    X_{\mu\nu} \equiv -f_RR_{\mu\nu} + \frac{1}{2}[f(R) - \nabla^{\lambda}\nabla_{\lambda}f_R]g_{\mu\nu} + \nabla_{\mu}\nabla_{\nu}f_R.
\end{equation}
Here $f_R = \frac{\mathrm{d}f(R)}{\mathrm{d}R}$ is the scalar field, that is, an additional scalar degree of freedom. Taking the trace of Eq.~\ref{eqn:modified-einstein-equation} gives:
\begin{equation}\label{eqn:scalaron-equation}
    \nabla^{\mu}\nabla_{\mu}f_R =\frac{1}{3}[R - f_RR + 2f(R) + 8\pi G\rho_m] = \frac{\partial V_{\mathrm{eff}}(f_R)}{\partial f_R},
\end{equation}
which governs the dynamical evolution of $f_R$, where $\rho_m$ is the density of non-relativistic matter. Thus, the modified Einstein equation, which contained fourth-order derivatives of $g_{\mu\nu}$, can now also be viewed as a second-order Einstein equation of GR with a scalar field.

It is reasonable to assume the quasi-static and weak-field approximations since we are interested in the non-linear regime of structure formation, and the time evolution of $f_R$ is negligible \citep{Bose2015}. Thus, Eq.~\ref{eqn:scalaron-equation} reduces to
\begin{equation}\label{eqn:MG-scalaron}
    \nabla^2f_R = -\frac{1}{3c^2}[\delta R + 8\pi G\delta \rho_m] a^2,
\end{equation}
where $\nabla^2$ is the three-dimensional Laplacian operator, $a$ is the scale factor, and $\delta \rho_m \equiv \rho_{\mathrm{m}} - \bar{\rho}_{\mathrm{m}}$ and $\delta R = R(f_R) - \bar{R}$ denote the density and curvature perturbations, respectively. Under this approximation, the modified Einstein equation, Eq.~\ref{eqn:modified-einstein-equation}, also reduces to a modified Poisson equation given by
\begin{equation}\label{eqn:MG-potential}
    \nabla^2\Phi = \frac{16}{3}\pi G a^2\delta\rho_{\mathrm{m}} + \frac{1}{6}a^2\delta R = 4\pi Ga^2\delta \rho_{\mathrm{m}} - \frac{1}{2}c^2\nabla^2f_R,
\end{equation}
where Eq.~\ref{eqn:MG-scalaron} is used for the second derivation. This equation relates the gravitational potential, $\Phi$, at a given position to the density and curvature perturbations.

These equations show that the introduction of the scalar field can change the background expansion of the Universe due to the new term added in Eq.~\ref{eqn:modified-einstein-equation}, and that the relation between matter and density is also modified. Thus, the clustering of matter and the evolution of density perturbations may differ from GR. When $\lvert f_R \rvert \ll \lvert \Phi \rvert$, $R \approx -8\pi G \rho_{\mathrm{m}}$ from Eq.~\ref{eqn:MG-scalaron}, then Eq.~\ref{eqn:MG-potential} reduces to the Poisson equation in GR. When $\lvert f_R \rvert \gg \lvert \Phi \rvert$, however, $\lvert \delta R \rvert \ll 8\pi G \lvert \delta \rho_{\mathrm{m}} \rvert$, so Eq.~\ref{eqn:MG-potential} reduces to the Poisson equation in GR but with G rescaled by $4/3$: this is the maximum enhancement possible in $f(R)$ gravity, regardless of the functional form of $f(R)$.

Fixing a functional form of $f(R)$, however, helps to specify the $f(R)$ model and quantifies when and on which scale the enhancement changes from unity to $4/3$. The choice of a plausible form of $f(R)$ is guided by observations: the $f(R)$ model must cause an accelerated cosmic expansion with a background evolution close to $\Lambda$CDM, and must be sufficiently parametrised to explain a range of observations. The specific form of $f(R)$ must also include $\Lambda$CDM as a limiting case in regions where GR is confirmed to high precision, such as the Solar System--some choices of $f(R)$ lead to the \emph{chameleon} screening mechanism to suppress the fifth force and pass this observational constraint. The form of $f(R)$ proposed by \citet{Hu2007} is the most studied form of $f(R)$ and has been found to satisfy these requirements. It follows a broken power law relation:
\begin{equation}
    f(R) = -m^2\dfrac{c_1(-R/m^2)^n}{c_2(-R/m^2)^n - 1},
\end{equation}
where $m^2 \equiv \frac{\kappa^2\bar{\rho}_0}{3}$, where $\bar{\rho}_0$ is the mean density today, i.e., at $a = 1$. $n, c_1, c_2$ are dimensionless parameters.

Relating $f_R$ to $R$ gives
\begin{equation}\label{eqn:scalar-field}
    f_R = -n\frac{c1}{c^2_2}\frac{(-R/m^2)^{n+1}}{[1 + (-R/m^2)^n]^2}.
\end{equation}
The scalar field corresponds to the minimum of its effective potential, $V_{\mathrm{eff}}(f_R)$, in background cosmology. Thus
\begin{equation}
    -\bar{R} \approx 8\pi G\bar{\rho}_m - 2\bar{f}(R) = 3m^2\left[a^{-3} + \frac{2}{3}\frac{c_1}{c_2}\right].
\end{equation}
To reproduce the $\Lambda$CDM background evolution, we must set
\begin{equation}
    \dfrac{c_1}{c_2} = 6\frac{\Omega_{\Lambda}}{\Omega_m},    
\end{equation}
where $\Omega_{\Lambda}$ and $\Omega_m$ are the present-day fractional energy density of dark energy (cosmological constant) and the present-day fractional density of non-relativistic matter, respectively. Choosing values of $\Omega_m$ and $\Omega_{\Lambda}$ from an appropriate cosmology suggests $\lvert \bar{R} \rvert \gg m^2$, which simplifies Eq.~\ref{eqn:scalar-field} to
\begin{equation}
    f_R \approx -n\frac{c1}{c^2_2}\left(\frac{m^2}{-R}\right)^{(n+1)}.
\end{equation}
Thus, $n$ and $\frac{c_1}{c^2_2}$ fully specify the $f(R)$ model. $\lvert f_{R0} \rvert$ can be derived from the latter.

In this work, we restrict our attention to $n = 1$, and $\lvert f_{R0} \rvert = 10^{-6}, 10^{-5}, 10^{-4}$ (hereafter referred to as F4, F5, and F6 respectively), which are the most widely studied cases in the literature.  $\lvert f_{R0} \rvert = 10^{-6}$ is found to be secure based on galaxy cluster abundance constraints, whereas $\lvert f_{R0} \rvert = 10^{-4}, 10^{-5}$ are in tension with these constraints and also those from the Solar System \citep[e.g.,][for a review of different constraints]{Lombriser2014,Koyama2016,Vogt2024}. Although some of these models have been ruled out by observational tests, they have been included in this study as a proof of concept to represent the broad characteristics of Chameleon-like MG models. We do not prefer any specific form of the $f(R)$ model, and all these models create distinguishable imprints compared to GR.

\subsection{The need for fast, field-level modified gravity emulators}\label{subsec:need-fast-MG-emulators}
As discussed in Sect.~\ref{sec:intro}, the modifications to gravity discussed above are scale-dependent since the fifth force must be strongly suppressed in high-density environments while enhancing gravity at larger scales. Eq.~\ref{eqn:MG-scalaron} is thus strongly non-linear. This non-linearity means that linear approximations become increasingly inaccurate at late times, making it imperative to use $N$-body simulations to study matter evolution in MG\@. Given an initial matter density field and the functional form of $f(R)$, cosmological simulations of $f(R)$ gravity solve for the scalar field, $f_R$, from Eq.~\ref{eqn:MG-scalaron} and use it in the modified Poisson equation in Eq.~\ref{eqn:MG-potential} to obtain $\Phi$. Differentiating $\Phi$ gives the modified gravitational force that governs the evolution of matter distribution. These equations can then be discretised on a mesh and solved numerically.

Simulations of $f(R)$ and other related MG models give rise to non-linear, elliptic partial differential equations that describe the scalar degree of freedom introduced by these models. This requires a multigrid relaxation method to improve the slow convergence often associated with the commonly employed Newton-Gauss-Seidel relaxation method. However, despite these efficient techniques, MG simulations can still introduce significant computational overhead, and mitigating this issue is generally not trivial. Since this paper does not aim to discuss or resolve these challenges, the reader is referred to previous works, such as \citet{Li2012,Llinares2018}, for more details. The inherent complexity of the MG equations and the necessity of solving them represent a fundamental limit on potential efficiency improvements. Thus, a major development in the MG community has been the development of fast MG emulators that bypass the need to explicitly solve these equations. Cosmological parameter inference based on Monte Carlo techniques may substantially benefit from such fast emulators, as they can require several evaluations \citep[on the order of tens of thousands, e.g.,][]{SpurioMancini2022}.
% Given the need to simulate several (hundred to thousand) realisations to conduct robust cosmological tests of MG, fast MG emulations will add minimal overhead to the execution times of their $\Lambda$CDM counterparts.

As mentioned in Sect.~\ref{sec:intro}, our emphasis is not only on the requirement for speed but also on emulators that generate the entire cosmic web. This capability is particularly sought because it allows accessing all available information in the data for field-level inference or even predefined higher-order statistics beyond the power spectrum. Since the amplitude of the fifth force is dependent on the environment, using such higher-order information could provide stringent constraints on MG models \citep[e.g.,][]{Barreira2017,Bose2018,Armijo2018,Peel2018,Davies2019,Bose2020}. Although $N$-body simulations are currently the most accurate means to study the non-linear evolution of matter, the enormous computational resources required for MG simulations render them less feasible for field-level inference. Therefore, developing a fast field-level MG emulator is an important advancement that will facilitate robust cosmological tests of MG.

% but this is a much milder issue for $\Lambda$CDM simulations. Based on this observation, our machine learning approach emulates $f(R)$ fields starting from $\Lambda$CDM simulated fields, and may help facilitate robust cosmological tests of MG.

While this paper focuses only on $f(R)$ gravity, the challenges encountered in $f(R)$ simulations are conceptually analogous to those in various other modified gravity models; therefore, the motivation to create an MG emulator is universal.

\subsection{Generative Adversarial Networks}
A brief background on generative adversarial networks (GANs) and conditional GANs is presented. Consequently, the \texttt{BicycleGAN} approach is introduced, which is a type of conditional GAN.

\subsubsection{Background}

Generative adversarial networks \citep[GANs,][]{Goodfellow2014} are part of the larger class of generative models that aim to capture the statistical distribution of training data and consequently be used to generate synthetic samples based on the learnt distribution. A GAN consists of a discriminator ($D$) and a generator ($G$), most commonly described as deep convolutional architectures in recent years. Assuming image data, the generator creates a synthetic (or fake) image from input noise; $G: \mathbf{z} \rightarrow G(\mathbf{z})$, where $\mathbf{z}$ is a random variable, typically chosen to be from a Gaussian distribution for ease of sampling, and the discriminator outputs the probability that a given image comes from the real rather than the fake data distribution; $D: \mathbf{x} \rightarrow D(\mathbf{x}) \in [0, 1]$. Both models are simultaneously trained using the backpropagation technique and try to fool each other by playing a minimax (or a zero-sum) game: the discriminator tries to assign high probabilities for real images and low probabilities for fake images, and the generator aims to generate image samples that fool the discriminator, i.e., making it output high probabilities to fake images. Training of both models is performed until neither model shows signs of improvement: the generator successfully generates realistic samples, and the discriminator cannot differentiate between real and fake images. Extensive studies over the past several years have concluded that such a `stable' state is challenging to achieve in practice, so the training is generally stopped using empirical guidelines, such as stopping when the quality of the generated images no longer improves.

\citet{Mirza2014} introduced conditional GANs and showed that conditioning the discriminator and generator with additional information, such as labels or images from a different domain, can advance the capabilities of traditional GANs. Unlike traditional (unsupervised) GANs, the image generation process in conditional GANs is controlled by the conditioning information, which is beneficial or naturally more suitable for many applications. An appealing task where conditional GANs have received attention is image-to-image translation, where a GAN is trained to translate an image from an input domain to an image from an output domain.

\subsubsection{BicycleGAN}\label{sec:bicyclegan}

An architecture analogous to BicycleGAN is the (more familiar) \texttt{pix2pix} framework. \texttt{pix2pix} is a general-purpose approach designed to address the image-to-image translation problem, where the GAN is conditioned on the image from the input domain. Although this approach excelled in perceptual realism (visually indistinguishable to the human eye from real-world scenarios) in various tasks, an unaddressed question was the expression of multiple modalities: the ability to map an input image to multiple possible outputs from the output domain, which is natural for many image-to-image translation tasks. However, \texttt{pix2pix} deterministically maps an input image to an output image. Modifying \texttt{pix2pix} to take in a random noise vector along with the input image to introduce non-determinism has been unsuccessful since the generator skilfully ignores the noise \citep{Isola2017}. BicycleGAN \citep{Zhu2017} was designed primarily to express multiple modalities as outputs while maintaining perceptual realism. In the context of this study, in the discussion that follows, $\mathbf{A}$ is the GR field, and $\mathbf{B}$ is the corresponding $f(R)$ field.

The BicycleGAN architecture contains three networks: a discriminator, $D$, a generator, $G$, and an encoder, $E$. BicycleGAN is a hybrid of two architectures: the conditional variational autoencoder GAN ({\sc cVAE-GAN}) and the conditional latent regressor GAN ({\sc cLR-GAN}). In {\sc cVAE-GAN}, $E$ encodes the ground-truth output domain image, $\mathbf{B}$, to $\mathbf{z}$, and consequently, $G$ uses $\mathbf{z}$ along with the input domain image, $\mathbf{A}$, to reconstruct $\mathbf{B}$. The {\sc cVAE-GAN} loss function consists of three terms: the conditional GAN loss modified for the VAE setting, $\mathcal{L}_{\textrm{VAE-GAN}}$, an L1 loss function to enforce closeness to the ground-truth, $\mathcal{L}_1$, and a regularisation term, $\mathcal{L}_{\textrm{KL}}$, to enforce the closeness of the encoded latent distribution to a standard normal distribution. The three objectives are as follows:
\begin{spreadlines}{8pt}
\begin{gather}
    \begin{multlined}
        \mathcal{L}_{\mathrm{VAE-GAN}}(G, D, E) = \mathbb{E}_{\mathbf{A}, \mathbf{B}} [\log(D(\mathbf{A}, \mathbf{B}))] + \\ \mathbb{E}_{\mathbf{A}, \mathbf{B}, \mathbf{z} \sim E(\mathbf{B})} [\log(1 - D(\mathbf{A}, G(\mathbf{A}, \mathbf{z})))]
    \end{multlined}\label{eqn:cvaegan-1},\\
    \mathcal{L}_1^\mathrm{VAE}(G, E) = \mathbb{E}_{\mathbf{A}, \mathbf{B}, \mathbf{z} \sim E(\mathbf{B})} \lVert\mathbf{B} - G(\mathbf{A}, \mathbf{z}) \lVert_1) \label{eqn:cvaegan-2}, \\
    \mathcal{L}_{\mathrm{KL}}(E) = \mathbb{E}_{\mathbf{B}} [\mathcal{D}_{\mathrm{KL}}(E(\mathbf{B}) \,\lVert\, \mathcal{N}(0, 1))].
\end{gather}
\end{spreadlines}
The final cVAE-GAN objective is thus:
\begin{multline}
    \min\limits_{G, E}\max\limits_{D}\mathcal{L}_{\mathrm{cVAE-GAN}} = \mathcal{L}_{\mathrm{VAE-GAN}}(G, D, E)  + \lambda_{\mathrm{image}}\mathcal{L}_1^\mathrm{VAE}(G, E) + \\ \lambda_{\mathrm{KL}}\mathcal{L}_{\mathrm{KL}}(E),
\end{multline}
$\lambda_{\mathrm{KL}}$ and $\lambda_{\mathrm{image}}$ being hyperparameters.

In {\sc cLR-GAN}, $G$ is given a random latent vector drawn from a standard normal distribution, and $G$ and $D$ are trained. This situation is similar to traditional unsupervised GANs. However, the encoder also aims to recover the latent vector from the generated image. The {\sc cLR-GAN} objective thus consists of two terms:
\begin{spreadlines}{8pt}
\begin{gather}
    \begin{multlined}
        \mathcal{L}_{\mathrm{GAN}}(G, D) = \mathbb{E}_{\mathbf{A}, \mathbf{B}} [\log(D(\mathbf{A}, \mathbf{B}))] + \\ \mathbb{E}_{\mathbf{A}, \mathbf{z} \sim p(\mathbf{z})} [\log(1 - D(\mathbf{A}, G(\mathbf{A}, \mathbf{z})))]
    \end{multlined}\label{eqn:clrgan-1}, \\
    \mathcal{L}_1^{\mathrm{latent}}(G, E) = \mathbb{E}_{\mathbf{A}, \mathbf{z} \sim p(\mathbf{z})} \lVert z - E(G(\mathbf{A}, \mathbf{z})) \lVert_1,
\end{gather}
\end{spreadlines}
so that the final cLR-GAN objective is given by:
\begin{multline}
    \min\limits_{G, E}\max\limits_{D}\mathcal{L}_{\mathrm{cLR-GAN}} = \mathcal{L}_{\mathrm{GAN}}(G, D) + \lambda_{\mathrm{latent}}\mathcal{L}_1^{\mathrm{latent}}(G, E),
\end{multline}
where $\lambda_{\mathrm{latent}}$ is a hyperparameter.
In this case, $G$ and $E$ may cooperate, so the generator could produce an unknown signal that the encoder may learn. Only $G$ is updated practically during training to alleviate this potential issue.

BicycleGAN provides an explicit way to prevent mode collapse (i.e., a many-to-one mapping from the latent representation to the output image) since by combining the two approaches discussed above, bijection is ensured between the latent space and the output space \citep{Zhu2017}. \texttt{pix2pix}, although a conditional GAN such as the BicycleGAN, may suffer from mode collapse since it does not possess an equivalent definition of latent space and, hence, does not have bijection.

\section{Method}\label{sec:method}
Having discussed the motivation for using GANs as modified gravity emulators and the scope of this study, we now discuss the implementation details.

\subsection{Dataset}
\subsubsection{\texorpdfstring{$N$}{N}-body simulations}\label{sec:sim-specifics}

Simulations are performed using MG-GLAM \citep{Ruan2022}, an efficient extension of the GLAM \citep{Klypin2018} parallel particle mesh (PPM) $N$-body code. The implementation of the $f(R)$ modified gravity model in MG-GLAM follows \citet{Hu2007}. The GR and $f(R)$ simulations are run based on a flag in MG-GLAM that disables the modified gravity solvers when not required so that the GR simulations default to the {\sc GLAM} implementation. For all simulations, a small box size, $L = 128$ $h^{-1}$ Mpc is used; the number of particles in one dimension, $N_{p} = 256$, and the number of grid points in one dimension, $N_g = 512$, giving a small force resolution of $\Delta x = L / N_g = 0.25 h^{-1}$ Mpc. Only simulation outputs at $z = 0$ are considered, and we consider $N$-point summary statistics defined in real space only.

The cosmological parameters used for all simulations are adopted from the Planck 2015 cosmology \citep{Planck2015}: $\{\Omega_{m}, \Omega_{\Lambda}, h, n_s, \sigma_8\} = \{0.3089, 0.6911, 0.6774, 0.9667, 0.8159\}$. Both GR and $f(R)$ simulations start with the same initial conditions at an initial redshift, $z_{i} = 100$ using the Zel'dovich approximation \citep{Li2011} and evolve to $z = 0$. The same assumption of initial condition is fair since the differences in matter clustering in GR and $f(R)$ are negligible at the initial redshift and allow accurate pinpointing of differences in matter clustering at late times \citep{Li2013}. Five distinct realisations of GR and F4 simulations are run each to generate sufficient training data for the neural network application. Two realisations of F5 and F6 simulations are run each: fewer simulations are needed for F5 and F6 since our model is only trained using GR and F4 simulations. For F5 and F6, only one realisation is used for validation and the other for testing (see Sect.~\ref{sec:latent-extrapolate}).

\subsubsection{Dataset preparation for machine learning}\label{subsubsec:datasetPrep}
Following the simulations, the Delaunay Tessellation Field Estimator (DTFE) code of \citet{Cautun2011} is used (see also \citealt{2000A&A...363L..29S}; \citealt{2009LNP...665..291V}) to estimate the density and velocity divergence fields and to interpolate them on a grid of size $512^3$. Volume-averaged fields are used because density estimates using volume averaging reduce the effects of Poisson noise, particularly in high-density regions. For velocity divergence, volume averaging does not assign zero velocity to empty cells, leading to an unbiased power spectrum and better noise properties than mass-weighting schemes \citep{Pueblas2009}.

Two-dimensional slices, each of thickness $0.25$ $h^{-1}$ Mpc, are extracted from each of the three axes of the 3D density and velocity divergence fields, yielding 1536 two-dimensional slices from each realisation of a simulation\footnote{Our internal experiments confirm that the expected differences in the 3D matter clustering of GR and $f(R)$ (quantified by the power spectrum) are qualitatively well reproduced even with these 2D slices, even though 2D slices contain lower number of available modes.}. The 2D slices are not randomly split into training, validation and testing sets since doing so can overestimate model performance \citep{Ribli2019}. Instead, slices extracted from three field realisations of GR and F4 are used for training; one is used for validation, and the other for testing. The resulting images have a spatial resolution of $128 \times 128$ $h^{-1}$ Mpc (or $512 \times 512$ pixels). We have opted to extract subdivisions of spatial size $256 \times 256$ pixels from each image for a lighter weight model.

The generator architecture contains a Tanh activation at the end (see Sect.~\ref{subsubsec:architecture}), so the images must be transformed to the $[-1, +1]$ range before being passed as input to the GAN\@. A reversible transformation scheme is sought so that summary statistics can be calculated on the inverse transformed images after generating the outputs from the GAN. The non-linear, log-like transformation of \citet{Rodriguez2018} is used for the density fields, given by: $s(x) = \dfrac{2x}{x + t} - 1$, where $x$ is the original density field, and $t$ is a free parameter with $t > 0$. Typical values of $t$ in the literature using this transformation have ranged from $t = 4$ to 45. Small (high) values of $t$ preserve low-density (high-density) features better \citep[see Appendix B of][]{Feder2020}. The differences between GR and $f(R)$ in the matter density distribution are most prominent in the low-density regime, where the scalar field is not screened \citep[e.g.,][]{Cataneo2022}, so a small $t$ value is preferred, e.g., $t = 4$. At the same time, high-density regions have the greatest impact on summary statistics such as the power spectrum, so a small $t$ value may lead to lower agreement in summary statistics. Since preserving low-density regions is more critical in our application, we have used $t = 7$ as a convenient trade-off between these two considerations. The above transformation does not restrict the range to $[-1, +1]$ for the velocity divergence fields since they contain negative values. Thus, for velocity divergence, we instead linearly scale the values by a constant factor, similar to \citet{Harrington2022}. An inspection of the distribution of velocity divergences suggested that they have heavy tails and an asymmetrical shape around zero. Thus, we use the following transformation:
\[
    s(x)= 
\begin{cases}
    -1,& \text{if } x\leq-900 \,\mathrm{km}$ $\mathrm{s}^{-1}$ $\mathrm{Mpc}^{-1}\\
    1,              & \text{else if } x\geq400 \,\mathrm{km}$ $\mathrm{s}^{-1}$ $\mathrm{Mpc}^{-1}\\
    \frac{x}{900},& \text{else if } x < 0 \,\mathrm{km}$ $\mathrm{s}^{-1}$ $\mathrm{Mpc}^{-1}\\
    \frac{x}{400},& \text{else if } x > 0 \,\mathrm{km}$ $\mathrm{s}^{-1}$ $\mathrm{Mpc}^{-1},
\end{cases}
\]
where the values -900 and 400 are found from visual inspection of the distribution of velocity divergences that cover almost the entire dynamic range of values. As a result, the inverse transformation will not preserve values beyond the range $[-900, 400]$ $\mathrm{km}$ $\mathrm{s}^{-1}$ $\mathrm{Mpc}^{-1}$. Although the consequence is that some information is lost, we have observed that the summary statistics are largely unaffected by our choice of transformation and inverse transformation.

Our experiments have suggested that, in general, the type of data transformation is crucial since the high dynamic range of values in these cosmological fields and their non-linearity must be normalised to a restricted range, which is a difficult task. Both transformation schemes described above have been settled after extensive experiments with different transformation approaches and their parameter choices.

\subsection{Model implementation}
\subsubsection{Architecture}\label{subsubsec:architecture}
The generator is a U-Net \citep{Ronneberger2015}, an encoder-decoder architecture with skip connections, which allows for recovering spatial details lost during downsampling due to the information flow from the encoder to the decoder. Eight downsampling and eight upsampling layers are used. Following the original BicycleGAN paper, the latent vector, $\mathbf{z}$, is added to each intermediate layer of the network. Dropout is used to prevent overfitting during training. The last layer of the generator is the Tanh operation to restrict the range of the images to $[-1, +1]$. The discriminator contains two PatchGAN discriminators that look at different spatial scales. The encoder is a ResNet with four residual blocks. Each convolutional layer has a kernel size of 4, leaky ReLU activation for the discriminator, the encoder, and the encoder branch of the generator with a negative slope of 0.2 and non-leaky for the decoder in the generator, and instance normalization \citep{Ulyanov2016} is used.
% (what scales both look at)

We have modified the overall architecture by introducing `attention', a modern machine learning technique that selectively focuses only on the relevant parts of the data based on their importance \citep{Bahdanau2014}. The first motivation to use attention is specific to our application: the input domain (GR) and the output domain ($f(R)$) density and velocity divergence fields have intricate differences that can be difficult to capture in the spatial domain\footnote{The GR and $f(R)$ fields appear very similar visually, and differences are only apparent when viewed closer. See Figs.~\ref{fig:f4-den}--\ref{fig:f6-veldiv} for examples.}. Therefore, attention can provide an effective solution to learning to pay attention to features that help distinguish GR from $f(R)$. Our second motivation is based on the fact that the inclusion of attention in GANs has previously shown that it improves the image quality generated \citep[e.g.,][]{Zhang2018}. The architecture has been modified to include the Convolutional Block Attention Module \citep[{\sc CBAM}; ][]{Woo2018}, which is a lightweight module that introduces channel and spatial attention modules in each convolutional block of the base architecture. The primary function of {\sc CBAM} is to refine the intermediate feature maps adaptively by learning `what' to attend to in the channels of an image and `which' regions to attend to in the spatial region of an image. We insert the CBAM into the bottleneck and the last two upsampling layers of the generator, and the first convolutional block of the discriminator and the encoder. In addition, we use spectral normalisation \citep{Miyato2018} along with instance normalisation in the generator, discriminator, and encoder to stabilise training; the implementation is taken from the \texttt{torchgan} Python library \citep{Pal2021}. The generator (the emulator that generates synthetic samples) contains $\mathcal{O}(10^9)$ parameters, which is 2-3 orders more than some previous GAN-based emulator applications in cosmology; however, the number of parameters may not necessarily be a good indicator of model speed or capacity \citep{Dehghani2021}.

\subsubsection{Training details}
We modify the open-source implementation of BicycleGAN\footnote{\url{https://github.com/junyanz/BicycleGAN}}. The Least Squares GANs variant is used such that Equations~\ref{eqn:cvaegan-1} and \ref{eqn:clrgan-1} are replaced to use the least squares objective instead of binary cross-entropy, mitigating vanishing gradient issues \citep{Mao2016}. The original BicycleGAN paper found that not conditioning the discriminator on input $\mathbf{A}$ provides better results, so we adopt this idea here. The following weights to the different loss terms are used: $\lambda_{\mathrm{image}} = 20$, $\lambda_{\mathrm{latent}} = 0.5$, $\lambda_{\mathrm{KL}} = 0.01$ -- these values are taken from the BicycleGAN paper except that we use two times larger $\lambda_{\mathrm{image}}$ to accentuate closeness of the generated image to the ground-truth. Although the weights of the generator and encoder model are shared, two separate discriminators are used across {\sc cVAE-GAN} and {\sc cLR-GAN}. Training is optimised using the Adam optimiser \citep{Kingma2014} with momentum parameters, $\beta_1 = 0.5$ and $\beta_2 = 0.999$, and a batch size of 1, with these hyperparameters again taken from the BicycleGAN paper. Training is performed for 75 epochs with the same initial learning rate of $2 \times 10^{-4}$ used for the generator, the discriminator, and the encoder. The learning rate is decayed by 50\% after every 25 epochs. The length of the latent code, $\lvert z \lvert = 128$, is used, which is higher than that used in the original paper. The weights of all models are initialised using a Xavier normal distribution. Models are saved after every five epochs of training.

Only two BicycleGANs are trained; one to map from GR to F4 for the density and the other for the velocity divergence. Evaluation on F5 and F6 is performed using an extrapolation of the learnt latent space. More details are presented in Sect.~\ref{sec:latent-extrapolate}. Other strategies, such as training a single BicycleGAN to learn a mapping from GR to F4, F5, and F6 simultaneously or fine-tuning the BicycleGAN to learn the mapping to F5 or F6 after pretraining it to learn F4, were experimented with, but did not perform as well. The former was reliably achieved by \citet{Saadeh2024} in which the strength of $f(R)$ gravity, $\lvert f_{R0} \rvert$, was used as a style parameter, allowing the emulator to handle multiple input domains; our approach achieves a similar objective but without using explicit style parameters.

We complement the (spatial-domain) L1 loss function in Equation~\ref{eqn:cvaegan-2} of {\sc cVAE-GAN} with the focal frequency loss \citep[FFL; ][]{Jiang2020}. FFL enforces frequency-correctness by adapting the loss to pay more attention to frequency components that are difficult to produce. Such a loss term is a natural choice, since cosmological fields carry essential frequency information (quantified commonly by the power spectrum). The FFL loss is given by:
\begin{equation}
    \mathrm{FFL}(G, E) = \frac{1}{MN}\sum_{u=0}^{M-1}\sum_{v=0}^{N-1} w(u, v) \lvert F_r(u, v) - F_f(u, v) \lvert^2,
\end{equation}
where the image's spatial size is $M \times N$ (here, $M = N = 256$; see Sect~\ref{subsubsec:datasetPrep}), $F_r(u, v)$ and $F_f(u, v)$ are the 2D discrete Fourier transforms of the ground-truth, $\mathbf{B}$, and that of the generated image, and $(u, v)$ are the frequency coordinates. Thus, the revised loss term is:
\begin{multline}
    \min\limits_{G, E}\max\limits_{D}\mathcal{L}_{\mathrm{cVAE-GAN}} = \mathcal{L}_{\mathrm{VAE-GAN}}(G, D, E) + \\ \lambda(\frac{1}{2}\mathcal{L}_1^\mathrm{VAE}(G, E) + \frac{1}{2}(m * \mathrm{FFL}(G, E))) + \lambda_{\mathrm{KL}}\mathcal{L}_{\mathrm{KL}}(E).
\end{multline}
The FFL loss term is scaled by $m$ to bring it in the same range as the L1 loss. Here, $m = 100$ is used. This frequency-based loss, combined with the attention mechanism introduced in the previous section, overall proved beneficial for the GR to $f(R)$ mapping.

\subsection{Latent space extrapolation}\label{sec:latent-extrapolate}

Diverse yet realistic and statistically consistent output domain images can be obtained using unconditional GANs by sampling random latent vectors from a prior distribution. It has also been found that interpolations in the generator's latent space show smooth and realistic transformations in the generated images, suggesting that these features actually capture the semantic properties of the data \citep[e.g.,][]{Radford2015}. However, such diversity is usually difficult to achieve using conditional GANs since the model is conditioned on input (i.e., a GR field in our case), so it is possible that the expression of diversity is affected as a result. As discussed in Sect.~\ref{sec:bicyclegan}, a naive attempt to introduce stochasticity in the outputs of \texttt{pix2pix} did not produce the expected diversity. BicycleGAN effectively addressed this problem partly because of its encoder network component and allowed for diversity (as commonly seen in unconditional GANs) while also conditioning on an input image. Thus, unlike conditional GANs such as \texttt{pix2pix} that focus only on the closeness of the generated image to the ground truth, BicycleGANs have more potential to be used for the fast emulation of random $f(R)$ fields consistent with the ground-truth (which can be viewed as random realisations of $f(R)$ fields) from a single GR field, which may facilitate studies that compare the two gravity models using large ensembles of fields.

However, the aim of this work is to emulate a cosmological field under $f(R)$, given its GR counterpart, which is a \emph{paired} image-to-image translation problem. This setup suggests that a random sampling of the latent space, although statistically consistent with $f(R)$ fields, may not exactly match the specific, desired $f(R)$ field. Since the encoder learns to encode the ground-truth images so that the generator uses this encoding to generate predictions that are close to the ground-truth, the latent space learnt by the encoder is, in that sense, `meaningful'. Thus, in this paper, we opt to use this encoded latent space instead of random sampling for the BicycleGAN approach. During testing, the learnt encoder network encodes the true $f(R)$ field, which is used along with the corresponding GR field to make the prediction\footnote{It is to be cautioned that the ground-truth may not be available when the machine learning model is deployed. This approach attempts to test the effectiveness of the encoded latent space and lay the foundation for future exploration.}.

As discussed in Sect.~\ref{sec:intro} and \ref{subsec:fr-theory}, three mappings are considered: GR to F4, GR to F5, and GR to F6. We hypothesise that training three independent BicycleGANs in each case is redundant since the input remains the same across these cases, and the output domains are not drastically different. The latter is because these gravity models describe the same modified gravity physics with only slightly different strengths. To test our hypothesis, we explore latent space extrapolation, where a BicycleGAN is trained to learn the mapping from GR to F4, and the learnt latent space is only extrapolated at test time while generating fields in F5 and F6, using the following rule:
\begin{equation}
    z\prime = z + p \times \dfrac{z - z_{\mathrm{F4}}}{\lVert z - z_{\mathrm{F4}} \rVert}
\end{equation}
where $z\prime$ is the extrapolated latent vector for F5/F6, $z$ is the latent vector originally derived from simulated (or ground-truth) F5/F6 field, $p$ is the scaling factor defining the extent of extrapolation, and $z_{\mathrm{F4}}$ is the encoded latent vector of the simulated F4 field. The idea is that the encoder network (which is trained to learn relevant features from simulated F4 fields) is used to derive latent vectors for simulated F5/F6 fields during prediction by a simple forward pass. However, our internal experiments suggested that simply using these derived latent vectors to predict F5/F6 fields produces inferior emulations, possibly because the encoder was only trained to extract latent vectors from F4 fields. As a result, we use these vectors derived from simulated F5/F6 fields as an initial estimate ($z$) and transform it towards the direction away from $z_{\mathrm{F4}}$ and towards $z$ (denoted by the unit vector $\frac{z - z_{\mathrm{F4}}}{\lVert z - z_{\mathrm{F4}} \rVert}$) to produce $z\prime$ which is used for emulation of F5/F6 fields. We select the value of $p$ based on trial and error, as is commonly required for approaches manipulating the latent space. Internal tests suggested that the optimal values of $p$ were 5 and 12 for F5 and F6 density and 12 and 25 for F5 and F6 velocity divergence, respectively.

\begin{figure*}
% \centering
\hspace{-1.7em}\includegraphics[keepaspectratio,width=0.85\linewidth]{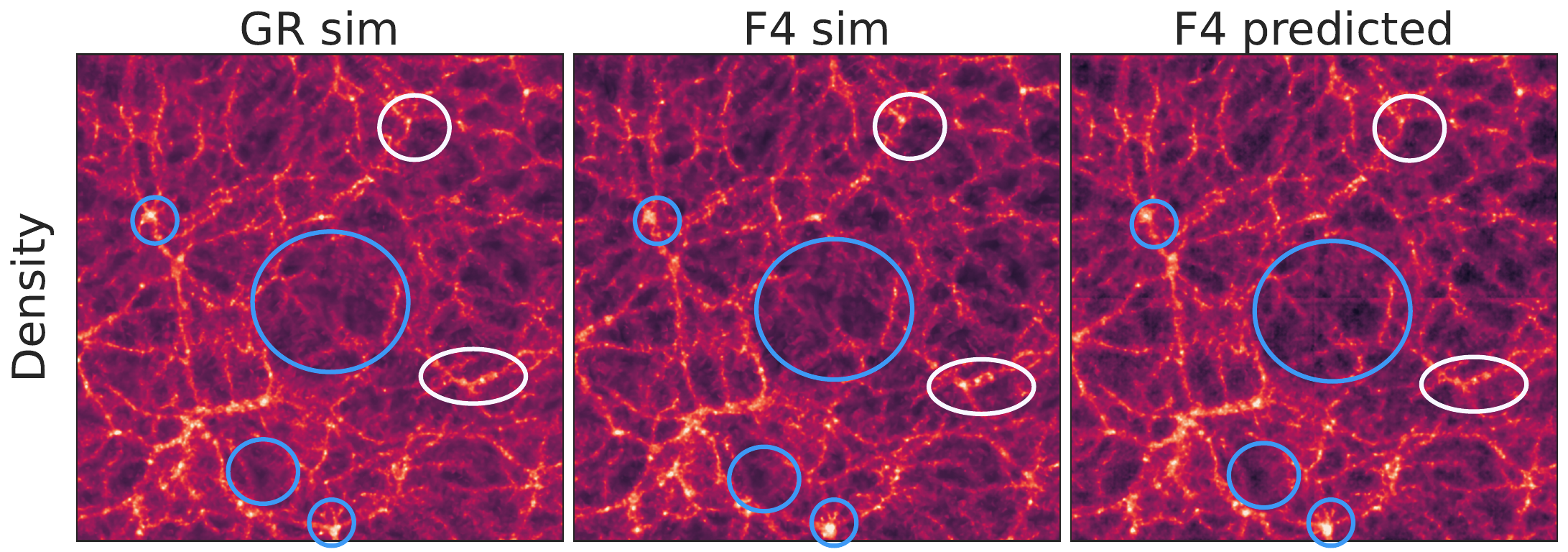}
\includegraphics[keepaspectratio,height=2in]{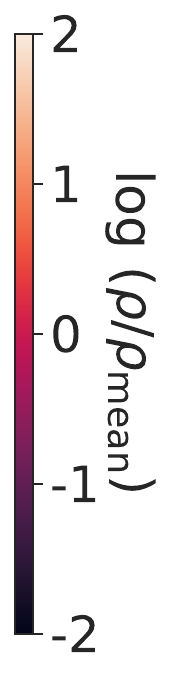}
\includegraphics[keepaspectratio,width=0.85\linewidth]{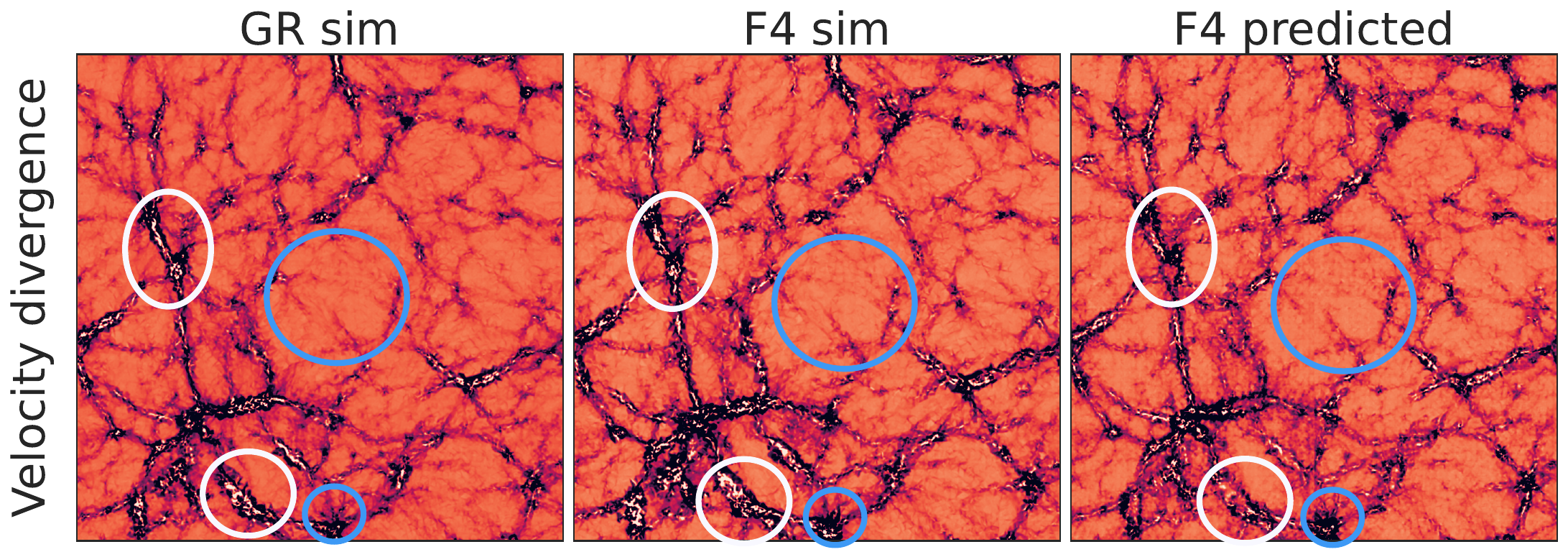}\hspace{0em}
\includegraphics[keepaspectratio,height=1.9in]{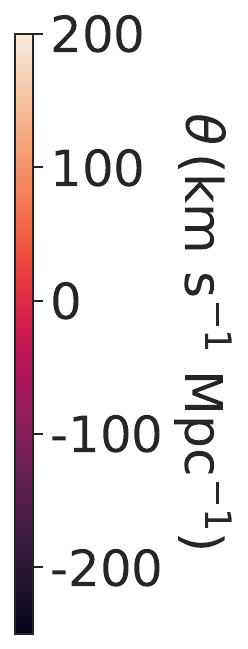}
\caption{Example visualisation of the density (first row) and the velocity divergence (second row) 2D slices for GR and F4 derived from $N$-body simulations (``GR sim'' and ``F4 sim'') and the GAN prediction (``F4 predicted''). The dimensions of all the maps shown are $128 \times 128$ $h^{-1}$ Mpc ($512 \times 512$ pixels) with a thickness of 0.25 $h^{-1}$ Mpc. The GAN model is trained on $256 \times 256$ pixels crops of these maps, so the crops are combined here only for visualisation. Hence, the artifacts at the crops' boundaries for the F4 prediction are not concerning. Our model uses a field from GR sim as input and predicts a corresponding F4 field (F4 pred), with the F4 sim field being the ground truth. Although the differences between GR and F4 sim are small, the F4 prediction qualitatively reproduces several features of interest (blue ellipses) and fails to reproduce certain features (white ellipses) in detail. For each row, the ellipses denote approximately the same region across the three columns. See the main text for a more detailed discussion. For velocity divergence, black (white) colours denote negative (positive) divergences.}
\label{fig:visual-comparison}
\end{figure*}

Thus, instead of training six BicycleGANs (for the density and velocity divergence fields in F4, F5, and F6), only two BicycleGANs are trained (density and velocity divergence fields in F4). F4 is chosen for training instead of F5 or F6 since it shows the largest deviation from GR. Although latent extrapolation is not commonly experimented with, the latent vectors here are expected to encode meaningful information about the output domain images, so extrapolation may not necessarily be unreliable. This idea differs from latent interpolation used previously in, for example, \citet{Tamosiunas2021}. Moreover, multiple rounds of retraining would be necessary if \texttt{pix2pix} were used since it cannot map from a single input domain to multiple output domains. This approach of using trained GANs on qualitatively similar yet different data than it was trained on directly for inference highlights one of the several novelties of our approach.

\section{Results}\label{sec:results}
% After training, all the saved model checkpoints are used on the validation set, and the model with the best performance (see Sect.~\ref{sec:evaluation-metrics} for the metrics) is selected for evaluation on the test set.
After training, all saved model checkpoints are evaluated on the validation set, and the model with the best performance on the validation set (see Sect.~\ref{sec:evaluation-metrics} for metrics) is selected for evaluation on the test set. The reported results are on the test set. For computational reasons, it was decided to only use 200 examples from the validation set during the validation phase.

\subsection{Visual comparison}\label{subsec:visual-comparison}

To present a qualitative description of the results and to appreciate the challenge of reproducing modified gravity from GR, a sample visualisation of the GR and F4 simulation and the F4 prediction of the density and velocity divergence fields is presented in Fig.~\ref{fig:visual-comparison}. Only F4 has been considered here for clarity, as it has the least efficient screening out of the other $f(R)$ models (F5 and F6), thus providing more visually apparent differences.

The large-scale differences between ``GR sim'' and ``F4 sim'' are negligible because both GR and $f(R)$ simulations share the same initial conditions. However, on smaller scales, clusters appear more compact and bulkier in F4 than in GR because of the enhanced gravity in F4, which is well reproduced by the GAN (the bottommost and the leftmost blue ellipses). Thus, voids are also expected to be emptier in F4 than in GR, which is also reproduced well (the other two blue ellipses). However, sometimes the GAN could not accurately reproduce high overdensities in filaments or clusters (white ellipses). The velocity divergence around the filaments and clusters is negative since matter inflows into these regions; however, the divergences become positive inside the filaments and clusters. The latter effect is not reproduced for the most extreme cases (the two white ellipses). The bottommost blue ellipse shows that the divergence is more negative for F4 than GR and that the cluster is bulkier; this is reproduced well by the GAN. It is noticeable that the velocity divergence in the voids is more positive in F4 than in GR due to the enhanced gravity in $f(R)$ \citep[which also empties voids faster, e.g.,][]{Li2013}. The middle blue ellipse (and other voids in the field) show that the divergences in the voids are similar for ``F4 sim'' and ``F4 predicted''\footnote{This difference may be challenging to notice at first; however, it becomes clearer if one tries to compare the overall contrast in these images.}. Thus, the GAN predictions show qualitatively many expected differences between F4 and GR. In the following subsections, we will introduce some quantitative metrics to help assess the quality of our GAN predictions.

% this paper said F4 and F6 was weaker clustering than GR, F5 has stronger than GR: https://academic.oup.com/mnras/article/479/4/4824/5050389.

\begin{figure*}
    \centering
    \begin{subfigure}{0.33\textwidth}
        \centering
        \includegraphics[keepaspectratio,width=\linewidth]{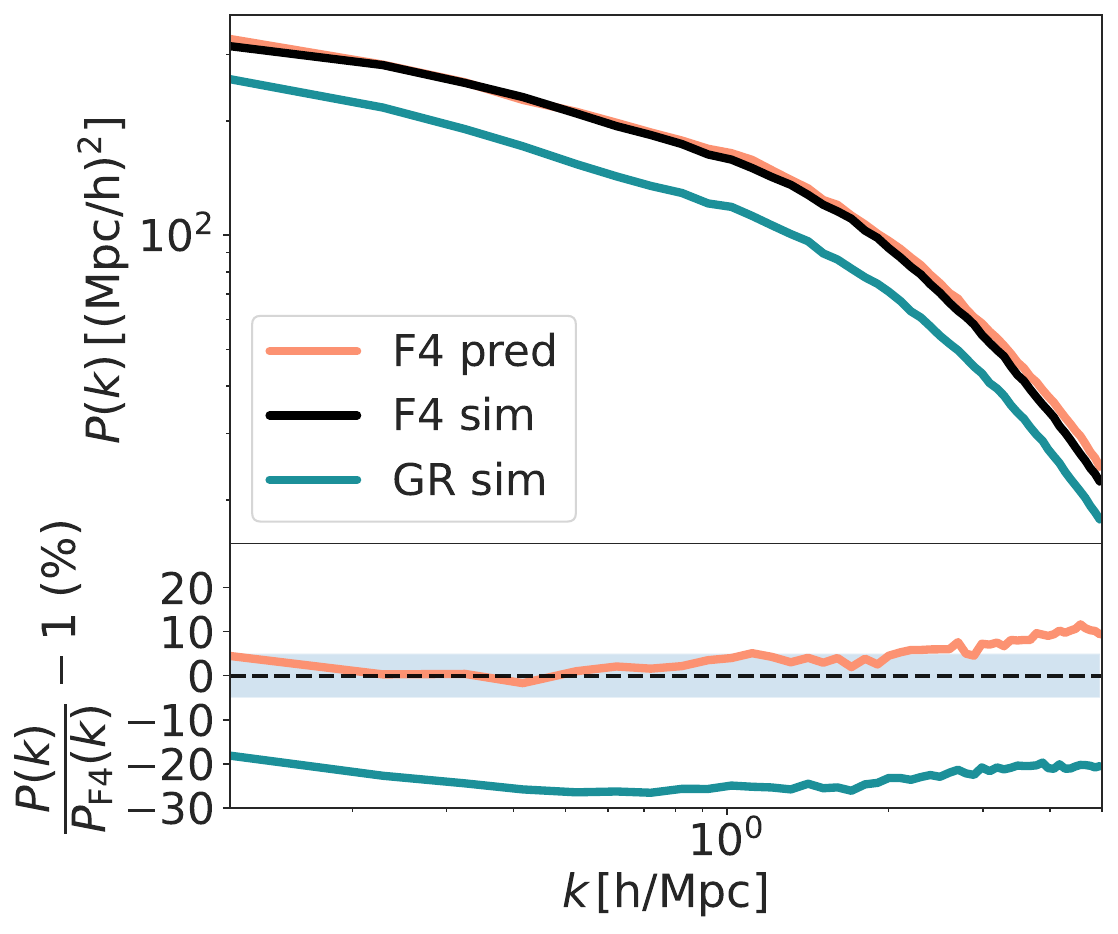}
        \caption{}
        \label{fig:ps-f4-den}
    \end{subfigure}
    \hspace{4em}
    \begin{subfigure}{0.33\textwidth}
        \centering
        \includegraphics[keepaspectratio,width=\linewidth]{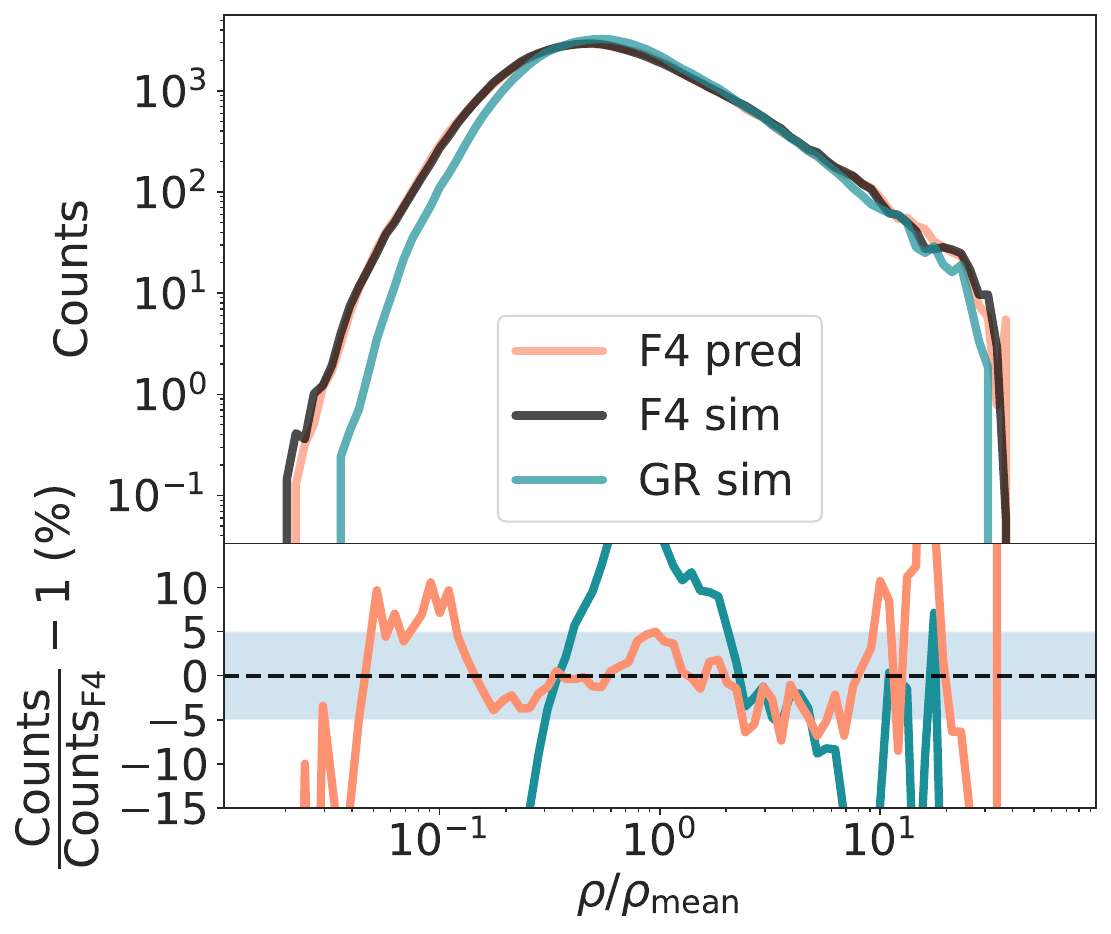}
        \caption{}
        \label{fig:mass-hist-f4-den}
    \end{subfigure}
    \begin{subfigure}{0.33\textwidth}
        \centering
        \includegraphics[keepaspectratio,width=\linewidth]{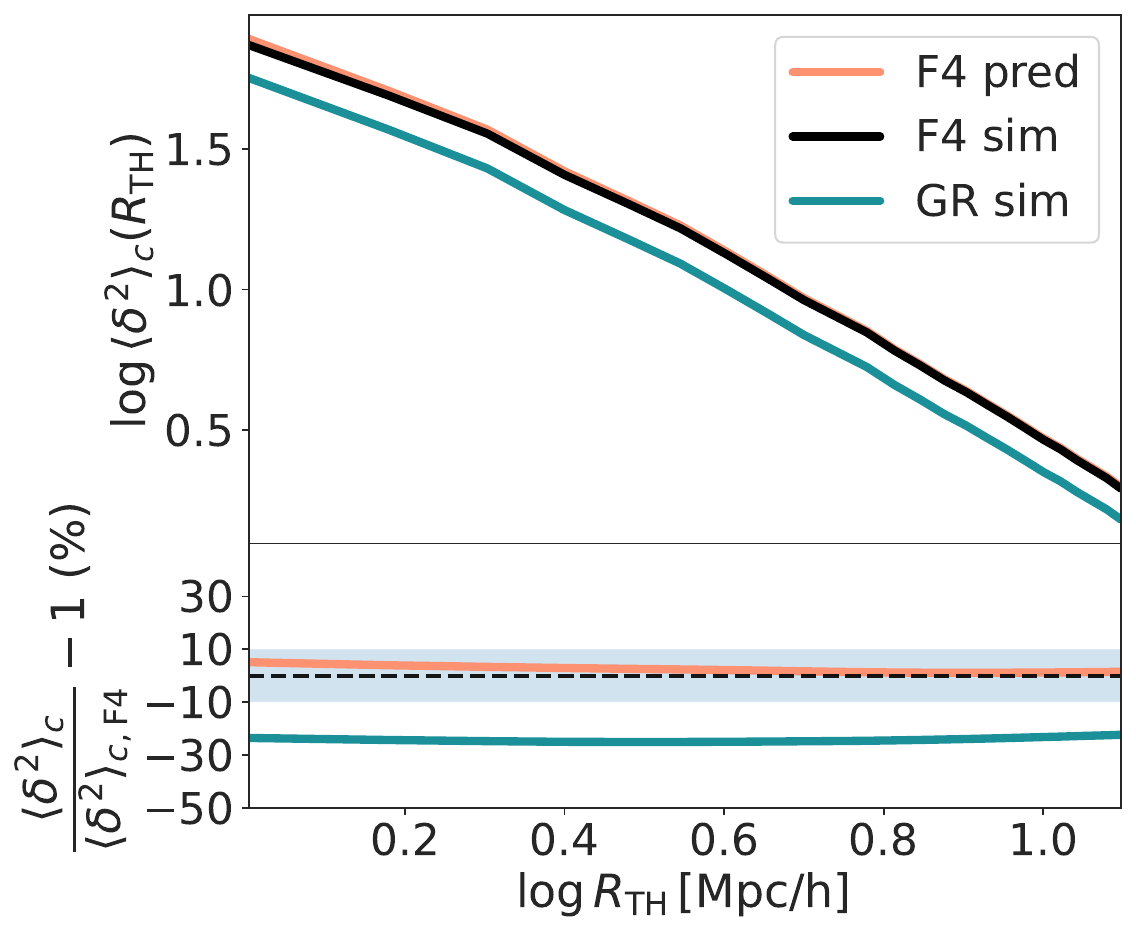}
        \caption{}
        \label{fig:cum-variance-f4-den}
    \end{subfigure}
    \hfill
    \begin{subfigure}{0.33\textwidth}
        \centering
        \includegraphics[keepaspectratio,width=\linewidth]{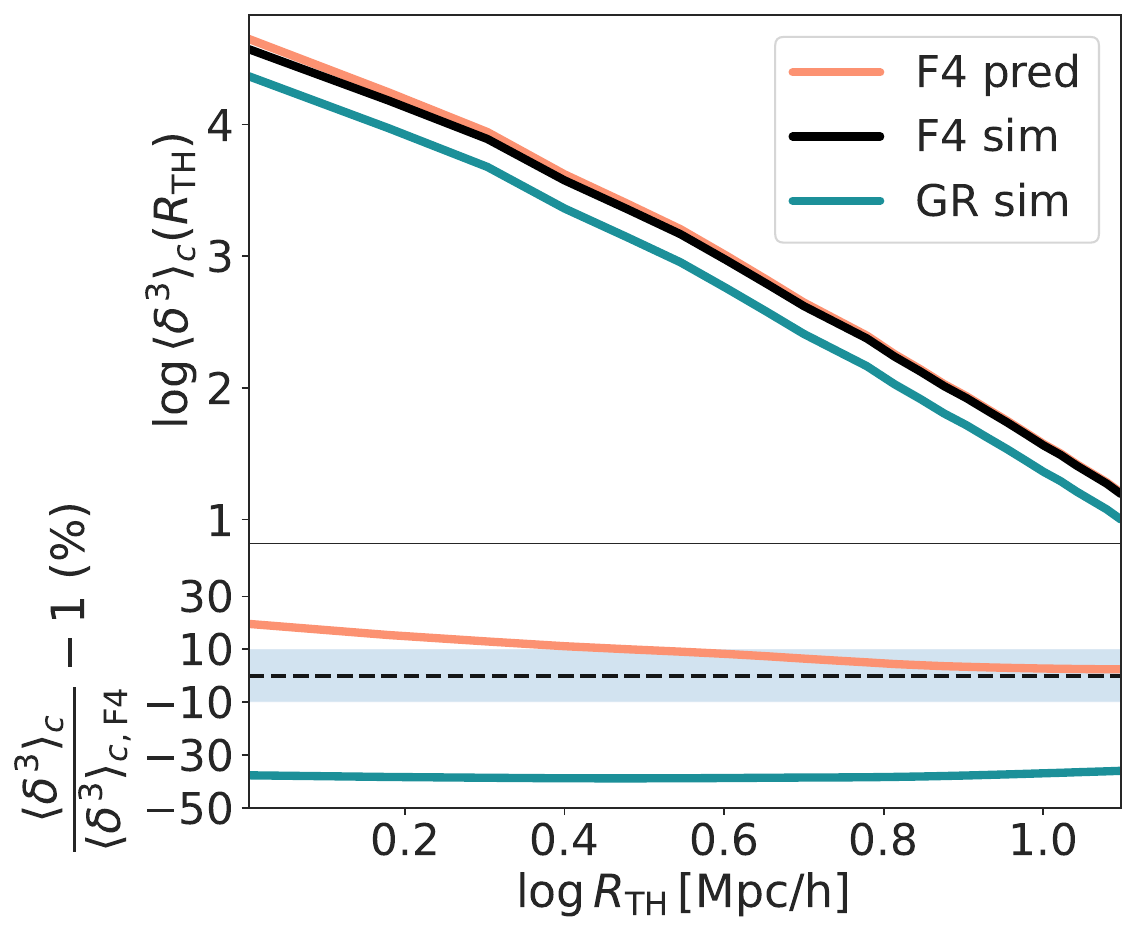}
        \caption{}
        \label{fig:cum-skewnes-f4-den}
    \end{subfigure}
    \hfill
    \begin{subfigure}{0.33\textwidth}
        \centering
        \includegraphics[keepaspectratio,width=\linewidth]{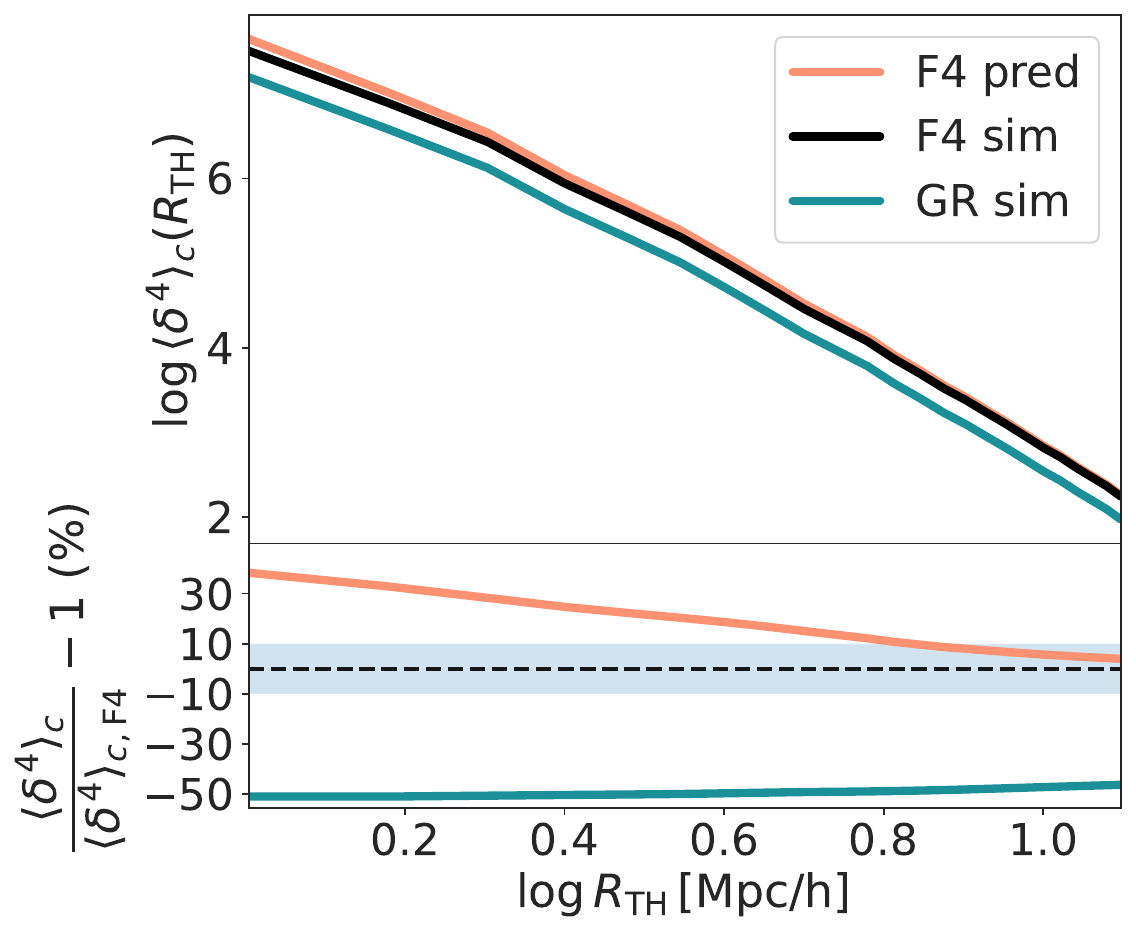}
        \caption{}
        \label{fig:cum-kurtosis-f4-den}
    \end{subfigure}
    \caption{\emph{F4 density}: Comparison of the evaluation metrics between the simulated GR and F4 density fields (GR sim and F4 sim) and the F4 density fields predicted by the GAN (F4 pred). The panels show the different evaluation metrics: (a) the 2D power spectrum, (b) the pixel histogram, (c) the second cumulant (variance), (c) the third cumulant (skewness), and (e) the fourth cumulant (kurtosis). The lower subplots of each panel show the relative difference of the metric with respect to the metric calculated using F4 sim fields. The black dashed horizontal lines denote perfect agreement with F4 sim. The shaded regions in (a) and (b) denote $\pm5\%$ relative difference, and in (c), (d), and (e) denote $\pm10\%$ relative difference. All metrics shown are averaged across 500 randomly selected examples from the test set. For (b), it must be cautioned that logarithmic scaling on the y-axis enlarges deviations in bins with low counts (extreme over- and under-densities).}
    \label{fig:f4-den}
\end{figure*}

\begin{figure*}
    \centering
    \begin{subfigure}{0.33\textwidth}
        \centering
        \includegraphics[keepaspectratio,width=\linewidth]{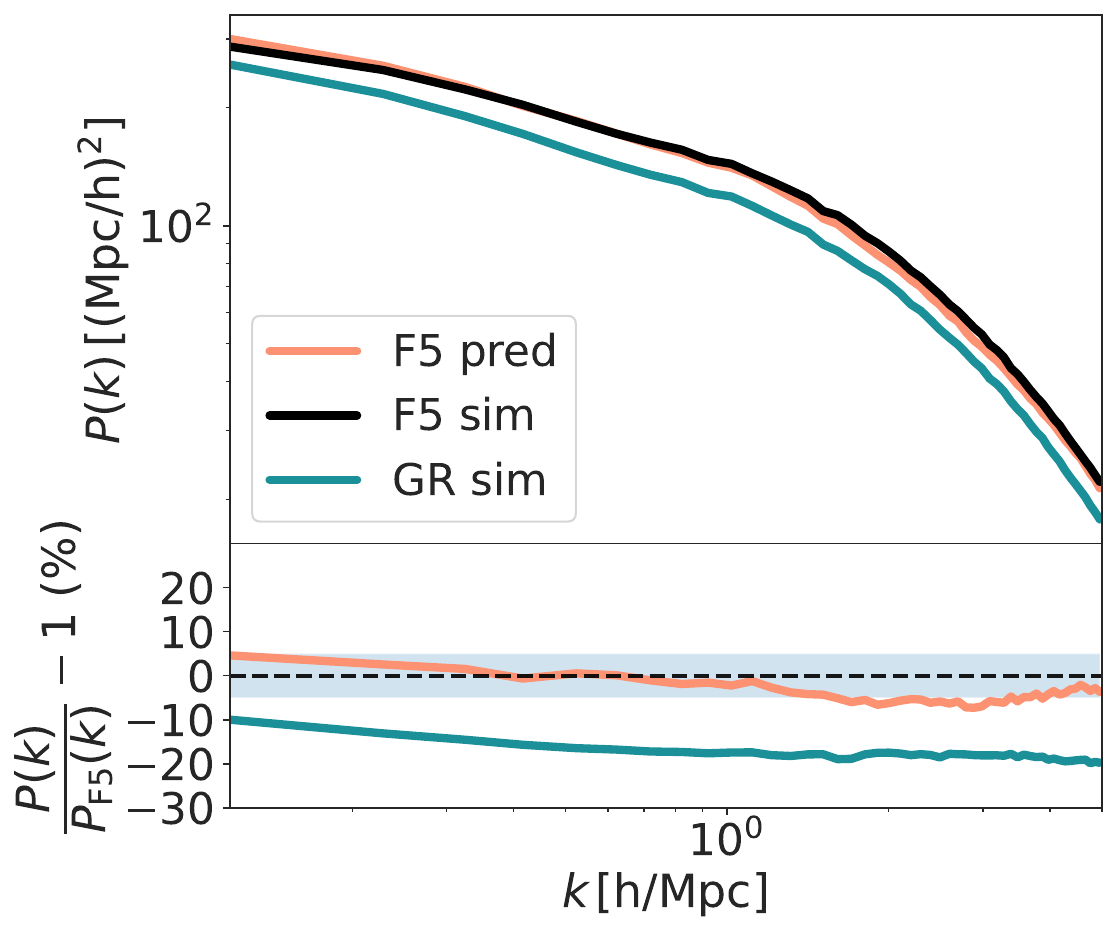}
        \caption{}
        \label{fig:ps-f5-den}
    \end{subfigure}
    \hspace{4em}
    \begin{subfigure}{0.33\textwidth}
        \centering
        \includegraphics[keepaspectratio,width=\linewidth]{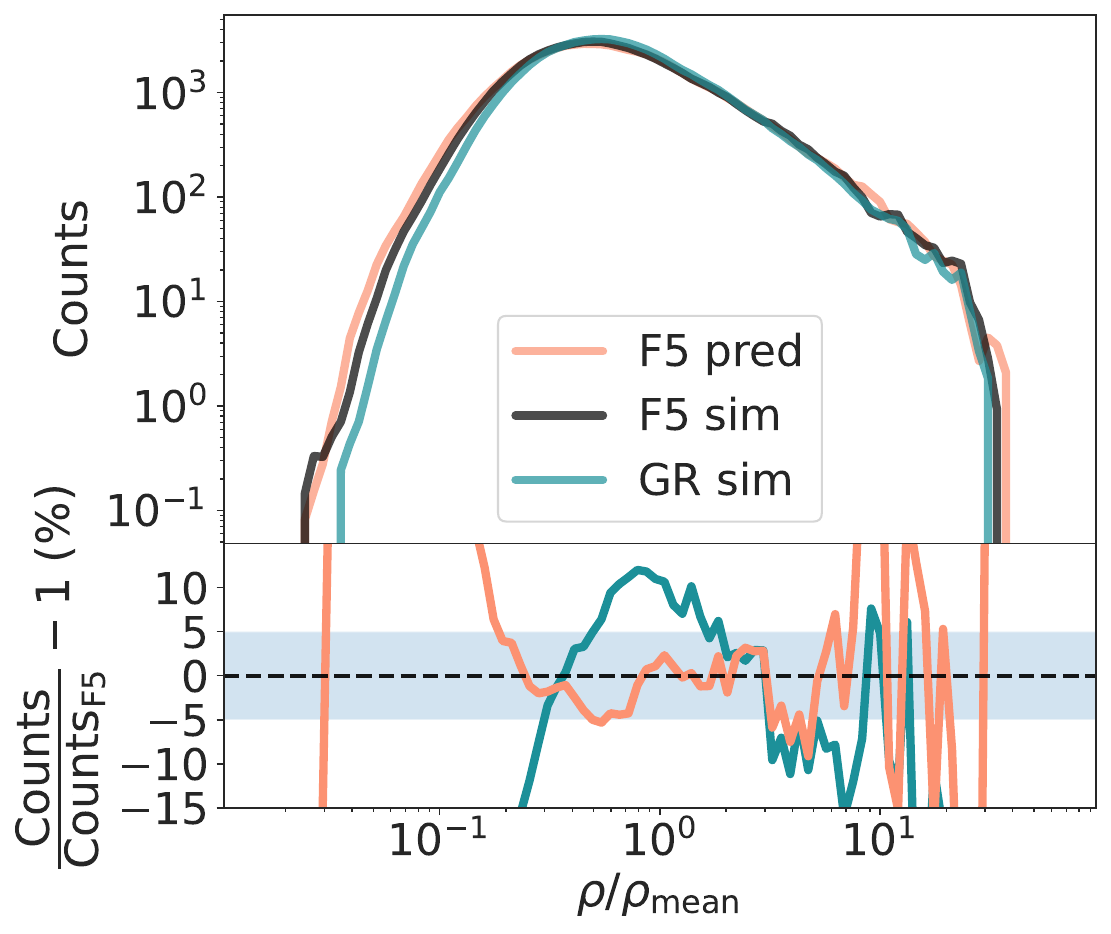}
        \caption{}
        \label{fig:mass-hist-f5-den}
    \end{subfigure}
    \begin{subfigure}{0.33\textwidth}
        \centering
        \includegraphics[keepaspectratio,width=\linewidth]{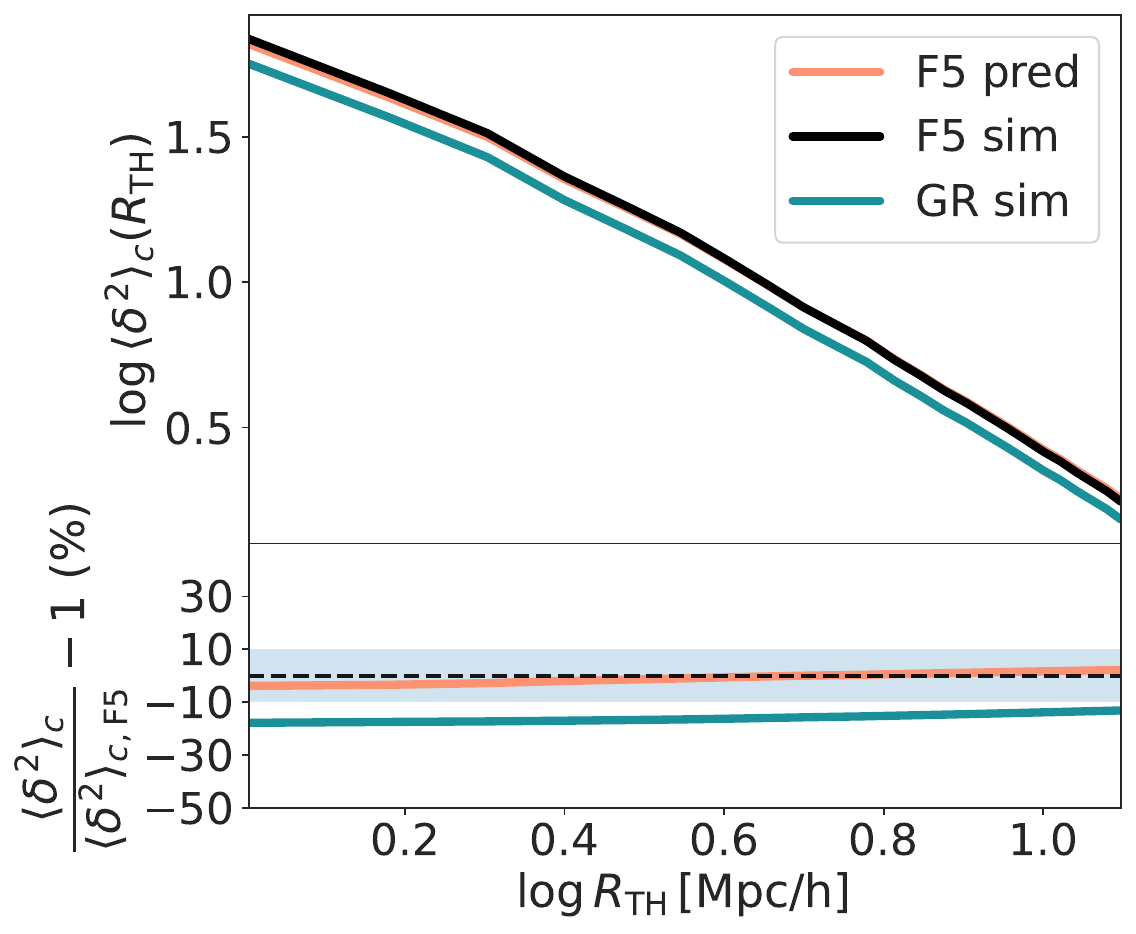}
        \caption{}
        \label{fig:cum-variance-f5-den}
    \end{subfigure}
    \hfill
    \begin{subfigure}{0.33\textwidth}
        \centering
        \includegraphics[keepaspectratio,width=\linewidth]{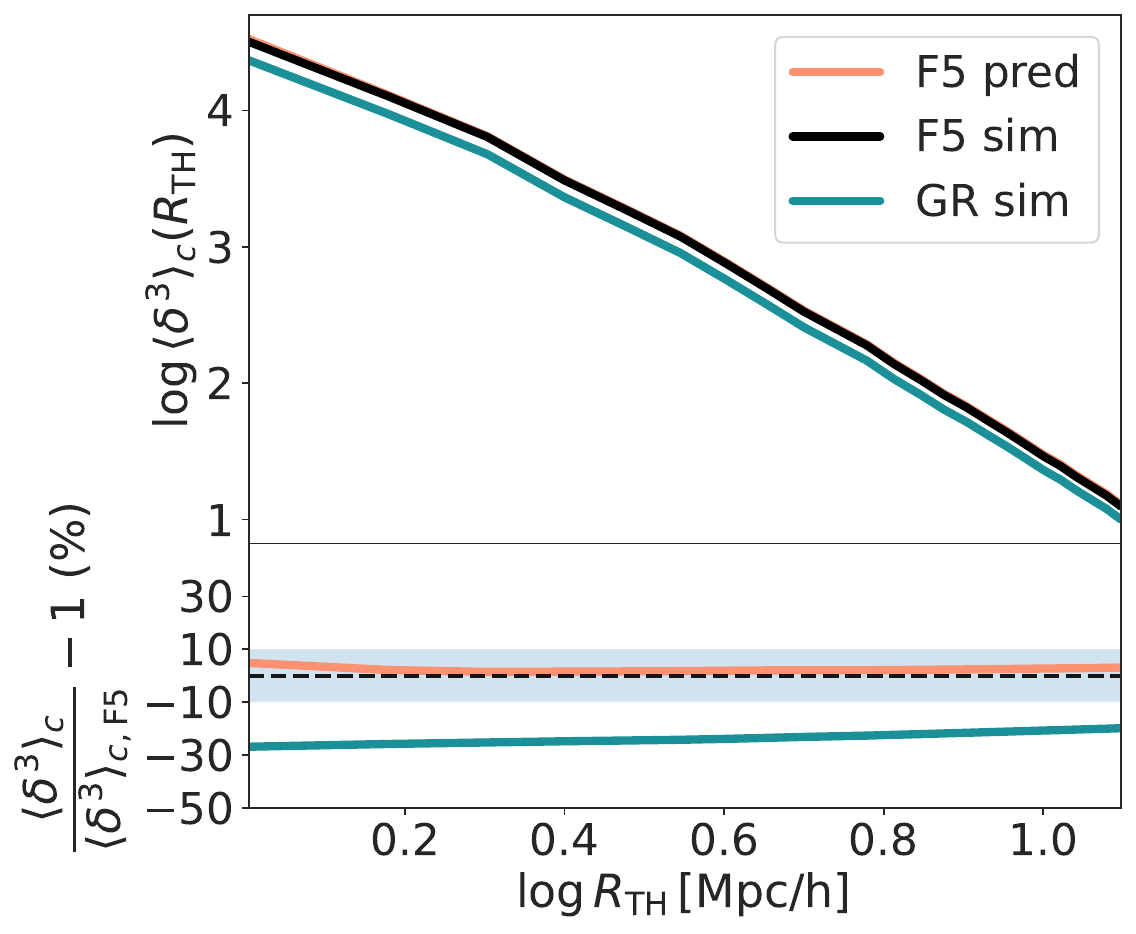}
        \caption{}
        \label{fig:cum-skewnes-f5-den}
    \end{subfigure}
    \hfill
    \begin{subfigure}{0.33\textwidth}
        \centering
        \includegraphics[keepaspectratio,width=\linewidth]{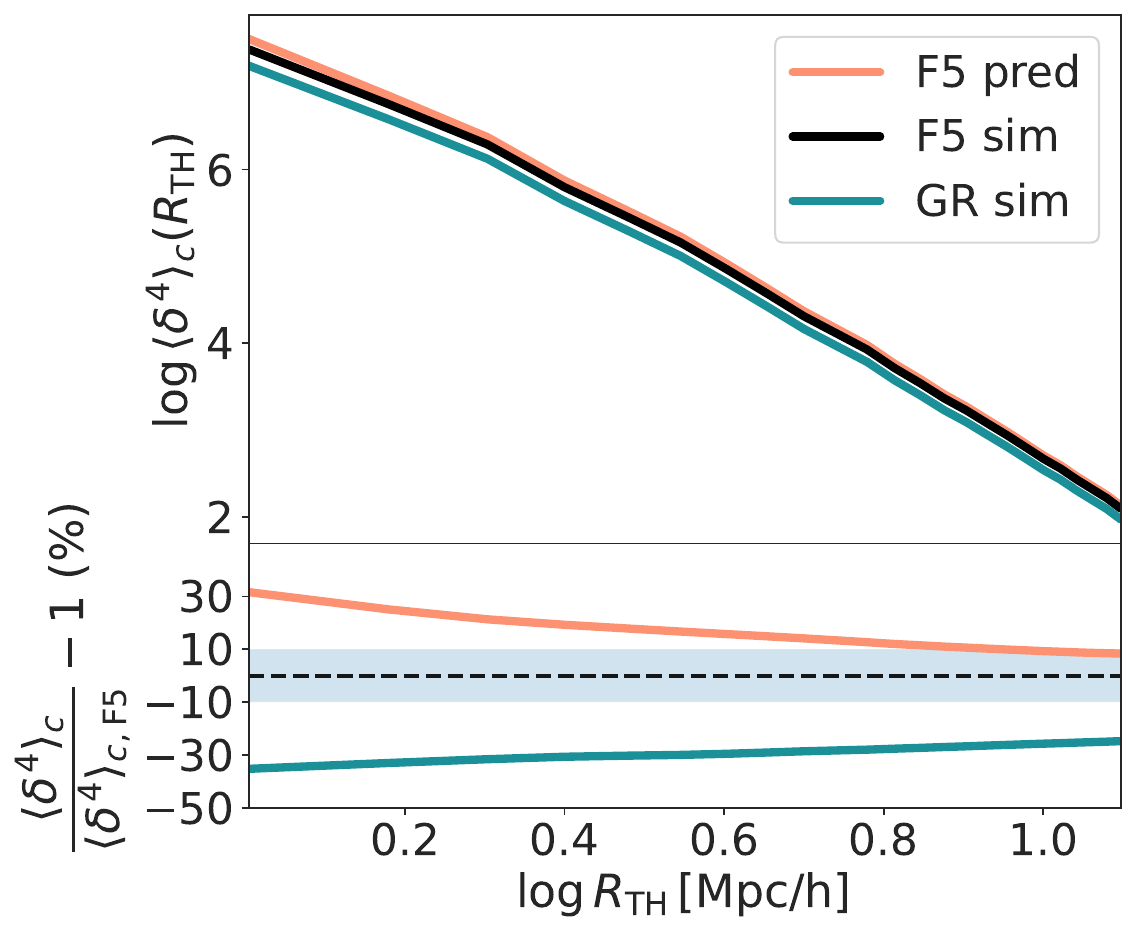}
        \caption{}
        \label{fig:cum-kurtosis-f5-den}
    \end{subfigure}
    \caption{\emph{F5 density}: Comparison of the evaluation metrics between the simulated GR and F5 density fields (GR sim and F5 sim) and the F5 density fields predicted by the GAN (F5 pred). See Fig.~\ref{fig:f4-den} for a description of the panels.}
    \label{fig:f5-den}

    \centering
    \begin{subfigure}{0.33\textwidth}
        \centering
        \includegraphics[keepaspectratio,width=\linewidth]{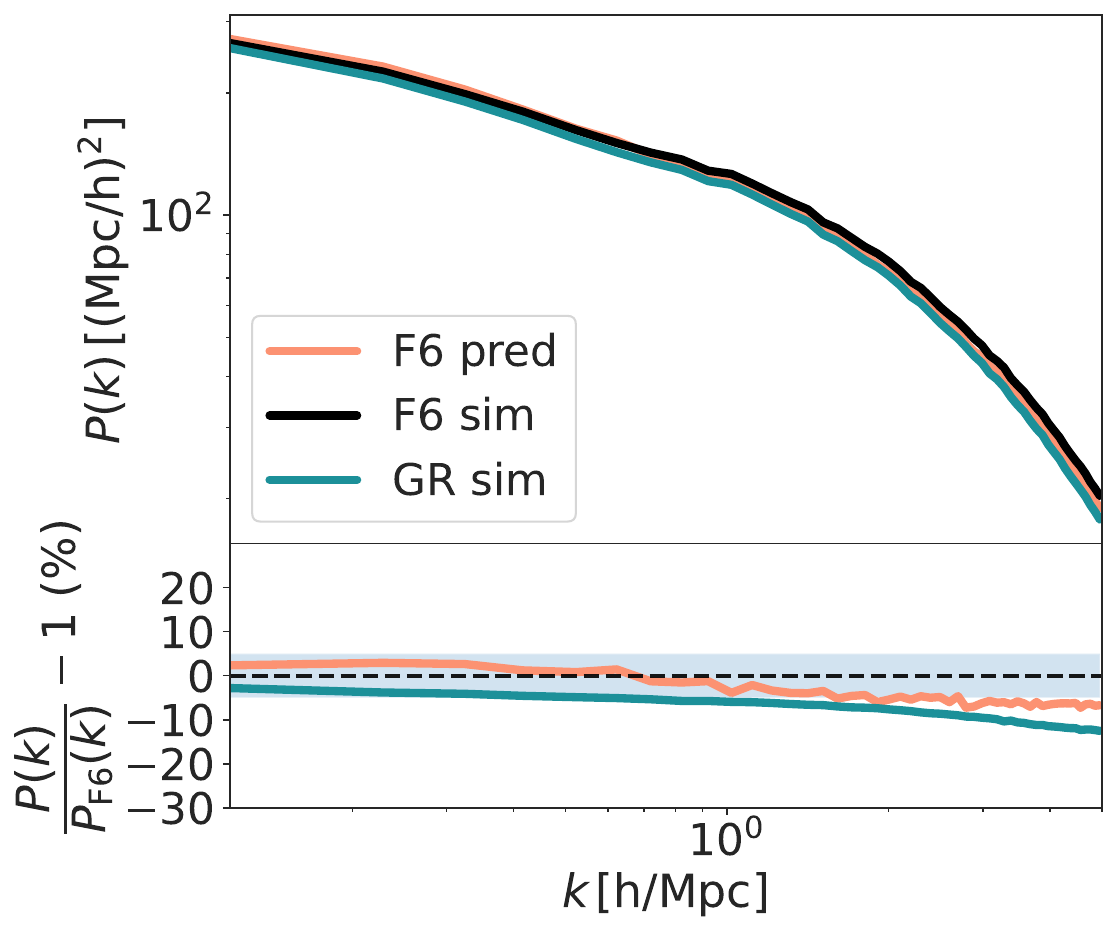}
        \caption{}
        \label{fig:ps-f6-den}
    \end{subfigure}
    \hspace{4em}
    \begin{subfigure}{0.33\textwidth}
        \centering
        \includegraphics[keepaspectratio,width=\linewidth]{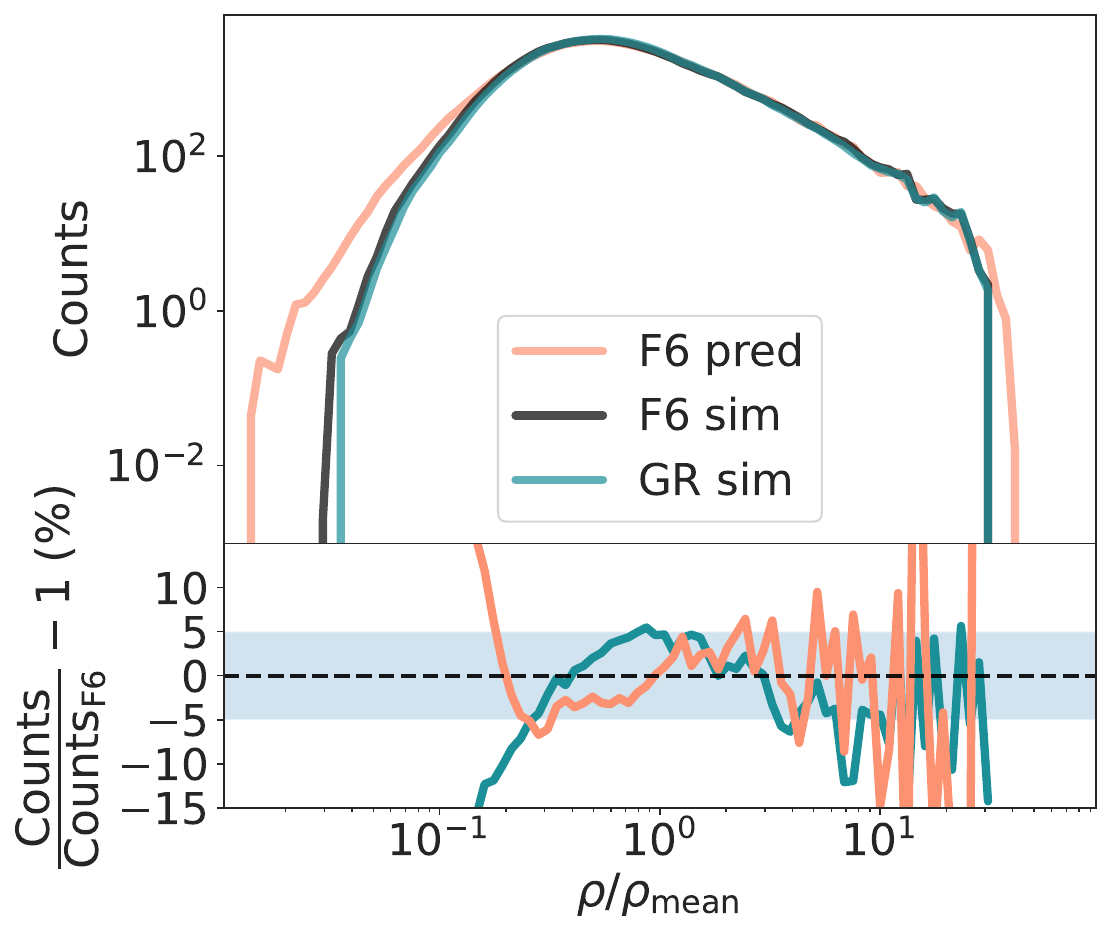}
        \caption{}
        \label{fig:mass-hist-f6-den}
    \end{subfigure}
    \begin{subfigure}{0.33\textwidth}
        \centering
        \includegraphics[keepaspectratio,width=\linewidth]{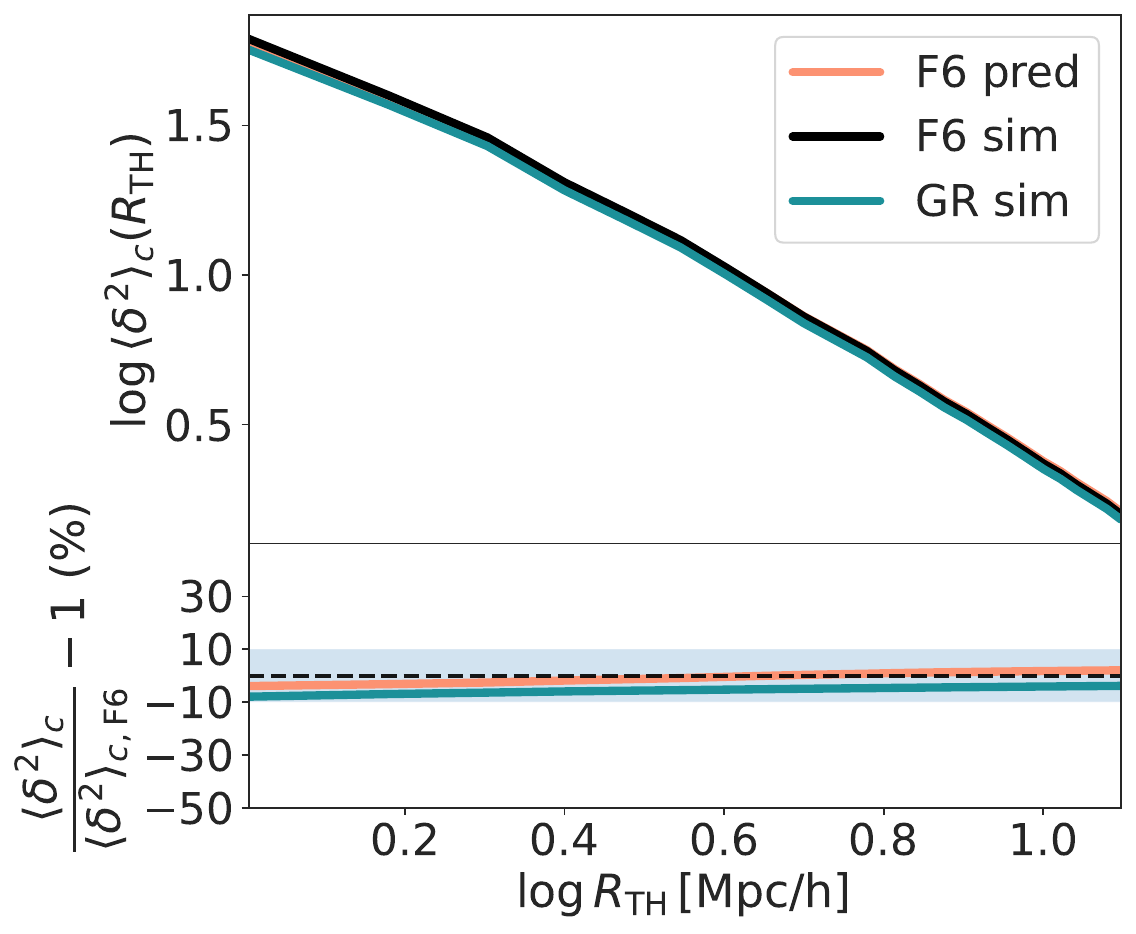}
        \caption{}
        \label{fig:cum-variance-f6-den}
    \end{subfigure}
    \hfill
    \begin{subfigure}{0.33\textwidth}
        \centering
        \includegraphics[keepaspectratio,width=\linewidth]{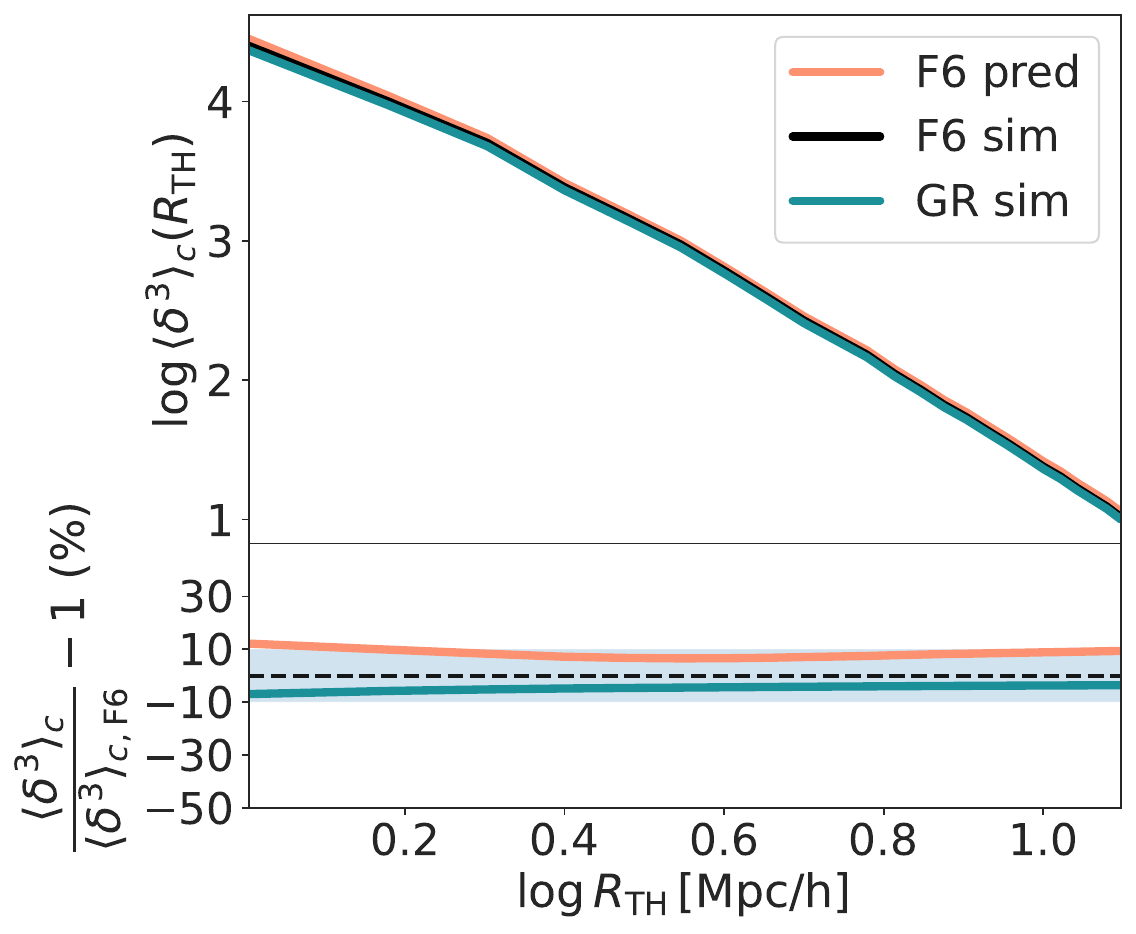}
        \caption{}
        \label{fig:cum-skewnes-f6-den}
    \end{subfigure}
    \hfill
    \begin{subfigure}{0.33\textwidth}
        \centering
        \includegraphics[keepaspectratio,width=\linewidth]{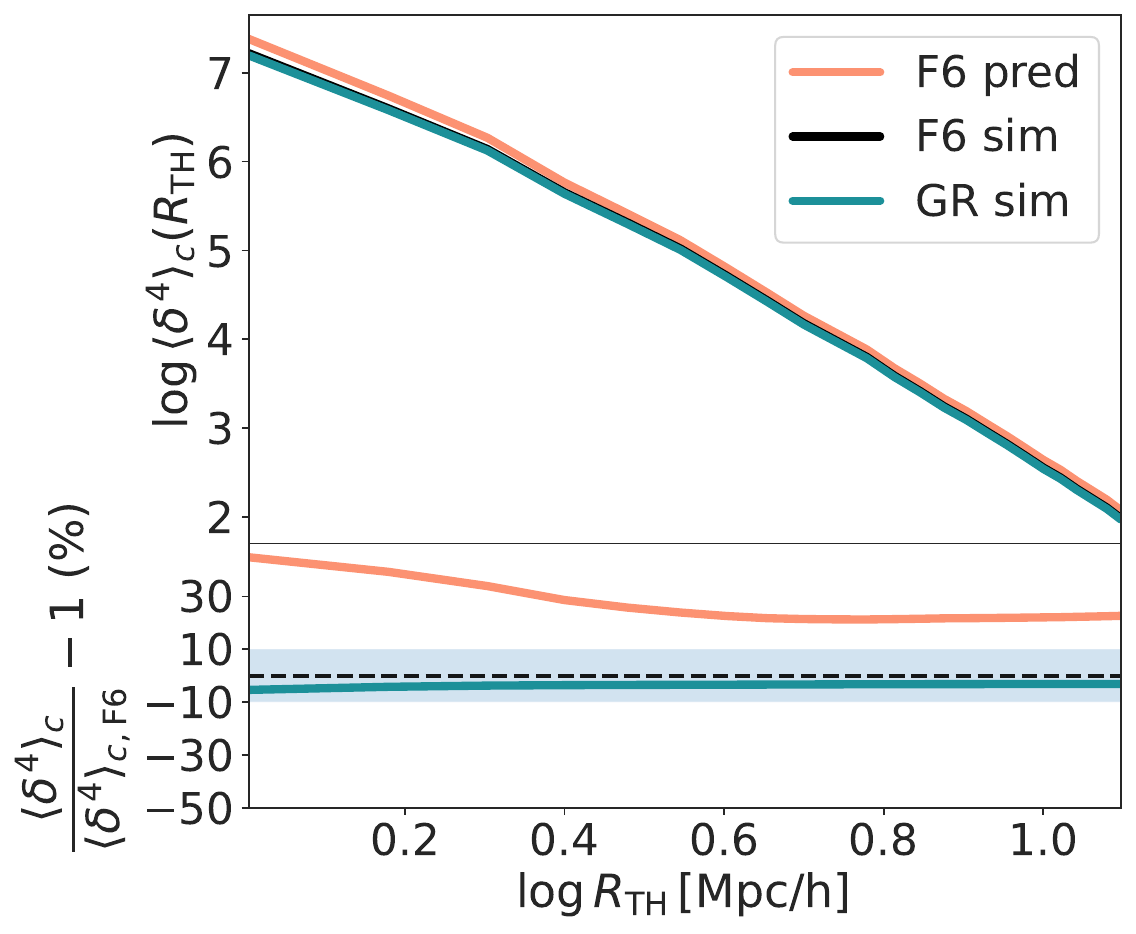}
        \caption{}
        \label{fig:cum-kurtosis-f6-den}
    \end{subfigure}
    \caption{\emph{F6 density}: Comparison of the evaluation metrics between the simulated GR and F6 density fields (GR sim and F6 sim) and the F6 density fields predicted by the GAN (F6 pred). See Fig.~\ref{fig:f4-den} for a description of the panels.}
    \label{fig:f6-den}
\end{figure*}

\subsection{Evaluation metrics}\label{sec:evaluation-metrics}
We use the following metrics for the statistical comparison in Sect.~\ref{sec:statistical-comparison}.
\begin{itemize}
    \item \emph{Two-dimensional (auto) power spectrum}: Defining the density contrast field as $\delta = \frac{\rho}{\rho_{\mathrm{mean}}} - 1$, the power spectrum is the Fourier transform of the two-point correlation function and is given by:
    \begin{equation}
        \langle\delta^\star(\mathbf{k})\delta(\mathbf{k^\prime})\rangle = (2\pi)^2P_{\delta\delta}(k)\delta^D(\mathbf{k}-\mathbf{k^\prime}),
    \end{equation}
where $\delta^D$ is the Dirac delta function. The implementation is taken from the {\sc PYLIANS} library \citep{Pylians}. The definition also holds for the power spectrum of the velocity divergence; thus $\delta$ can be replaced by $\theta$.
    % Equation ref: https://lss.fnal.gov/archive/2021/pub/fermilab-pub-21-306-t.pdf

    \item \emph{Pixel histogram}: The procedure is similar to \cite{Cataneo2022}. The fields are first convolved with a spherical top-hat filter of radius $R_{\mathrm{TH}} = 10$ $h^{-1}$ Mpc. For density fields, the histograms of smoothed overdensities ($\frac{\rho}{\rho_{\mathrm{mean}}}$) are calculated, and 99 logarithmically spaced bins in the range $[0.01, 100]$ are used. For velocity divergence, linearly spaced bins in the range $[-2500, +2500]$ $\mathrm{km}$ $\mathrm{s}^{-1}$ $\mathrm{Mpc}^{-1}$ are used.

    \item \emph{Cumulants}: Cumulants (or reduced moments) quantify deviations from Gaussianity, so it is a helpful measure to characterise non-linear matter distribution. They have been used in previous works that studied higher-order statistics of matter clustering in MG \citep[e.g.,][]{Hellwing2013,Hellwing2017} and were found to be useful in illustrating differences compared to $\Lambda$CDM. The procedure we use to obtain the cumulants is as follows. First, the density contrast field is calculated. The resulting fields are smoothed using the same approach as the pixel histograms but with different smoothing scales: 24 equally spaced values in $[1.01, 12.8]$ $h^{-1}$ Mpc\footnote{The smoothing scale, $R_{\mathrm{TH}}$, must be small enough to avoid finite volume effects and be larger than the Nyquist sampling limit (see \citet{Hellwing2013}).}. The second, third, and fourth cumulants are then calculated (i.e., variance, skewness, and kurtosis, respectively) using a similar approach described in \citep{Hellwing2013}. In summary, first, the central moments of the distribution function, $p(\delta)$, are calculated as $\langle\delta^n_R\rangle = \frac{1}{N_g}\sum_{i=0}^{i=N_g}(\delta^i_R - \langle\delta_R\rangle)^n$, where $n$ is the cumulant order and $\delta_R$ is the smoothed density field. Cumulants are derived from these central moments and are given by \citep[see, e.g.,][]{Bernardeau1994,Gaztanaga1994,Lokas1995}:
    \begin{equation}
        \begin{aligned}
            \langle\delta^2\rangle_c &= \langle\delta^2\rangle \qquad \qquad \qquad \mathrm{(variance)}\\
            \langle\delta^3\rangle_c &= \langle\delta^3\rangle \qquad\qquad\qquad \mathrm{(skewness)}\\
            \langle\delta^4\rangle_c &= \langle\delta^4\rangle - 3\langle\delta^2\rangle^2_c \,\qquad \mathrm{(kurtosis)}.
        \end{aligned}
    \end{equation}
    The same definition can be extended to the velocity divergence fields, except we skip the contrast field calculation, i.e., replacing $\delta$ by $\theta$.

\end{itemize}
These metrics have quantified the expected differences between GR and modified gravity in previous studies and thus motivate our choice here.

\subsection{Statistical comparison}\label{sec:statistical-comparison}

Different fields in the data set sample cosmic regions (such as halos) differently, so a statistically representative sample containing multiple examples is required to obtain a robust evaluation of the GAN, as opposed to the visual comparison in Sect.~\ref{subsec:visual-comparison}. Thus, to demonstrate the agreement between the simulated and predicted $f(R)$ density and velocity divergence fields, the averaged results are presented using multiple examples from the test set. Appendix~\ref{appn:supplementary-scatter-reproduction} provides supplementary results on how the GAN reproduces the variation in the statistical metrics, where it is found that the scatter is reproduced overall with reasonable accuracy.  %see https://iopscience.iop.org/article/10.1086/377574/pdf

\subsubsection{Density fields}
Figs.~\ref{fig:f4-den}--\ref{fig:f6-den} show the results.

\emph{Power spectrum}: As shown by the relative difference in the power spectra of the simulated GR and $f(R)$ fields (blue vs. black dotted lines in the lower subplot), the enhancement in the clustering of matter in $f(R)$ depends on the scale. Although the enhancement is monotonic with respect to $k$ for F5 and F6, the relation is non-monotonic for F4, which corroborates the observation in \citet{Ruan2022}. For F4, the enhancement is maximum at $k \sim 0.5-0.7\,h$ $\mathrm{Mpc}^{-1}$, where the GAN prediction matches within $\sim$1\% with the ground truth. The agreement is within 5\% up to $k \sim 2\,h$ $\mathrm{Mpc}^{-1}$. On smaller scales than this, the predicted power is overestimated by $\sim$10\%, suggesting that the GAN predicts more under and over-dense structures at these smaller scales. For F5 and F6, the prediction agrees within $\sim$1\% with the ground-truth for $k \sim 0.3-1\,h$ $\mathrm{Mpc}^{-1}$ and within $\sim$5\% up to $k \sim 5\,h$ $\mathrm{Mpc}^{-1}$.

We caution that since our simulations use a coarse force resolution ($\Delta x = 0.25 h^{-1}$ Mpc), the power spectrum enhancement of $f(R)$ compared to GR may be less precise at small scales ($k \gtrsim 3-5\,h$ $\mathrm{Mpc}^{-1}$). % see: "If the force resolution is too low, the density field tends to be underestimated3 and the fifth force overestimated, resulting in more clustering of matter in these simulations compared to the ones which have higher resolution."

\emph{Pixel histogram}: It is known that $f(R)$ gravity increases the skewness of the distribution of densities compared to GR, where the peak of the distribution in $f(R)$ shifts towards lower overdensities with a small excess in the most overdense regions \citep{Hellwing2013,Cataneo2022}. This effect is most evident in the pixel histogram for the F4 density (the black and blue lines in Fig.~\ref{fig:mass-hist-f4-den}; see also Fig.~3 of \citealt{Cataneo2022}). For all three cases (F4, F5, and F6), the GAN fairly reproduces the range of densities ($0.2 \lesssim \rho/\rho_{\mathrm{mean}} \lesssim 10$) fairly well, with a relative difference $\lesssim5\%$. The emptiest ($\rho/\rho_{\mathrm{mean}} \lesssim 0.2$) and densest ($\rho/\rho_{\mathrm{mean}} \gtrsim 10$) regions are also accurately reproduced for F4 and F5, but slightly overestimated for F6. % was present earlier: "Since the power spectrum for F5 and F6 showed a decrease in the power decreases at smaller scales, the overestimation of high overdensities may have produced a ``smoother'' small-scale structure than the corresponding ground-truths."

\emph{Cumulants}: All cumulants are enhanced in $f(R)$ compared to GR, with the most (least) prominent enhancement in F4 (F6). The figure also shows that the higher the order of the cumulant, the higher the relative difference between GR and $f(R)$. The relative difference in the cumulants of GR and $f(R)$ is, in general, not strongly dependent on the smoothing scale, $R_{\mathrm{TH}}$, as is apparent by comparing the blue and black dotted lines in the lower subplots. This may result from the (restricted) range of smoothing scales used, as shown on the x-axis of these plots. However, for F5 and F6, the differences with GR decrease by up to a few per cent with increasing smoothing scales. The relative differences between the GAN predictions and the ground truth, in general, have decreased compared to the differences between the GR and the ground truth.  A $\sim$20-25\% relative difference between the variance of GR and F4 is reduced to $\sim$5\% difference between the GAN prediction and F4. A $\sim$35\% difference in skewness is reduced to $\lesssim20\%$, and a $\sim$50\% difference in kurtosis is reduced to $\lesssim30\%$. For F5, the agreement of variance and skewness between the GAN prediction and the ground truth is within $\lesssim5\%$, while for kurtosis, the difference is only marginally lower than the difference between GR and the ground truth. For F6, the differences of $\gtrsim5\%$ between the variance of GR and the ground truth are reduced to $\lesssim5\%$, but GAN struggles to reproduce skewness and kurtosis accurately: $\sim$5-10\% differences increased to $\sim$10\% for skewness and $\sim$5\% differences increased to $\sim$20-40\%.

% was present earlier: Trends suggest that the enhancements are more prominent at smaller smoothing scales, except for the variance and skewness in F4, where the enhancement is maximum at an intermediate scale

Higher-order cumulants of the density and velocity divergence distribution of $f(R)$ gravity show larger deviations from GR than the power spectrum, with this effect becoming less pronounced as we go from F4 to F6. This indicates that higher-order statistics might be helpful in improving constraints on the $f(R)$ model beyond existing limits.

\subsubsection{Velocity divergence fields}
\begin{figure*}
    \centering
    \begin{subfigure}{0.33\textwidth}
        \centering
        \includegraphics[keepaspectratio,width=\linewidth]{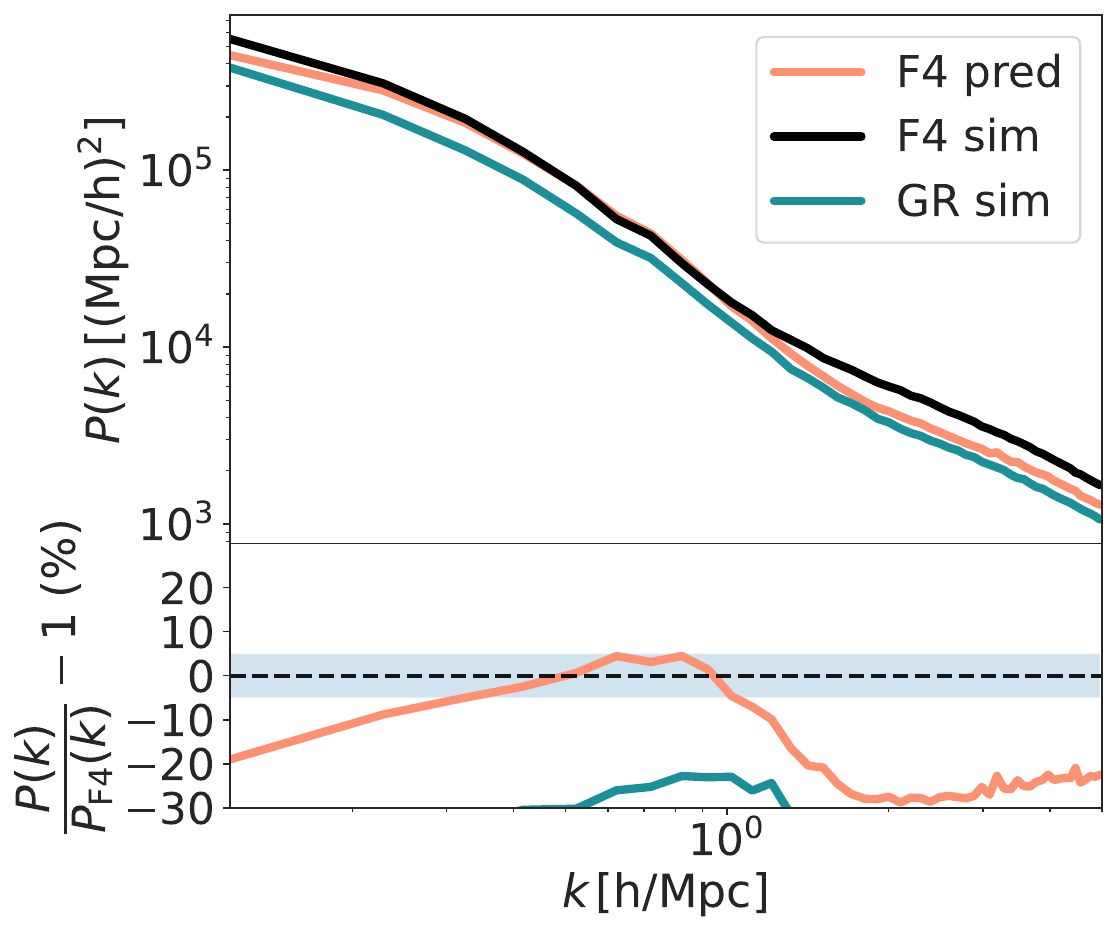}
        \caption{}
        \label{fig:ps-f4-veldiv}
    \end{subfigure}
    \hspace{4em}
    \begin{subfigure}{0.33\textwidth}
        \centering
        \includegraphics[keepaspectratio,width=\linewidth]{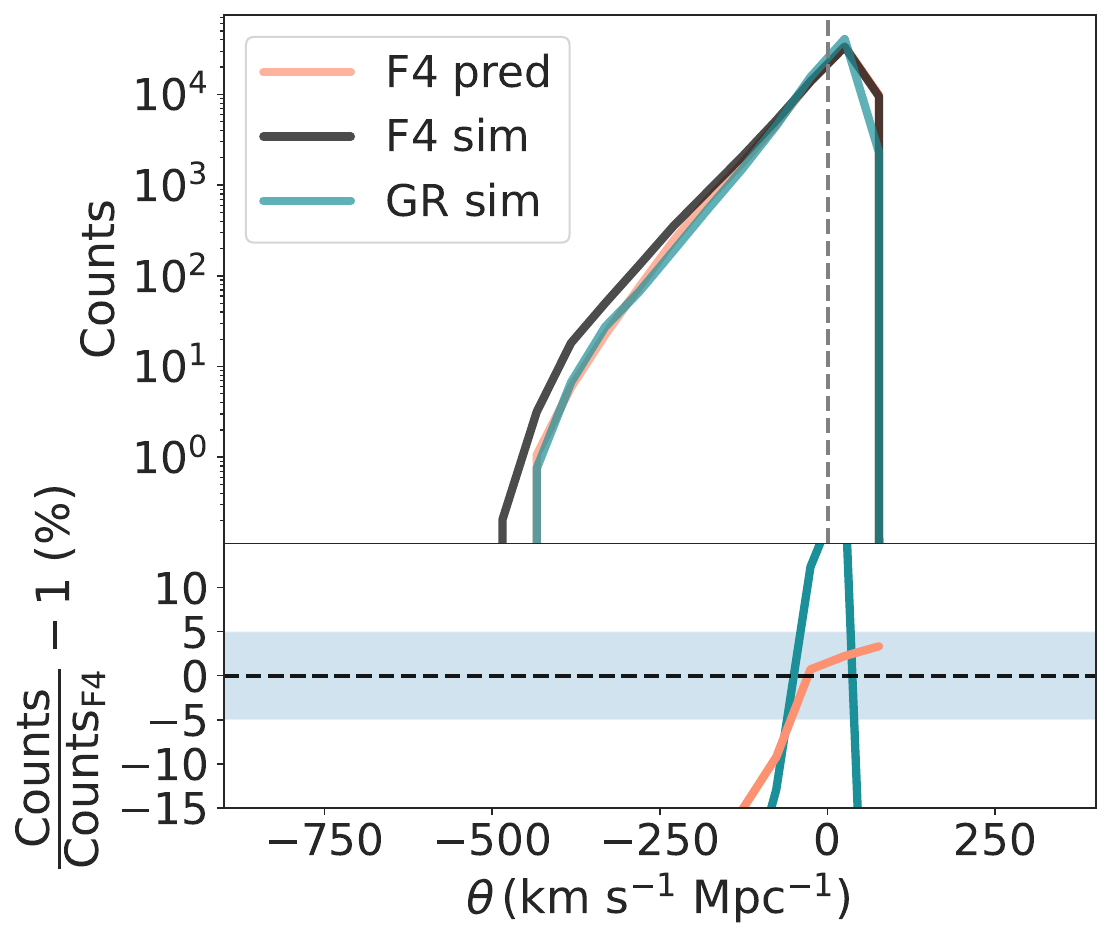}
        \caption{}
        \label{fig:mass-hist-f4-veldiv}
    \end{subfigure}
    \begin{subfigure}{0.33\textwidth}
        \centering
        \includegraphics[keepaspectratio,width=\linewidth]{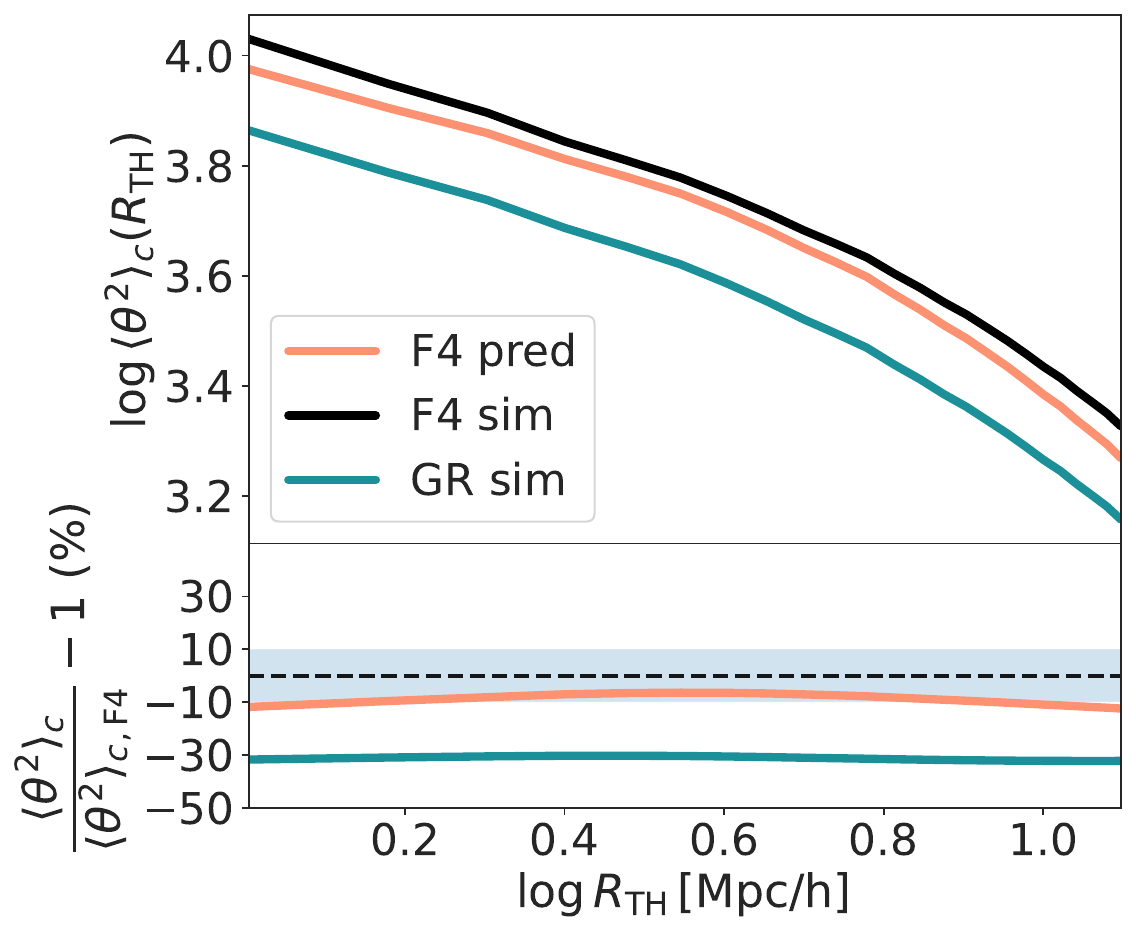}
        \caption{}
        \label{fig:cum-variance-f4-veldiv}
    \end{subfigure}
    \hfill
    \begin{subfigure}{0.33\textwidth}
        \centering
        \includegraphics[keepaspectratio,width=\linewidth]{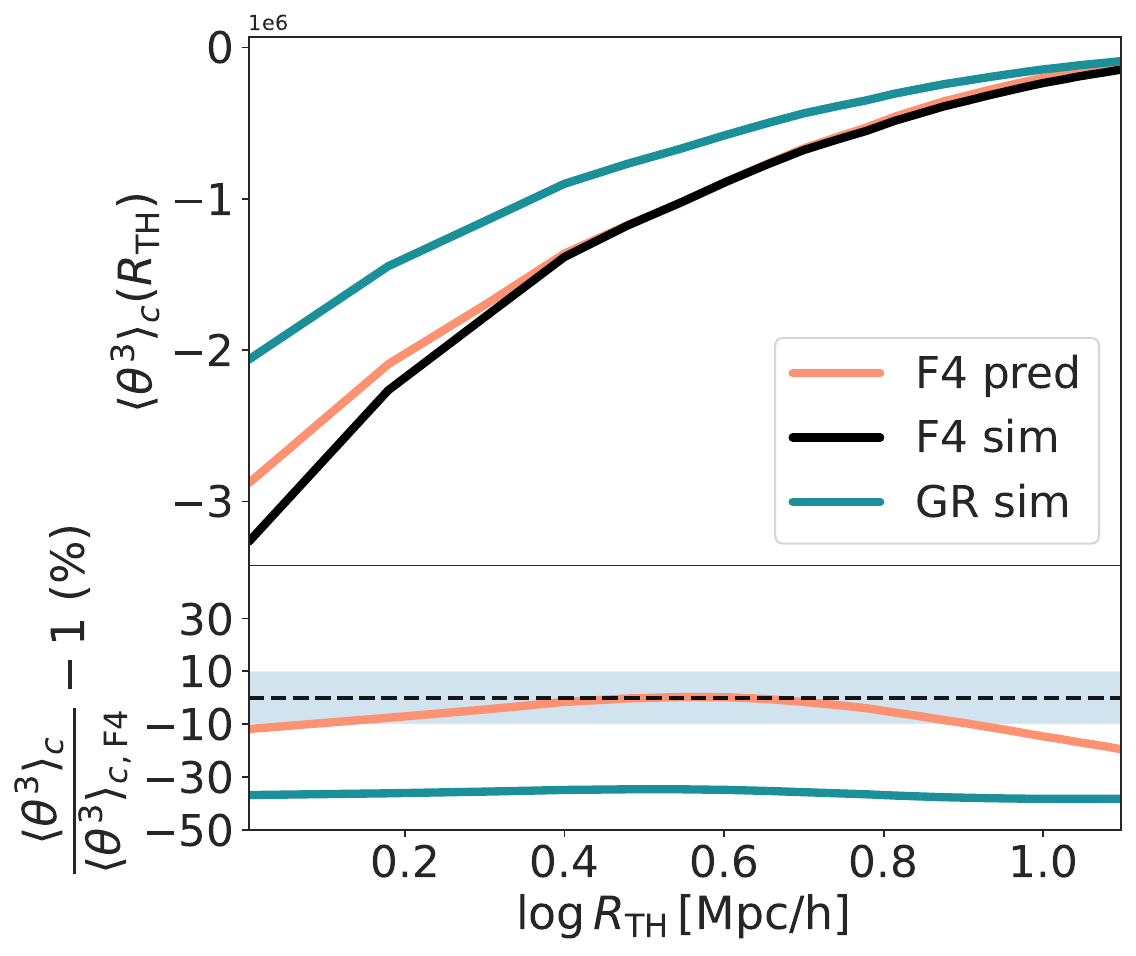}
        \caption{}
        \label{fig:cum-skewnes-f4-veldiv}
    \end{subfigure}
    \hfill
    \begin{subfigure}{0.33\textwidth}
        \centering
        \includegraphics[keepaspectratio,width=\linewidth]{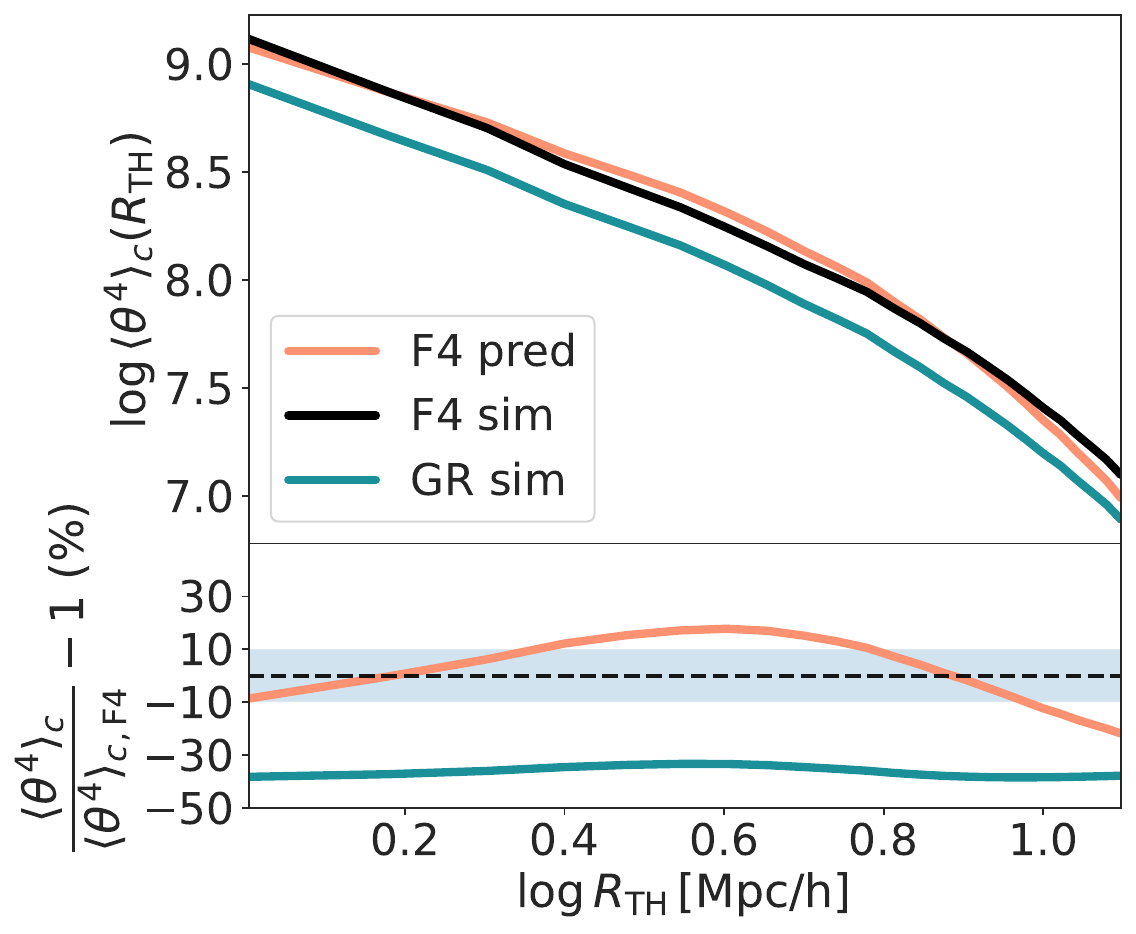}
        \caption{}
        \label{fig:cum-kurtosis-f4-veldiv}
    \end{subfigure}
    \caption{\emph{F4 velocity divergence}: Comparison of the evaluation metrics between the simulated GR and F4 velocity divergence fields (GR sim and F4 sim) and the F4 velocity divergence fields predicted by the GAN (F4 pred). See Fig.~\ref{fig:f4-den} for a description of the panels. Since velocity divergence can be negative, odd-ordered cumulants ($n = 1, 3, 5, ...$) may contain negative values, and hence we show $\langle\theta^3\rangle_c$ instead of $\log{\langle\theta^3\rangle_c}$. The gray dashed vertical lines in (b) are marked for reference and denote $\theta = 0$. %$\theta$ is dimensionless since it is the divergence of the velocity field normalised by the Hubble constant.
    }
    \label{fig:f4-veldiv}
\end{figure*}

\begin{figure*}
    \centering
    \begin{subfigure}{0.33\textwidth}
        \centering
        \includegraphics[keepaspectratio,width=\linewidth]{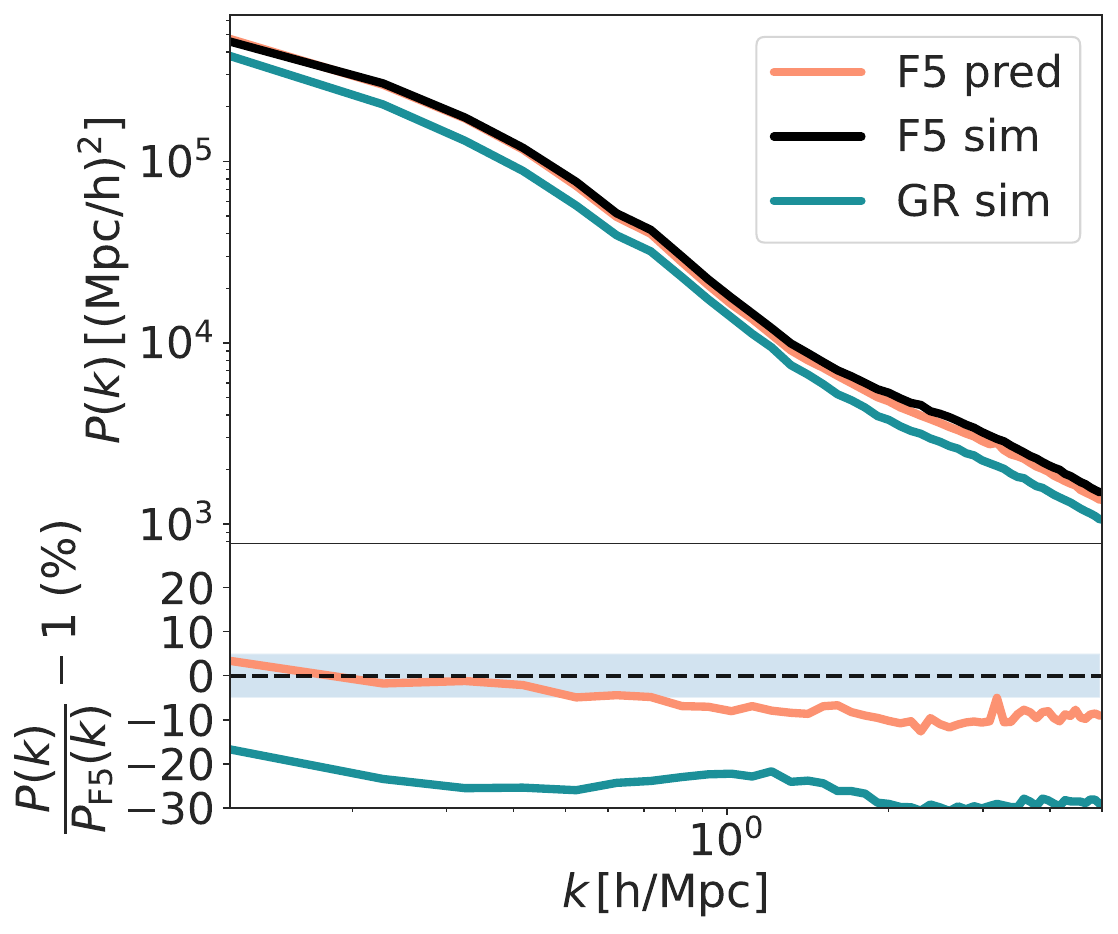}
        \caption{}
        \label{fig:ps-f5-veldiv}
    \end{subfigure}
    \hspace{4em}
    \begin{subfigure}{0.33\textwidth}
        \centering
        \includegraphics[keepaspectratio,width=\linewidth]{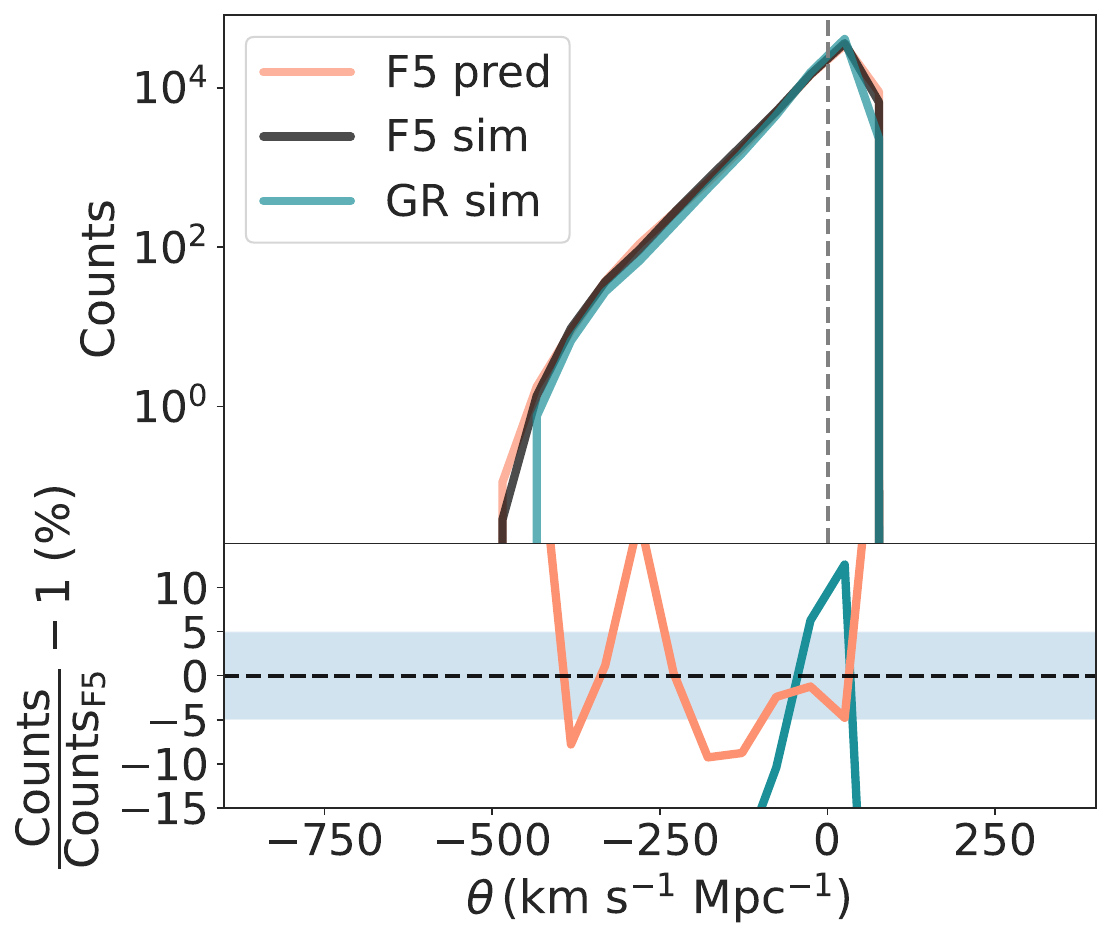}
        \caption{}
        \label{fig:mass-hist-f5-veldiv}
    \end{subfigure}
    \begin{subfigure}{0.33\textwidth}
        \centering
        \includegraphics[keepaspectratio,width=\linewidth]{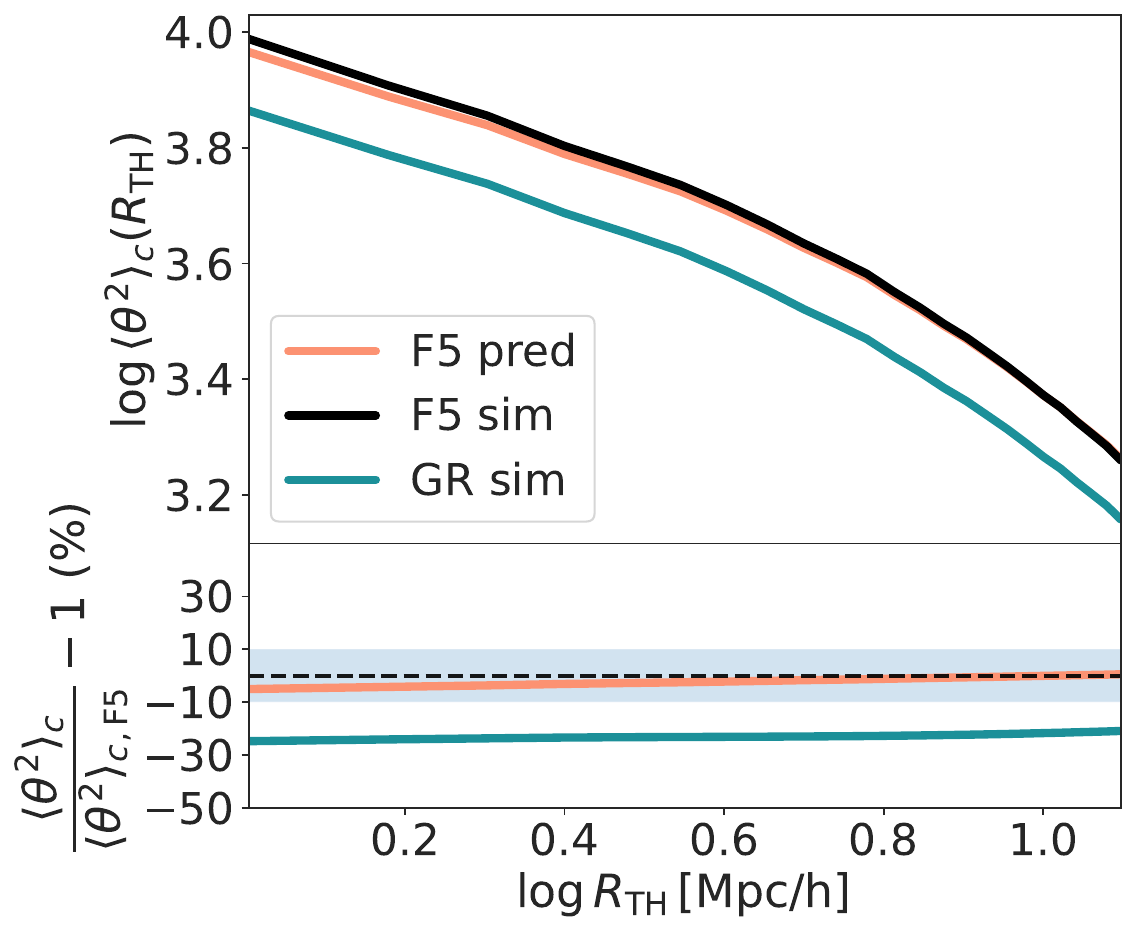}
        \caption{}
        \label{fig:cum-variance-f5-veldiv}
    \end{subfigure}
    \hfill
    \begin{subfigure}{0.33\textwidth}
        \centering
        \includegraphics[keepaspectratio,width=\linewidth]{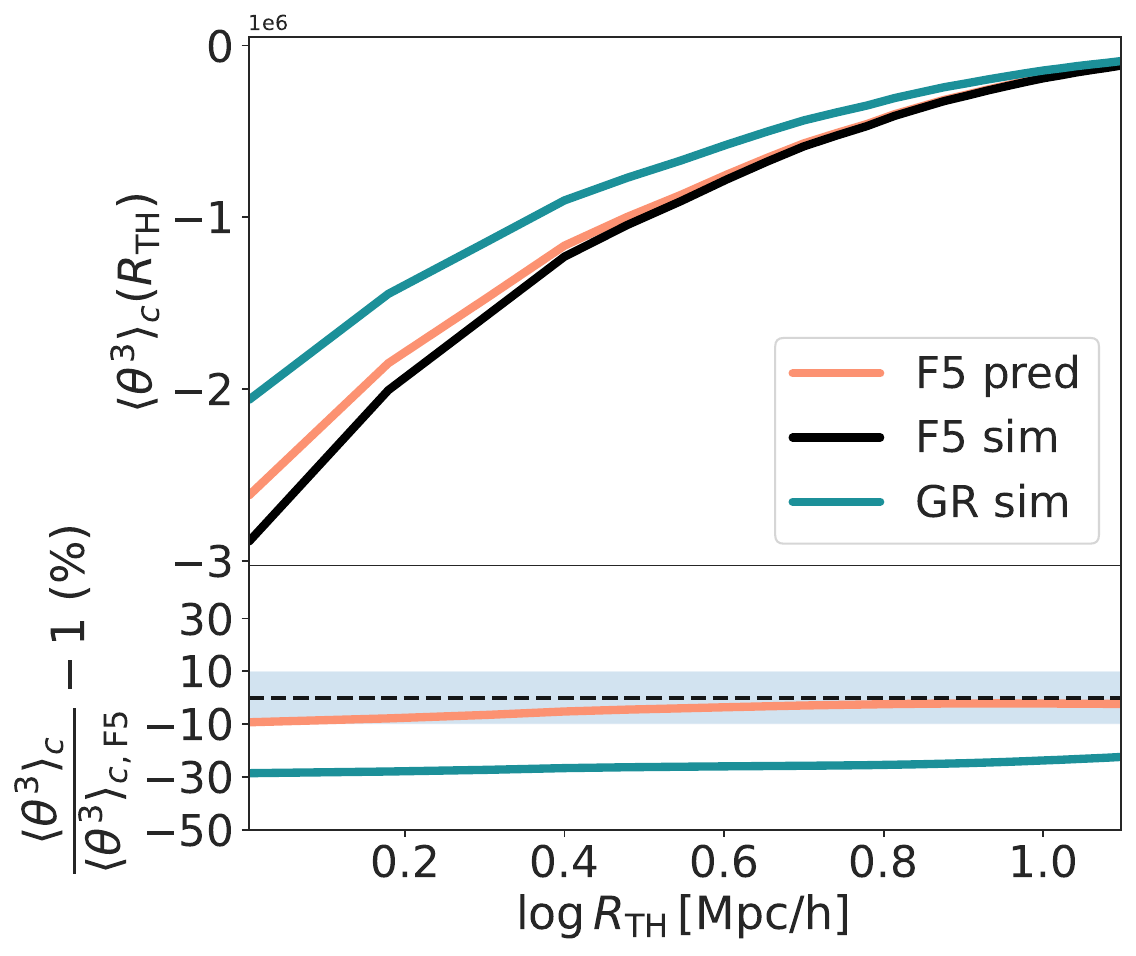}
        \caption{}
        \label{fig:cum-skewnes-f5-veldiv}
    \end{subfigure}
    \hfill
    \begin{subfigure}{0.33\textwidth}
        \centering
        \includegraphics[keepaspectratio,width=\linewidth]{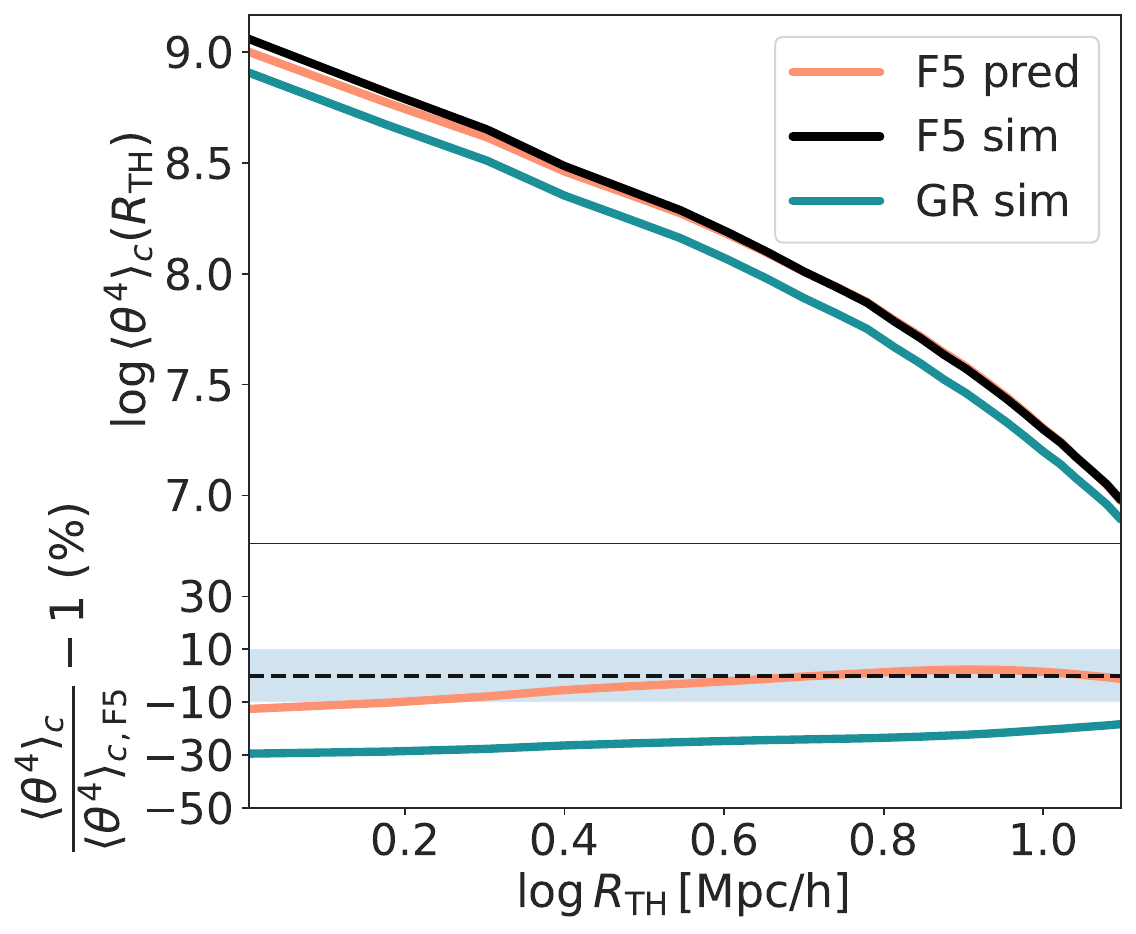}
        \caption{}
        \label{fig:cum-kurtosis-f5-veldiv}
    \end{subfigure}
    \caption{\emph{F5 velocity divergence}: Comparison of the evaluation metrics between the simulated GR and F5 velocity divergence fields (GR sim and F5 sim) and the F5 velocity divergence fields predicted by the GAN (F5 pred). See Fig.~\ref{fig:f4-den} and \ref{fig:f4-veldiv} for a description of the panels.}
    \label{fig:f5-veldiv}
% \end{figure*}

% \begin{figure*}
    \centering
    \begin{subfigure}{0.33\textwidth}
        \centering
        \includegraphics[keepaspectratio,width=\linewidth]{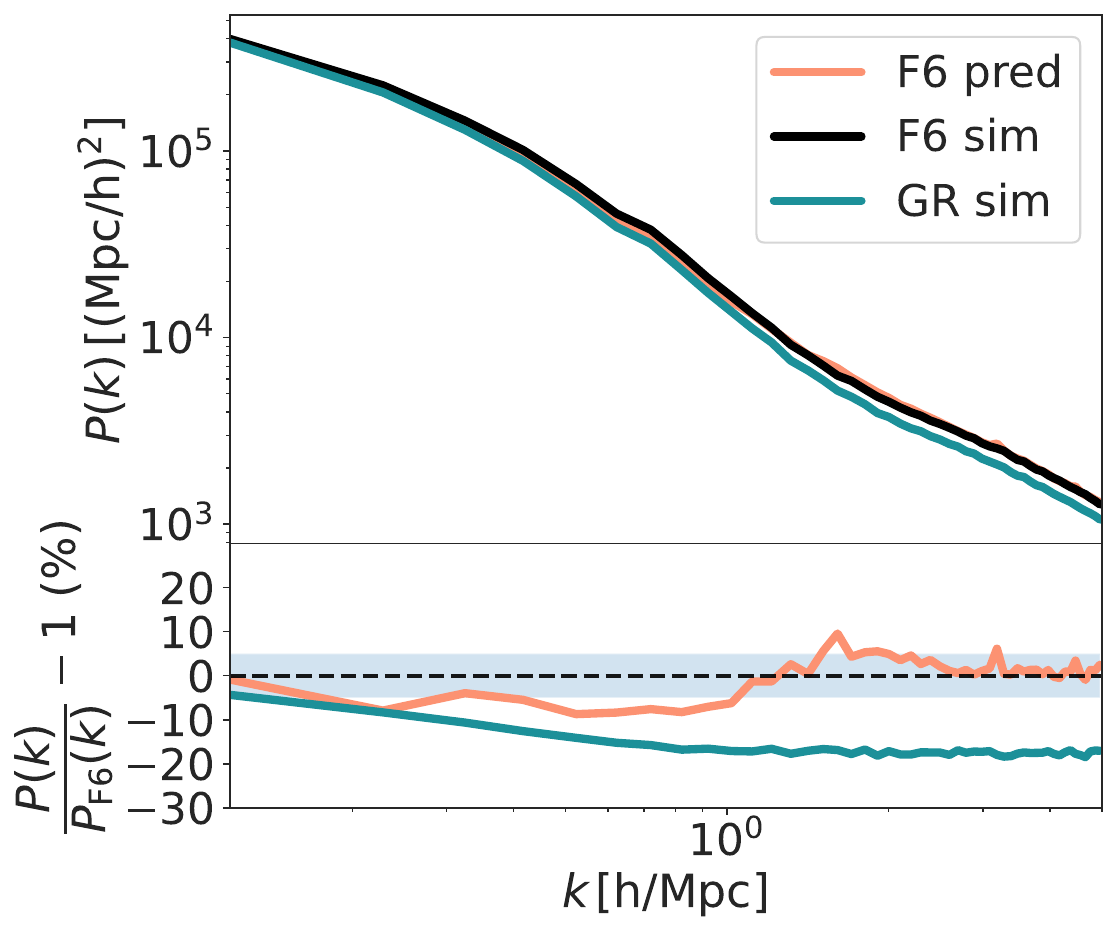}
        \caption{}
        \label{fig:ps-f6-veldiv}
    \end{subfigure}
    \hspace{4em}
    \begin{subfigure}{0.33\textwidth}
        \centering
        \includegraphics[keepaspectratio,width=\linewidth]{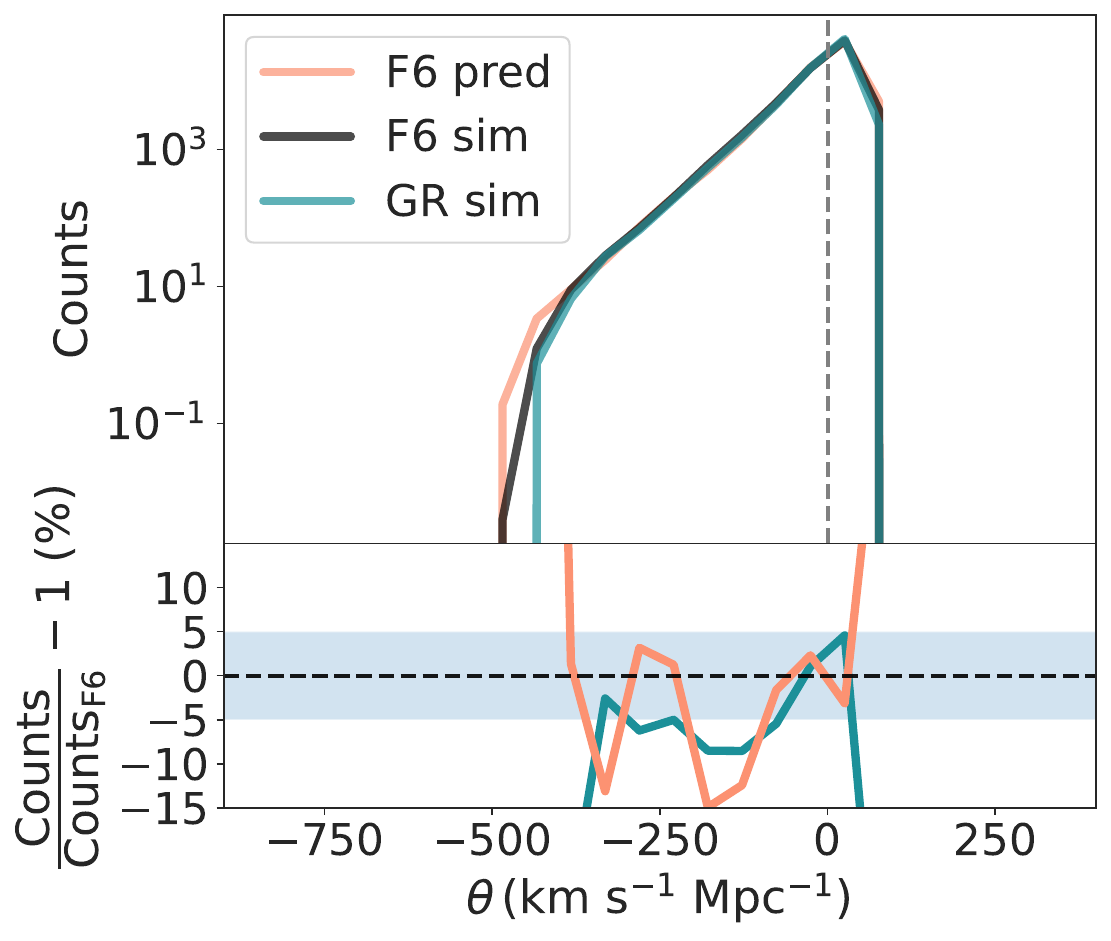}
        \caption{}
        \label{fig:mass-hist-f6-veldiv}
    \end{subfigure}
    \begin{subfigure}{0.33\textwidth}
        \centering
        \includegraphics[keepaspectratio,width=\linewidth]{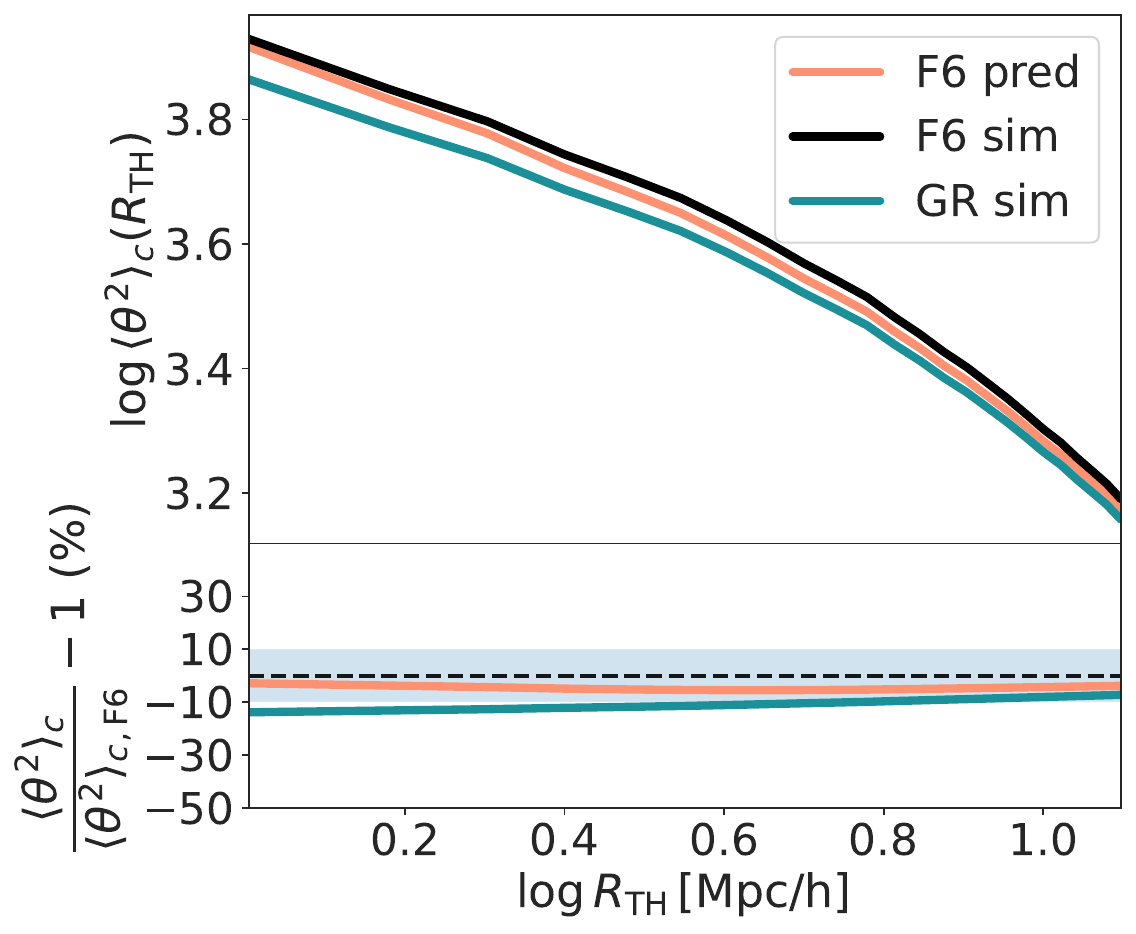}
        \caption{}
        \label{fig:cum-variance-f6-veldiv}
    \end{subfigure}
    \hfill
    \begin{subfigure}{0.33\textwidth}
        \centering
        \includegraphics[keepaspectratio,width=\linewidth]{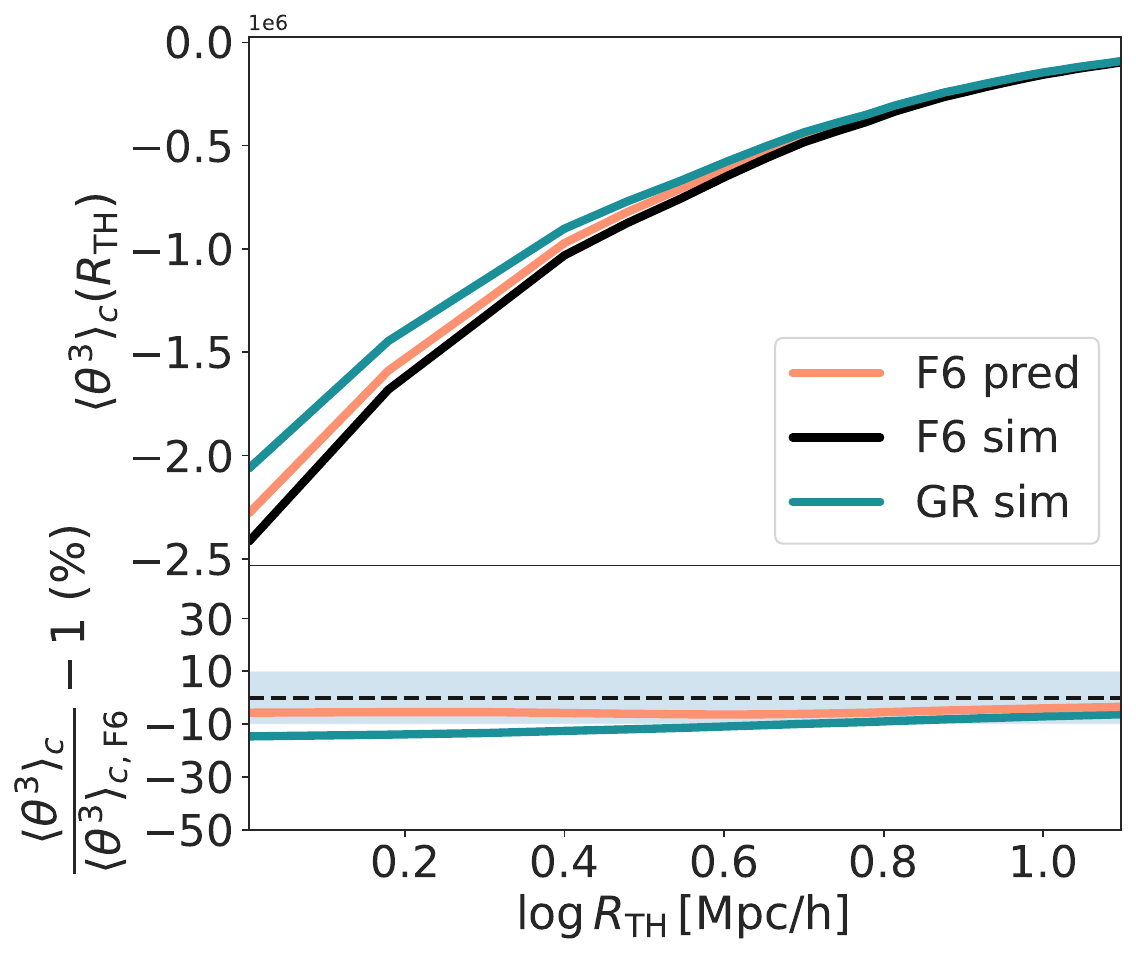}
        \caption{}
        \label{fig:cum-skewnes-f6-veldiv}
    \end{subfigure}
    \hfill
    \begin{subfigure}{0.33\textwidth}
        \centering
        \includegraphics[keepaspectratio,width=\linewidth]{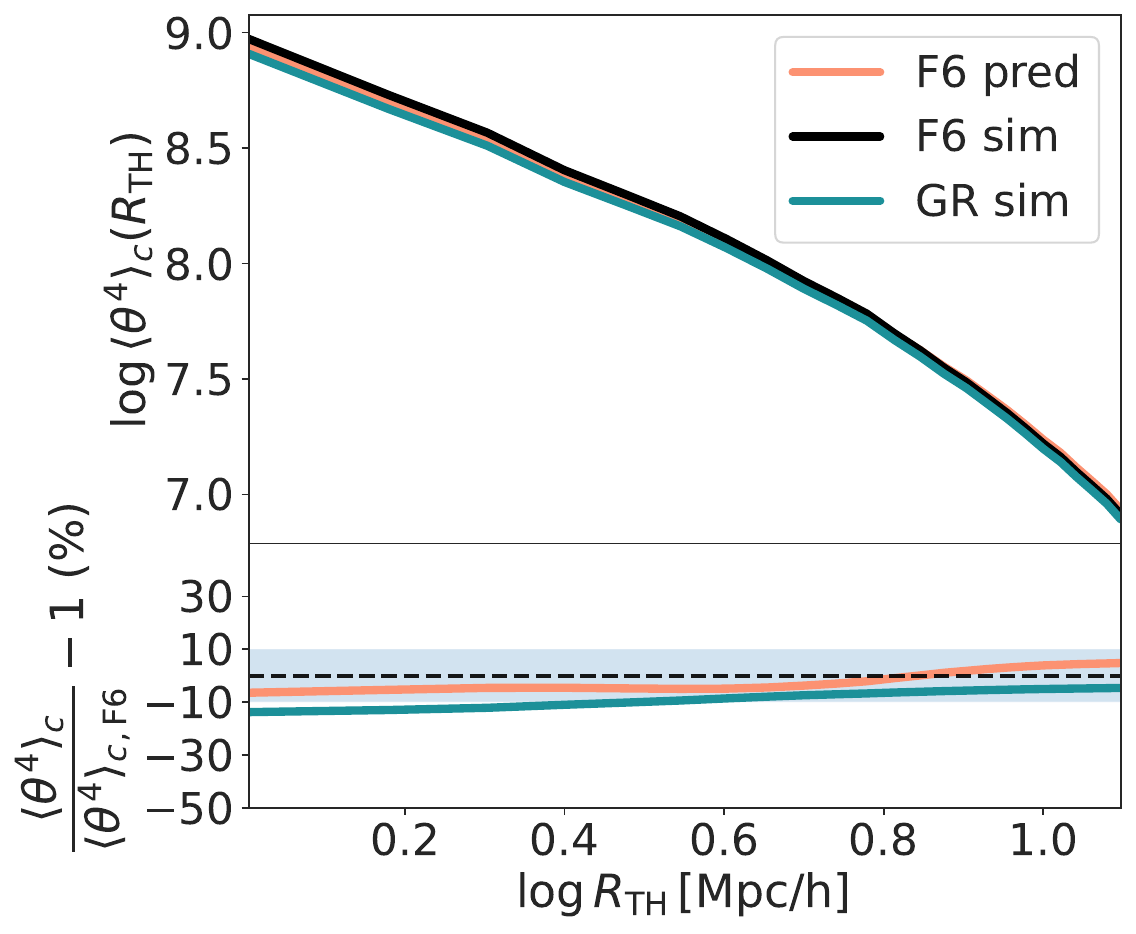}
        \caption{}
        \label{fig:cum-kurtosis-f6-veldiv}
    \end{subfigure}
    \caption{\emph{F6 velocity divergence}: Comparison of the evaluation metrics between the simulated GR and F6 velocity divergence fields (GR sim and F6 sim) and the F6 velocity divergence fields predicted by the GAN (F6 pred). See Fig.~\ref{fig:f4-den} and \ref{fig:f4-veldiv} for a description of the panels.}
    \label{fig:f6-veldiv}
\end{figure*}

Figs.~\ref{fig:f4-veldiv}--\ref{fig:f6-veldiv} show the results.

\emph{Power spectrum}: The velocity divergence power spectrum of GR and $f(R)$ shows a gradual decline at large scales, a steeper decline at intermediate scales, and again a gradual decline at small scales. This pattern is analogous to and more clearly observed by the `peak-dip-peak' pattern in dimensionless power spectra in \citet{Li2013} (their Fig.~6). The figure shows that the velocity divergence power spectra are enhanced relative to GR by orders of magnitude more than the density power spectra. For F4, the GAN prediction is closer to ground truth than GR with $\sim$20\% differences compared to the $\sim$30-40\% differences between simulated GR and F4, but the prediction still does not agree closely with ground truth. %or $k \sim 0.3-1\,h$ $\mathrm{Mpc}^{-1}$, and the predicted power is not sufficiently enhanced compared to GR at extremely large and small scales.
For F5, the agreement on extremely large and small scales becomes better, with $\lesssim$5\% differences up to $k \sim 0.7\,h$ $\mathrm{Mpc}^{-1}$ and $\lesssim10\%$ differences up to $k \sim 5\,h$ $\mathrm{Mpc}^{-1}$. For F6, the agreement of prediction is $\sim$5-7\% on all scales.%, with overall $\lesssim10\%$ difference and better agreement at smaller scales than the F4 and F5 cases.

\emph{Pixel histogram}: As hinted earlier in Sect.~\ref{subsubsec:datasetPrep}, the distribution of the velocity divergences has a heavy tail toward negative divergences. The bulk of the distribution consists of positive divergences (which correspond to `void-like' regions). For F4, F5, and F6, the GAN reproduces such regions with $\lesssim5\%$ difference from the ground truth. F4 shows a strong departure from GR for negative divergences with $\theta \lesssim -100$ $\mathrm{km}$ $\mathrm{s}^{-1}$ $\mathrm{Mpc}^{-1}$, representing regions around clusters and filaments; however, the GAN cannot reproduce this effect. The difference in these negative divergences is diminished for F5 and F6, and the GAN could reasonably reproduce those differences.

\emph{Cumulants}: The cumulants of $f(R)$ are enhanced compared to GR by orders of magnitude larger in velocity divergence compared to density. As observed with density fields, here also the relative difference between GR and F5 and GR and F6 decreases by a few per cent as the smoothing scale increases. %For F4, unlike the density fields, the enhancement decreases instead of increasing at an intermediate scale.
The relative differences between predicted and ground truth are reduced compared to the difference between GR and ground truth. A $\sim$30\% difference in the variance of GR and F4 is reduced to a $\sim$10\% difference between GAN prediction and F4. The difference in skewness improves from $\sim$40\% to $\sim$10-20\% and kurtosis from $\sim$35\% to $\lesssim$20\%. These are modest improvements, expected due to the mediocre power spectrum and pixel histogram agreement. For F5, the agreement is better than F4: a $20-25\%$ difference between the variance of GR and F5 improves to $\lesssim5\%$, and differences of $\sim$25-30\% in skewness and kurtosis improve to $\lesssim$10\%. For F6, the difference between the predicted and ground truth for all three cumulants is $\sim$5\%, which is an improvement compared to the $\sim$10\% difference between GR and the ground truth.

\subsection{Execution time}

We compare the execution times of the proposed approach with the $f(R)$ simulation in Table~\ref{tab:time-comparison} to quantify the improvement in computational load. Since our emulator predicts 2D fields, $N_g$ 2D fields are required to be predicted in order to emulate the entire 3D field. Predictions are made in batches of size 16 for execution efficiency.

% Our emulator adds a minor overhead in execution time over the GR simulations ($\sim$0.08 minutes on a GPU, which is $\sim$60 times faster than GR simulations), requiring $\sim$600 times less time than $f(R)$ simulations, and thus requiring the proposed approach almost the same time as the GR simulations, i.e., $\sim$10 times faster than f(R) simulations

The computational speed-up by our emulator approach leads to $\sim$9 times less CO$_2$ emission, and so is an overall greener alternative to $N$-body simulations of $f(R)$ gravity. These estimations were conducted using the `MachineLearning Impact calculator'\footnote{\url{https://mlco2.github.io/impact\#compute}} presented in \cite{lacoste2019quantifying}. It must be noted that our emulator code is entirely written in Python, which is widely considered an energy-inefficient language. Although using libraries such as NumPy and PyTorch has made it more efficient than native Python, future applications may focus on more thoughtful Python code optimisation or use optimised, compiled programming languages (for example, Julia) to provide further environmental benefits \citep{PortegiesZwart2020}.

Since these simulations used the {\sc MG-GLAM} code, which already offers a significant speed-up compared to several other full $N$-body codes (such as {\sc ECOSMOG}), the improvements in time and energy efficiency using the proposed approach can be further enlarged when compared to other (slower) full $N$-body codes, especially with simulations larger than the one used in this work. Larger batch sizes for the emulator predictions would also prove much faster than the current estimates. In this sense, the gain in efficiency enabled by our approach represents a lower limit than full simulations.

\begin{table}
\renewcommand{\arraystretch}{1.4}
\centering
\caption{Comparison of execution times of $f(R)$ simulations and our proposed approach. The F4 gravity model is used for the $f(R)$ simulations.  $t_{\mathrm{GAN}, \mathrm{CPU}}$ denotes the elapsed wall-clock time on the CPU and $t_{\mathrm{GAN}, \mathrm{GPU}}$ on the GPU for emulating a 3D field with dimensions $N_g^3$, which includes sampling a random latent vector and the forward pass of the GAN architecture in inference mode. $N_g = 512$ here. The execution time using the CPU is only mentioned for reference, and we consider the GPU execution time of our emulator for comparison with simulations. For this experiment, we make our emulator predict 2D fields with dimensions $N_g^2$. The mentioned execution times include the time to predict $N_g$ 2D fields processed in batch mode with a batch size of 16. $N_g = 512$ is the grid size used in the simulations. The values are shown after averaging across a few independent runs. $t_{\mathrm{f(R)}}$ and $t_{\mathrm{GR}}$ are the wall-clock time required for the $f(R)$ and GR simulations, and both values shown are averaged across a few independent runs. Both simulations and the GAN inference used 32 threads. The time required for DTFE to interpolate the particle positions from the GR and $f(R)$ simulations is assumed to be the same and is not included in the comparison. Python 3.12.4 and PyTorch 2.3.1 were used.}
\label{tab:time-comparison}
\begin{tabular}{lll}
\hline
                               & F4 simulation & Proposed approach                \\ \hline
\multirow{2}{*}{Time required (mins)} & $t_{\mathrm{f(R)}} \approx 49$    & $t_{\mathrm{GR}} \approx 5$                  \\ 
                               &               & $t_{\mathrm{GAN}, \mathrm{CPU}} = 4.9$, $t_{\mathrm{GAN}, \mathrm{GPU}} = 0.08$                \\ \hline
\textbf{Total time (mins)}                     & $t_{\mathrm{f(R)}} \approx \mathbf{49}$    & $t_{\mathrm{GR}} + t_{\mathrm{GAN}, \mathrm{GPU}} \approx \mathbf{5.08}$ \\ \hline
\end{tabular}
\end{table}

\subsection{A physical interpretation of the discriminator} \label{subsec:interpretation-disc-attention}
Since our emulator approximates the mapping from GR to $f(R)$ empirically, it is desirable to understand whether the mapping is connected to the underlying physics. The emulator may be practically useful if the mapping is mostly physical. In the context of $f(R)$ gravity, a convenient proxy of physics for such an application is the scalar field, which mediates the fifth force and, hence, modifications to standard GR gravity. For this experiment, the discriminator in the {\sc cVAE-GAN} branch is used, and as mentioned in Sect.~\ref{subsubsec:architecture}, it contains two PatchGAN discriminators looking at different scales. We therefore compare the scalar field and the weights of the spatial attention layer of the {\sc CBAM} added to the two PatchGAN discriminators and inspect possible correlations. For this, we calculate the cross-power spectrum and the cross-correlation coefficient, $r(k)$ between the scalar field and the spatial attention layer weights, where $r(k) = \dfrac{P_{\mathrm{X}}(k)}{\sqrt{P_{\mathrm{SF}}(k) P_{\mathrm{SA}}(k)}}$. The scalar field resolution is degraded to match the spatial attention map resolution for this calculation.

\begin{figure*}
    \centering
    \begin{subfigure}{0.95\textwidth}
        \centering
        \includegraphics[keepaspectratio,width=\linewidth]{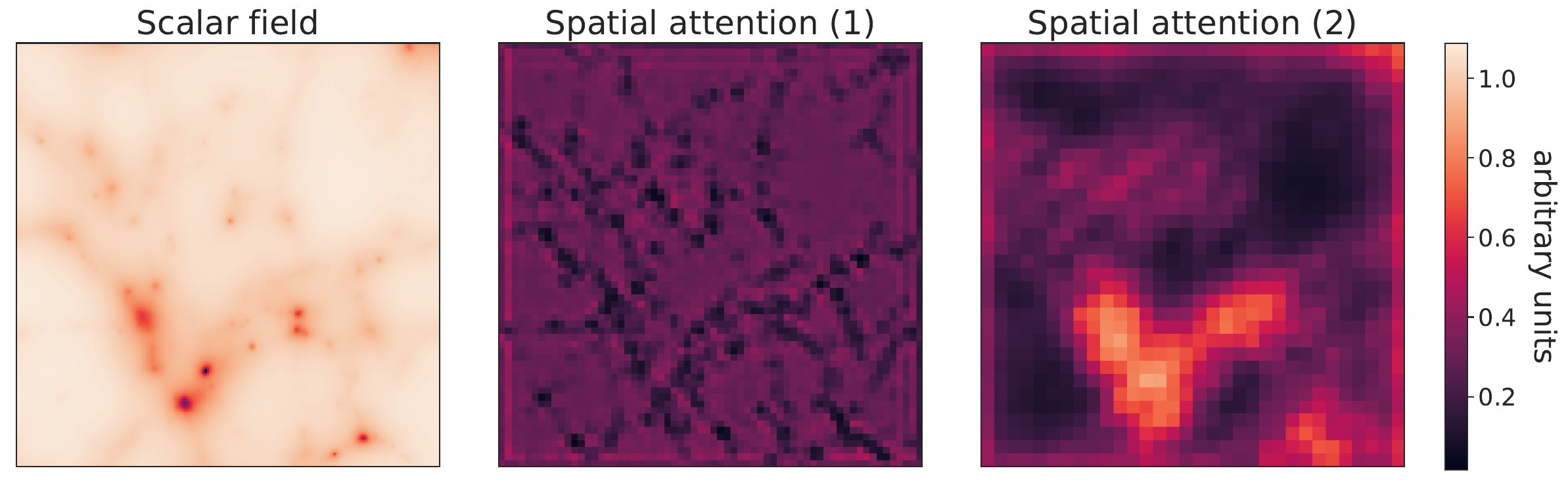}
        % \caption{}
        \label{fig:interpret-disc}
    \end{subfigure}
    \hfill
    \begin{subfigure}{0.95\textwidth}
        \centering
        \includegraphics[keepaspectratio,width=\linewidth]{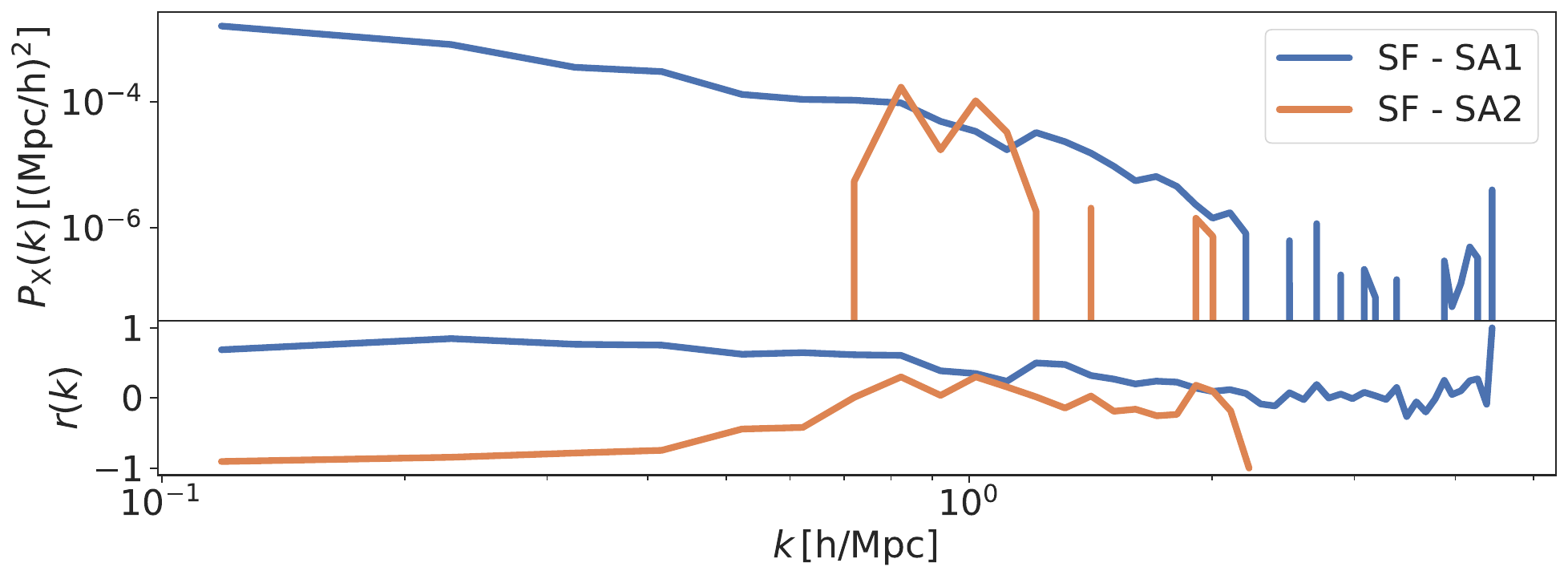}
        % \caption{}
        \label{fig:interpret-disc-cross-spec}
    \end{subfigure}
    \caption{Visual comparison of the scalar field (first column, obtained from the {\sc MG-GLAM} simulations, is in the code units, with dimensions $256 \times 256$ pixels, and each pixel spanning $0.25 h^{-1}$ Mpc) for the model with $\lvert f_{R0} \rvert = 10^{-6}$, and the spatial attention layer weights of the discriminator obtained using a random example from the test set (second and third columns; dimensions for SA1 and SA2 are $64 \times 64$ pixels and $32 \times 32$ pixels, respectively), and the cross-power spectrum of the scalar field and the spatial attention weights (bottom row). `SF' denotes the scalar field, and `SA1' and `SA2' are the spatial attention layer weights for the two PatchGAN discriminators of the full discriminator architecture, with darker (brighter) colours denoting low (high) attention. The values in SA1 and SA2 are in arbitrary units, and what matters more is how these values change spatially in coordination with the spatial changes in the scalar field. The bottom panel in the cross-power spectrum plot shows the cross-correlation coefficient, $r(k)$. The spatial attention maps show a non-negligible spatial correlation with the scalar field, discussed further in the main text. The artifacts in the borders of the attention map may arise due to padding in the convolution layers of the GAN.}
    \label{fig:spat-attn-vs-scalar-field}
\end{figure*}

Fig.~\ref{fig:spat-attn-vs-scalar-field} compares the scalar field and the spatial attention layer weights of the two PatchGAN discriminators. The scalar field shows the chameleon screening mechanism at play (the corresponding density field is not shown for clarity): its value decays exponentially to approach zero near high-density environments and is fairly uniform in void-like regions. The first spatial attention map (SA1), which focuses on relatively smaller scales, captures finer-scale structures, whereas the second spatial attention map (SA2) extracts larger-scale features. SA1 assigns low attention to regions where the scalar field is suppressed (which are the high-density regions), whereas SA2 imposes high attention on such regions. The cross-correlation coefficients shown in the plot in the bottom row quantify that while SF and SA1 show a strong positive correlation, SF and SA2 show a strong anti-correlation\footnote{Due to the low resolutions of these maps, we focus on power on relatively large scales.}. Thus, the discriminator corresponding to SA1 prioritises regions that are relatively different from GR for its prediction, and so it may be capable of distinguishing GR from the $f(R)$ field; this may have forced the generator to produce plausible $f(R)$ fields. On the other hand, the discriminator corresponding to SA2 prioritises regions similar to GR and may have captured more general features about the cosmic web. In this sense, SA1 and SA2 complement each other.

Overall, the presence of a non-trivial correlation between SA1/SA2 and SF suggests that the discriminator focuses on different regions of the cosmic web guided by the scalar field, which is desirable, even though this connection was not explicitly imposed during training. 

\section{Discussion}\label{sec:discussion}

We have introduced a $N$-body emulator that reproduces matter density and velocity divergence fields in $f(R)$ gravity. The emulator is conditioned on the corresponding fields from the $\Lambda$CDM cosmological model because cosmological tests of gravity require $\Lambda$CDM (GR) and $f(R)$ simulations to be run in conjunction. Our emulator is based on the BicycleGAN architecture, which we have modified to include attention mechanisms and a frequency-based loss function during its training. These modifications are made to handle the training challenges that arise due to the intricate nature of the mapping from GR to $f(R)$: the visual (Fig.~\ref{fig:visual-comparison}) differences between the simulated versions are not immediately obvious, and the statistical differences are generally not huge: $\mathcal{O}(10)$\% (Figs.~\ref{fig:f4-den}--\ref{fig:f6-veldiv}). The primary motivation for developing a field-level emulator is to alleviate the enormous computational times required for $f(R)$ simulations and thus enable the use of higher-order summary statistics beyond the power spectrum (and potentially field-level inference) to improve constraints $f(R)$ gravity models in future applications. We have demonstrated emulation on three variants of the $f(R)$ model, which today have the scalaron value $\lvert f_{R0} \rvert = 10^{-6}, 10^{-5}, 10^{-4}$ (called F6, F5 and F4, respectively). These variants represent the same parameterisation of modified gravity physics, and thus we train the emulator to learn just one mapping (from GR to F4) and then use an extrapolation of the latent space to generate F5 and F6 fields. This approach eliminated the need to retrain or fine-tune the emulator for each variant individually, and our empirical results have shown that it works well.

We have quantified the agreement between the predicted and the ground-truth $f(R)$ fields obtained from $N$-body simulations using various statistical metrics. For density fields, the power spectrum of the predicted fields agreed within 5\% up to $k \sim 2-5\,h$ $\mathrm{Mpc}^{-1}$ (depending on whether the model is F4, F5 or F6) and within 1\% at larger scales $k \sim 0.3-0.7\,h$ $\mathrm{Mpc}^{-1}$. This level of agreement is similar to the 1\% agreement up to $k \sim 1\,h$ $\mathrm{Mpc}^{-1}$ observed in the recent $f(R)$ emulator approach by \citet{Saadeh2024}. Since the differences in the power spectra between the simulations of GR and $f(R)$ are $\mathcal{O}(10)$\%, the predictions from the emulator can moderately constrain $f(R)$ using cosmological observables forecasted by the power spectrum. It is important to note that the scales $k \gtrsim 2\,h$ $\mathrm{Mpc}^{-1}$ show perceptible effects of baryonic feedback \citep{Arnold2019}; thus, the emulator, in its current form, may be unsuitable to use for these scales. For velocity divergence, the power spectrum of the predicted fields agreed $\lesssim$10\% up to $k \sim 5\,h$ $\mathrm{Mpc}^{-1}$ for F5 and F6, whereas the agreement was poorer for F4. Since the differences between the velocity divergence of the GR and $f(R)$ simulations are enlarged than those for density, this level of agreement for F5 and F6 models can still be considered satisfactory.

In terms of non-Gaussian statistics, such as the shape of the distribution or the higher-order cumulants of the distribution of densities and velocity divergences, the overall agreement was also reasonably good (relative differences of $\sim$10\%  or lower and also lower than the typical differences between GR and $f(R)$ simulations). However, the kurtosis of the density distribution showed unsatisfactory differences, and the GAN underpredicted regions with negative velocity divergence (i.e., near clusters and filaments of the cosmic web) for F4. Finally, even though the GAN did not explicitly learn a mapping from GR to F5 or F6, the agreement of the predictions for F5 and F6 remained robust (like for F4) with respect to most evaluation metrics, except for the kurtosis of the density fields. Overall, our GAN reproduces higher-order statistics somewhat less accurately than the power spectrum. This suggests that accurately reproducing two-point statistics does not necessarily imply that higher-order statistics will also be reproduced well. This highlights the importance of considering statistics beyond the simple two-point statistics when evaluating the quality of an emulator's predictions.

We conducted an interpretability test using the discriminator from the BicycleGAN architecture to understand which regions it focuses on for its predictions. The test revealed that the discriminator does not randomly attend to different regions in the input, but rather, its focus favourably traces the scalar field that governs the fifth force introduced in $f(R)$ gravity theory. This shows that even though the emulator was not explicitly guided by physical principles, the biases within the neural network may still be able to mimic them.

Previous approaches to accelerate $f(R)$ $N$-body simulations that emulate the enhancement of the matter power spectrum \citep[e.g.,][see Sect.~\ref{sec:intro} for a more exhaustive list of references]{Cataneo2019,Fiorini2023,Mauland2023,Casares2024} achieve an accuracy of $\lesssim3\%$ (and in many cases sub-per cent) up to scales as small as $k \sim 5-7\,h$ $\mathrm{Mpc}^{-1}$. This high level of accuracy level has been possible because these methods are explicitly tailored for the power spectrum statistic. Approximate $N$-body methods based on {\sc COLA} \citep[e.g.,][]{Valogiannis2017,Winther2017} yield a similar level of agreement as these power spectrum emulators; however, this is only up to relatively larger scales, $k \sim 2-3\,h$ $\mathrm{Mpc}^{-1}$. These approaches predict the enhancement of the power spectrum rather than the MG power spectrum directly, as the enhancement is statistically more robust and easier to predict. Our emulator resembles approximate $N$-body simulations more closely than power spectrum emulators, as it `approximates' the full matter distribution instead of a summary statistic (although it must be noted that the underlying mechanisms are entirely different; our emulator learns to make predictions empirically from data rather than solving equations prescriptively). However, our emulator is slightly less accurate than these approximate $N$-body simulations. The velocity divergence power spectrum derived from our emulated fields is also slightly less accurate than the approximate simulations of \citet{Winther2017}. It has been shown that modified gravity solvers used in various full $N$-body simulations of $f(R)$ gravity agree within $\lesssim$1\%, which is better than what our emulator is able to achieve. Thus, there is potential for improving the accuracy of our emulator, the exploration of which is left for future work.

In contrast to these previous approaches, our emulator predicts the entire matter field, and our predicted power spectrum is only one possible derivation from the predicted fields. This permits field-level inference using non-conventional summary statistics \citep[e.g.,][]{White2016,Cheng2020}, which allows extracting more information, such as phase interactions, from these fields. A natural consequence of this leads to another important difference with such past approaches, which is that we derive the MG power spectrum from our predictions directly, rather than just the enhancement compared to $\Lambda$CDM\footnote{The GAN only learns a mapping from GR to $f(R)$, which may indicate that the GAN is also only predicting the enhancement; however, this may not necessarily be the case since the entire, high-dimensional distribution of the field is being approximated rather than a summary statistic.}.

An advantage of our emulator is that, once it is trained, it can generate the entire 3D $512^3$ fields in $f(R)$ gravity in approximately five seconds on a GPU. This is $\sim$600 times faster than $f(R)$ simulations using MG-GLAM. Since our emulator is designed to be used after GR simulations, we note that the overhead due to the emulator is negligible, requiring only $\sim$1-2\% of the time required for GR simulations. This allows us to emulate $f(R)$ simulations in nearly the same amount of time as an equivalent GR simulation. For example, the full $N$-body simulations of $f(R)$ using the efficient {\sc ECOSMOG} code described in \citet{Bose2017} incurred an overhead of 180\% compared to $\Lambda$CDM for medium-resolution simulations (like the one in this work), which makes our emulator $\sim$40-80 times faster. Further, it is about 3-4 times faster than the {\sc MG-GLAM} code and the {\sc COLA}-based {\sc MG-PICOLA} code, resulting in one of the fastest speed-ups achieved for $f(R)$ simulations to date. We expect these improvements represent a lower limit, as {\sc MG-GLAM}, which was used for simulations in this paper, is already around 100 times faster than traditional $f(R)$ $N$-body simulation codes. It may also be possible to combine the approximate {\sc COLA} simulations of $\Lambda$CDM with the emulator presented here to provide even greater time improvements; there may be potential treatments available to correct these approximate methods \citep{Kaushal2022}.

Since this study analysed the performance of the emulator on effectively six cases (three $f(R)$ variants and density and velocity divergence fields), for clarity, we have omitted the discussion of how the emulator might perform across different cosmologies, redshifts, or other physically-viable factors. However, it is known that the fifth force in $f(R)$ models with smaller $\lvert f_{R0} \rvert$ is suppressed until later times so that comparing F4, F5, and F6 at a given redshift (which has been performed in this study) qualitatively mimics the comparison of a given $f(R)$ variant across different redshifts; see, for example, the discussion of the time evolution of power spectra in \citet{Li2013}. Modified gravity is known to affect the matter power spectrum in a way that is degenerate with massive neutrinos and baryonic feedback \citep[e.g.,][]{Mead2016,SpurioMancini2023,EuclidCollaboration2024}, for example, and thus, a future compelling application might be to observe if alternative statistical measures, allowed by our field-level predictions, can be used to break these degeneracies \citep[e.g.,][]{Baldi2014,Arnold2019-2,Wright2019}. The execution time comparison is based on our low-to-moderate resolution simulations ($L = 128$ $h^{-1}$ Mpc and $N_{p} = 256^3$), and we envision that the GAN would scale to higher-resolution simulations much better than $f(R)$ simulations. This is partly because, unlike most $N$-body simulations, the runtime of the emulator is almost unaffected by the box size and the number of particles and will depend only on the grid size since the emulator presented here learns from patterns within the data instead of explicitly solving the modified gravity equations.

% Although we have performed extrapolation of the latent space to adapt itself to a (slightly) different parameterization of the $f(R)$ model, this paper has skipped a detailed explanation and thus does not necessarily suggest this idea can be reliably used for generalising to different physics. We rather advocate that equipping conditional GANs with \emph{feature} learning (the ability to learn latent representations from output domain distributions; in BicycleGAN, this is mainly due to the encoder network) along with its traditional task of distribution learning (the ability to learn output domain distributions from latent representations) might be useful \citep[see also, for e.g., ][]{Donahue2016}; BicycleGAN (the approach used in this paper) may be one such promising solution. This paper shows how feature learning could help extrapolate to the F5 and F6 variants once trained to emulate F4. However, it is natural to expect it to emulate any other intermediate variant, such as F4.5 or F5.5.
Although we have performed extrapolation of the latent space to adapt itself to a (slightly) different parameterization of the $f(R)$ model, this paper has skipped a detailed explanation and thus does not necessarily suggest this idea can be reliably used for generalising to different physics. Instead, we would like to highlight that emulators based on GANs that can learn inverse mapping from data to latent representations, along with their traditional task of learning to map latent representations to data, would be more useful than those based on other GANs. There are at least two reasons for this claim. First, latent representations of even unconditional GANs have been shown to encode semantic information about the data (discussed in Sect.~\ref{sec:latent-extrapolate}), and the same is expected for conditional GANs. Second, the capacity of inverse mapping helps GANs become inherently unsupervised feature learners (see the BiGAN paper by \citealt{Donahue2016} where this was shown for unconditional GANs), so the learnt features may elucidate differences between two sets of data without supervision in a data-driven manner. This may ultimately allow such emulators to be flexible to data with different physics without requiring retraining the entire emulator on each data set. We have shown how this exact aspect of BicycleGAN helped it generalise to F5 and F6 modified gravity once trained to emulate F4, but it is natural to expect it to emulate any other intermediate variant, such as F4.5 or F5.5. There could also be different applications where the inverse mapping idea can be beneficial, for example, to generalise to $f(R)$ gravity with different values of its dimensionless parameters instead of $\lvert f_{R0} \rvert$. The latent codes possess the ability to additionally incorporate or implicitly learn information about other conditioning variables (e.g., cosmological parameters, redshift), and if this latent space can be `disentangled' so that a smooth transition is ensured in the outputs while traversing the latent space, a highly effective emulator may emerge in the future. Finally, since the latent codes are compressed feature representations of the images, these may also serve as a `summary statistic', potentially making them useful for various scientific downstream tasks. %A few advantages of (i) are that it makes the model's image predictions more understandable and allows the use of the learned representations for a variety of downstream tasks, such as supervised classification.

% Due to the structure of the $\Lambda$CDM to $f(R)$ mapping problem dealt with in this paper and as an essential first step, the latent code was derived from an encoding of the ground truth.
As a first step, the aim of this work was to perform image-to-image translation from GR to $f(R)$ fields where only the strength of $f(R)$ gravity is changed, and other physical factors like simulation realisation or redshift are unchanged, so the latent code was derived from an encoding of the ground-truth. For practical applications, the mapping from $\Lambda$CDM to $f(R)$ can be made stochastic (i.e., a one-to-many mapping) by sampling random vectors from a prior distribution since the ground truth may be unavailable. The emulator design ensures that it will still yield realistic emulations, so it may be useful for sampling multiple realisations on small scales and statistical inference \citep[e.g.,][for a similar application]{Li2021}. If indeed the emulator can produce stochastic yet realistic $f(R)$ fields from a given GR field, it may provide a promising solution to the challenge of realism--diversity tradeoff seen in generative approaches for conditional image generation \citep{Astolfi2024}. Also, this would mean that it does not necessarily require running different realisations of $\Lambda$CDM simulations to use the emulator in practice, which may further alleviate some computational bottlenecks. %note: "the small-scale structure is not unique (do diff simulations or even just random seeds). Hence, using a deterministic model also does not solve the problem. Hence, a generative model (not deterministic) is better."

While there is some scope to improve the emulator's accuracy, the benefits of the presented emulator are evident for confronting modified gravity theories with the increasing amount of observational data, since they are much more scalable in generating numerous mock samples than relying on simulations alone. Studies focused on investigating a specific MG theory, where scalability is not the main focus, may prefer other fast and more accurate alternatives discussed in this paper until the presented approach improves in accuracy in the future.

\section{Conclusion}\label{sec:conclusion}

% In this paper, we have presented a $f(R)$ gravity model emulator to alleviate prohibitively expensive runtimes of $f(R)$ simulations. The emulator is a conditional generative model based on the BicycleGAN architecture, additionally equipped with attention mechanisms and a frequency-based loss function for training. It uses outputs of an equivalent $\Lambda$CDM simulation as input and generates matter density and velocity divergence fields in the $f(R)$ model. The emulator presented in this paper is extremely fast (on GPU, it is $\sim$600 times faster than full $N$-body simulations of $f(R)$ gravity), making it one of the fastest speed-ups compared to other approaches that predict the entire matter distribution in $f(R)$. It is generally accurate to within 5\% for the density field and within 10\% for the velocity divergence field up to $k \sim 2\,h$ $\mathrm{Mpc}^{-1}$ in $P(k)$. Higher-order statistics (quantifying the entire distribution and its cumulants) also show good agreement (within $\sim$10\%), except for the kurtosis of the density distribution. Although fairly competitive, further improvements in the emulator might be required to fully realise the $\lesssim$1\% differences up to small scales required for ongoing and future cosmological surveys.

In this paper, we have presented a $f(R)$ gravity model emulator (a generative model) that translates fields from relatively inexpensive $\Lambda$CDM simulations into $f(R)$ simulations. This emulator significantly reduced the prohibitively long runtimes of $f(R)$ simulations and agreed fairly well with two-point and higher-order summary statistics of the simulated $f(R)$ fields. While the emulator is competitive, additional improvements may be required to fully realise the $\lesssim$1\% differences up to small scales required for ongoing and future cosmological surveys.

Unlike many previous conditional generative models, such as the \texttt{pix2pix}, the emulator presented here is multimodal. This is achieved by using a latent code, allowing it to map an input image to a distribution of plausible outputs. Since the learnt latent spaces of GANs can encode meaningful semantic information, we demonstrated how manipulating this learnt latent space allowed the emulator to generalise to different $f(R)$ models after being trained on one variant. We also emphasised that GANs that can learn bidirectionally, i.e., not only learn to model matter distribution but also learn to extract encoded representations from matter fields, similar to the one presented here, may have high practical value. For example, such models can generalise to slightly different physics (e.g., different parameterisations of the same physical theory) by simply using the latent space. However, further studies are required to investigate the latent spaces of such conditional GANs.

Emulators designed to reproduce specific summary statistics unavoidably lose some information and thus do not fully use observational surveys. Field-level emulators such as the one introduced here can be viewed as instead optimising over all possible (including unknown) summary statistics and are thus key to obtaining tighter constraints on cosmological parameters. As cosmological tests of gravity in upcoming cosmological surveys are expected to become more demanding, several simulation realisations will be necessary. These tests will also aim to use high-order and non-traditional summary statistics to better constrain deviations from standard GR gravity. Fast predictions from such emulators may help address these challenges. Since the emulator does not assume a specific physical model, it can also be trained to emulate other modified gravity theories to potentially bypass their computationally expensive simulations. Our work demonstrates the capabilities of machine learning-based techniques to accelerate full $N$-body simulations, which may be extremely valuable in cosmological tests of gravity and could ultimately lead to the discovery of new physics from future cosmological observations.

\section*{Acknowledgements}

We thank the referee for useful feedback that has improved the clarity of this work. We acknowledge useful discussions with Cheng-Zong Ruan, and for giving us access to the MG-GLAM code. SB is supported by the UK Research and Innovation (UKRI) Future Leaders Fellowship (grant number MR/V023381/1). BL is supported by STFC via Consolidated Grant ST/X001075/1 . This work used the DiRAC@Durham facility managed by the Institute for Computational Cosmology on behalf of the STFC DiRAC HPC Facility (\url{www.dirac.ac.uk}). The equipment was funded by BEIS capital funding via STFC capital grants ST/P002293/1, ST/R002371/1 and ST/S002502/1, Durham University and STFC operations grant ST/R000832/1. DiRAC is part of the National e-Infrastructure.
 
%%%%%%%%%%%%%%%%%%%%%%%%%%%%%%%%%%%%%%%%%%%%%%%%%%
\section*{Data Availability}

The code for designing and training the emulator architecture, and other associated scripts are available at \url{https://github.com/Yash-10/modified_gravity_emulation}. Other data presented in this paper can be made available upon reasonable request to the authors.

%The inclusion of a Data Availability Statement is a requirement for articles published in MNRAS. Data Availability Statements provide a standardised format for readers to understand the availability of data underlying the research results described in the article. The statement may refer to original data generated in the course of the study or to third-party data analysed in the article. The statement should describe and provide means of access, where possible, by linking to the data or providing the required accession numbers for the relevant databases or DOIs. \Sownak{Need to add.}



\bibliographystyle{mnras}
\bibliography{ref} % if your bibtex file is called example.bib

%\begin{thebibliography}{99}
%\end{thebibliography}

%%%%%%%%%%%%%%%%%%%%%%%%%%%%%%%%%%%%%%%%%%%%%%%%%%

%%%%%%%%%%%%%%%%% APPENDICES %%%%%%%%%%%%%%%%%%%%%

\appendix

\section{Reproduction of the scatter in simulated fields}\label{appn:supplementary-scatter-reproduction}
\begin{figure*}
    \centering
    \begin{subfigure}{0.33\textwidth}
        \centering
        \includegraphics[keepaspectratio,width=\linewidth]{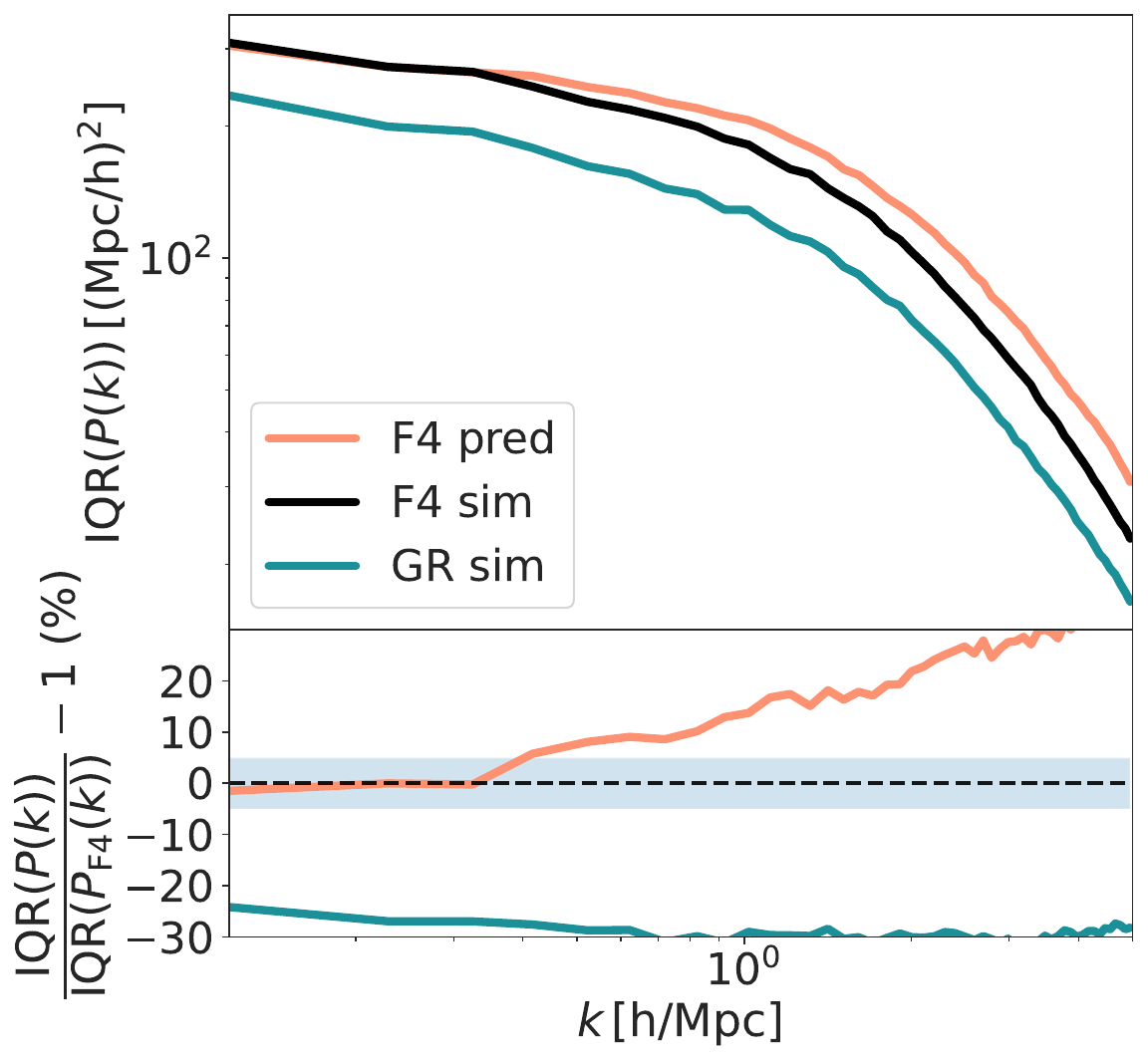}
        \caption{}
        \label{fig:ps-f4-den-scatter}
    \end{subfigure}
    \hspace{4em}
    \begin{subfigure}{0.33\textwidth}
        \centering
        \includegraphics[keepaspectratio,width=\linewidth]{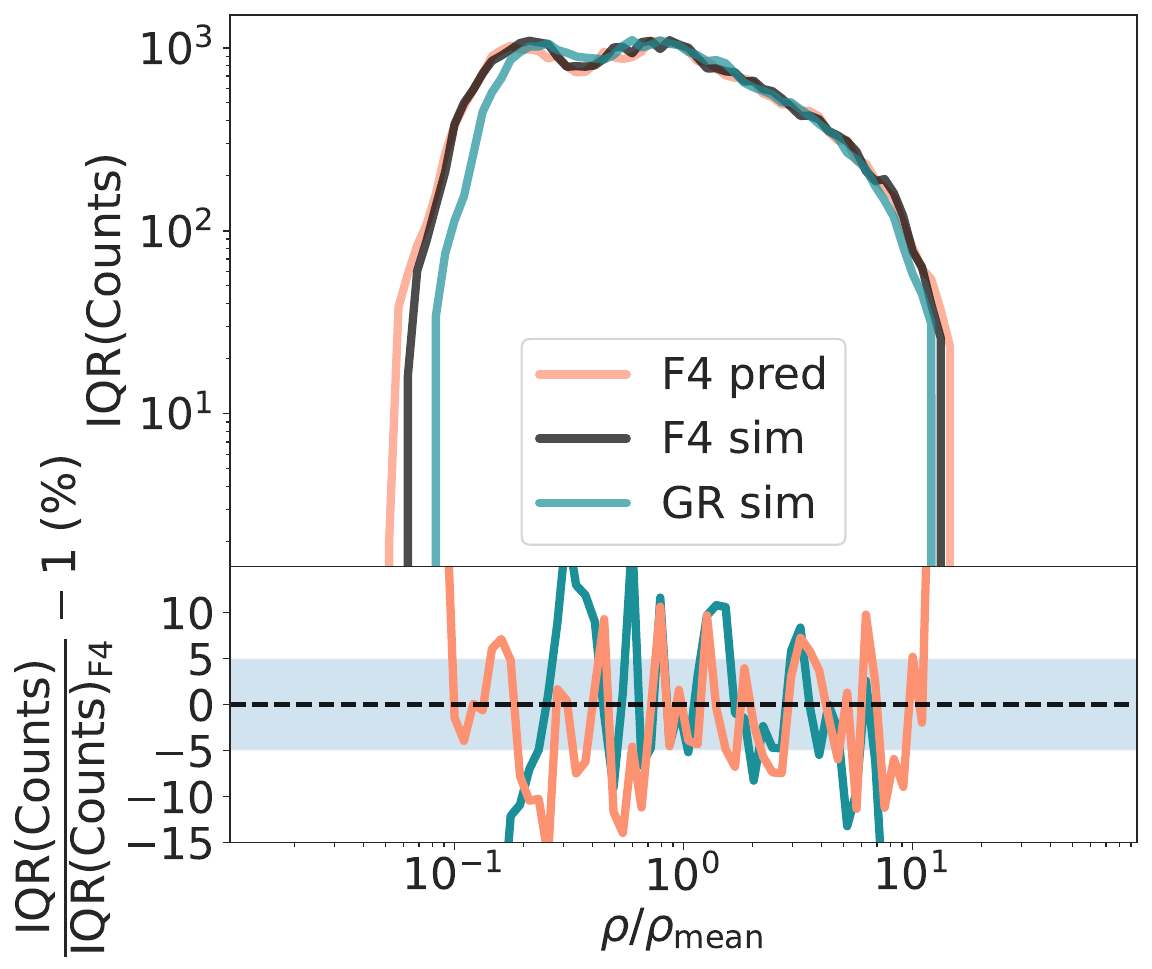}
        \caption{}
        \label{fig:mass-hist-f4-den-scatter}
    \end{subfigure}
    \begin{subfigure}{0.33\textwidth}
        \centering
        \includegraphics[keepaspectratio,width=\linewidth]{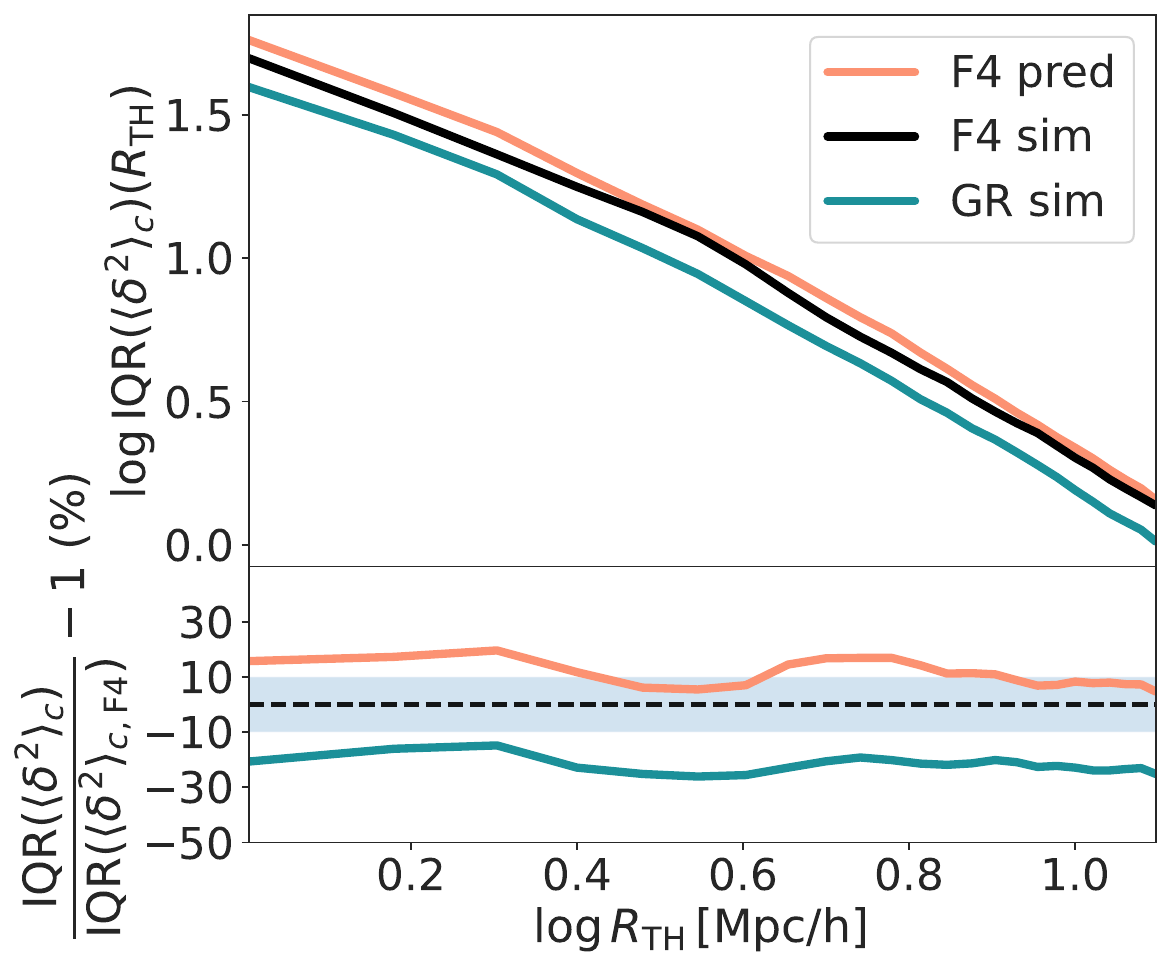}
        \caption{}
        \label{fig:cum-variance-f4-den-scatter}
    \end{subfigure}
    \hfill
    \begin{subfigure}{0.33\textwidth}
        \centering
        \includegraphics[keepaspectratio,width=\linewidth]{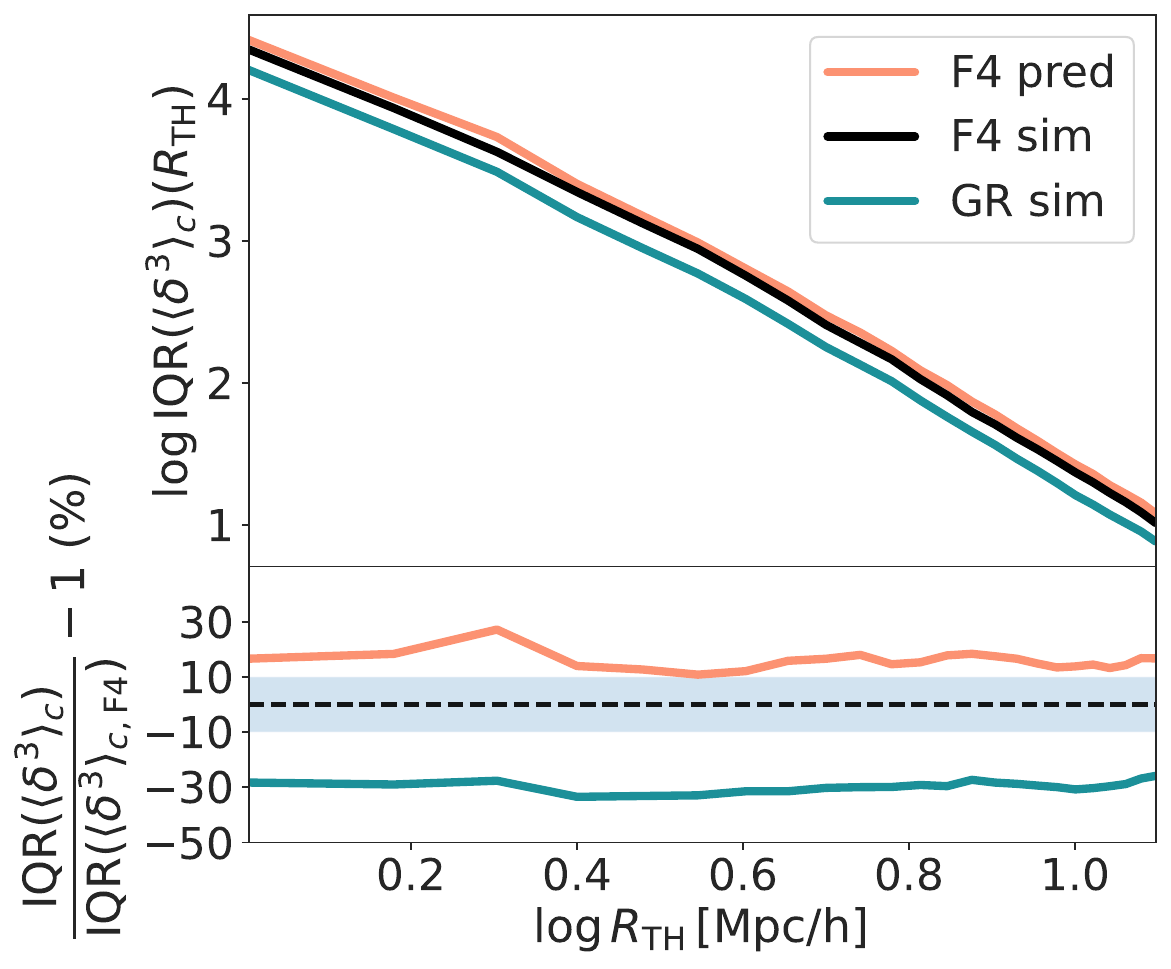}
        \caption{}
        \label{fig:cum-skewnes-f4-den-scatter}
    \end{subfigure}
    \hfill
    \begin{subfigure}{0.33\textwidth}
        \centering
        \includegraphics[keepaspectratio,width=\linewidth]{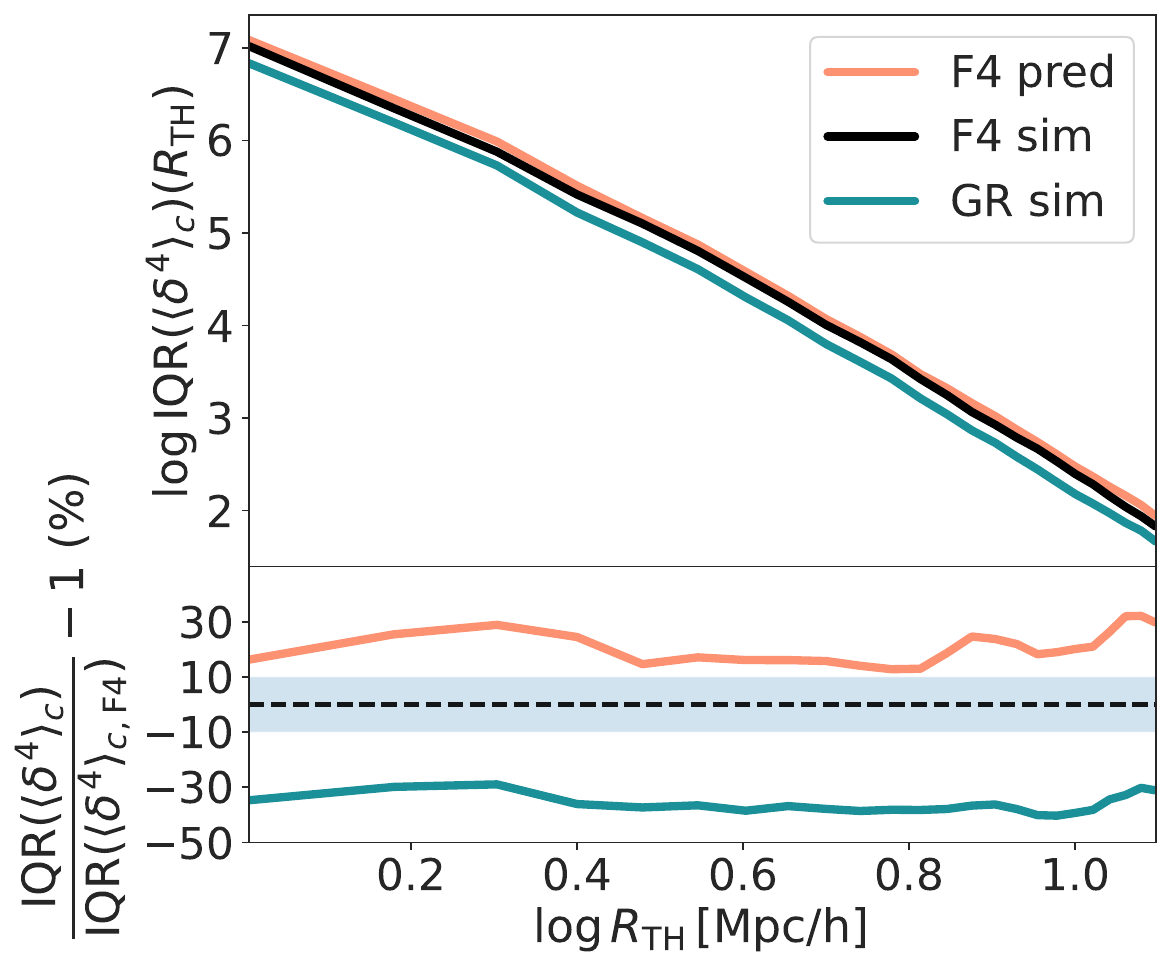}
        \caption{}
        \label{fig:cum-kurtosis-f4-den-scatter}
    \end{subfigure}
    \caption{\emph{F4 density (scatter)}: Comparison of the scatter in the evaluation metrics between the simulated GR and F4 density fields (GR sim and F4 sim) and the F4 density fields predicted by the GAN (F4 pred). The panel descriptions remain the same as in Fig.~\ref{fig:f4-den}, except that we now show the scatter instead of the mean. The scatter is measured using the interquartile range (IQR), the difference between the 75th and 25th percentiles of the evaluation metrics across the test set.}
    \label{fig:f4-den-scatter}
\end{figure*}

\begin{figure*}
    \centering
    \begin{subfigure}{0.33\textwidth}
        \centering
        \includegraphics[keepaspectratio,width=\linewidth]{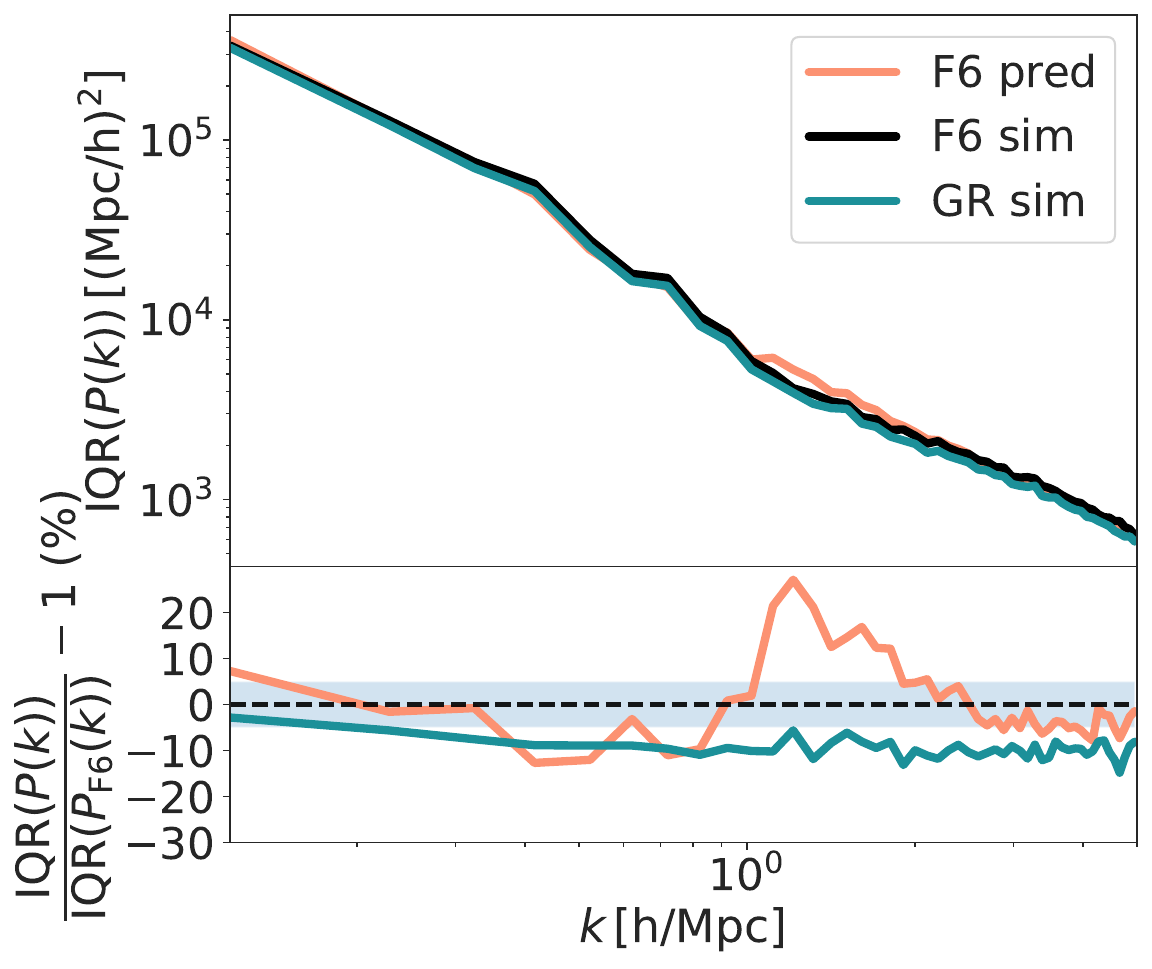}
        \caption{}
        \label{fig:ps-f6-veldiv-scatter}
    \end{subfigure}
    \hspace{4em}
    \begin{subfigure}{0.33\textwidth}
        \centering
        \includegraphics[keepaspectratio,width=\linewidth]{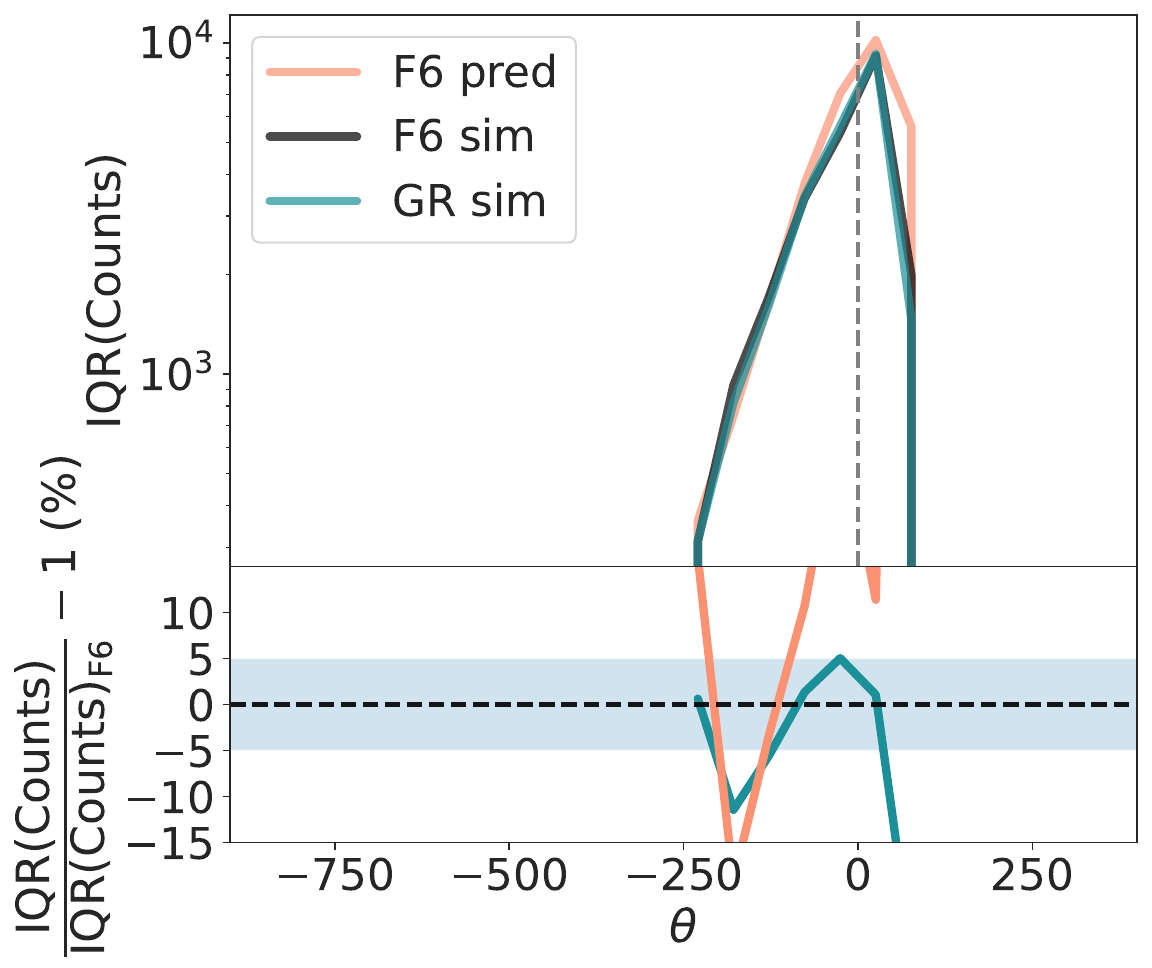}
        \caption{}
        \label{fig:mass-hist-f6-veldiv-scatter}
    \end{subfigure}
    \begin{subfigure}{0.33\textwidth}
        \centering
        \includegraphics[keepaspectratio,width=\linewidth]{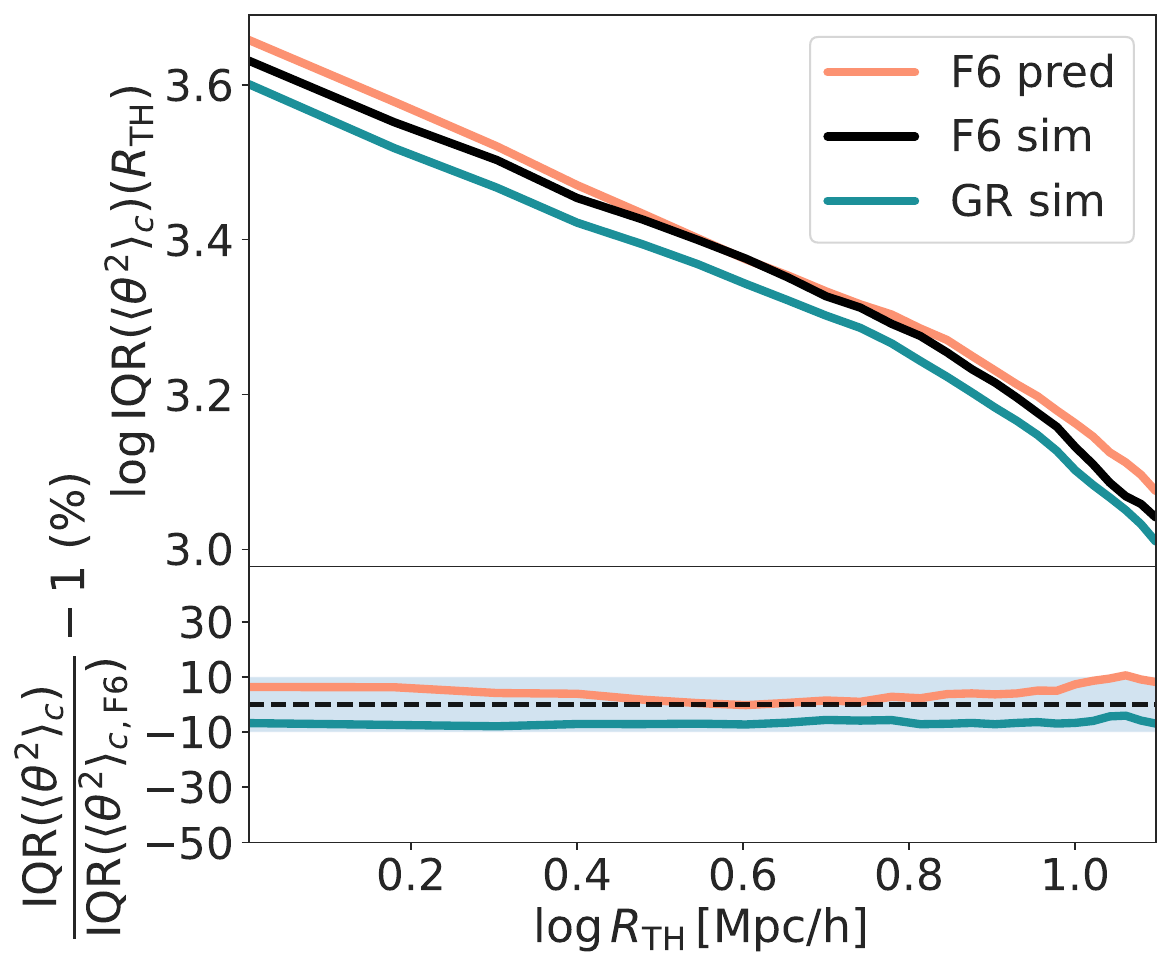}
        \caption{}
        \label{fig:cum-variance-f6-veldiv-scatter}
    \end{subfigure}
    \hfill
    \begin{subfigure}{0.33\textwidth}
        \centering
        \includegraphics[keepaspectratio,width=\linewidth]{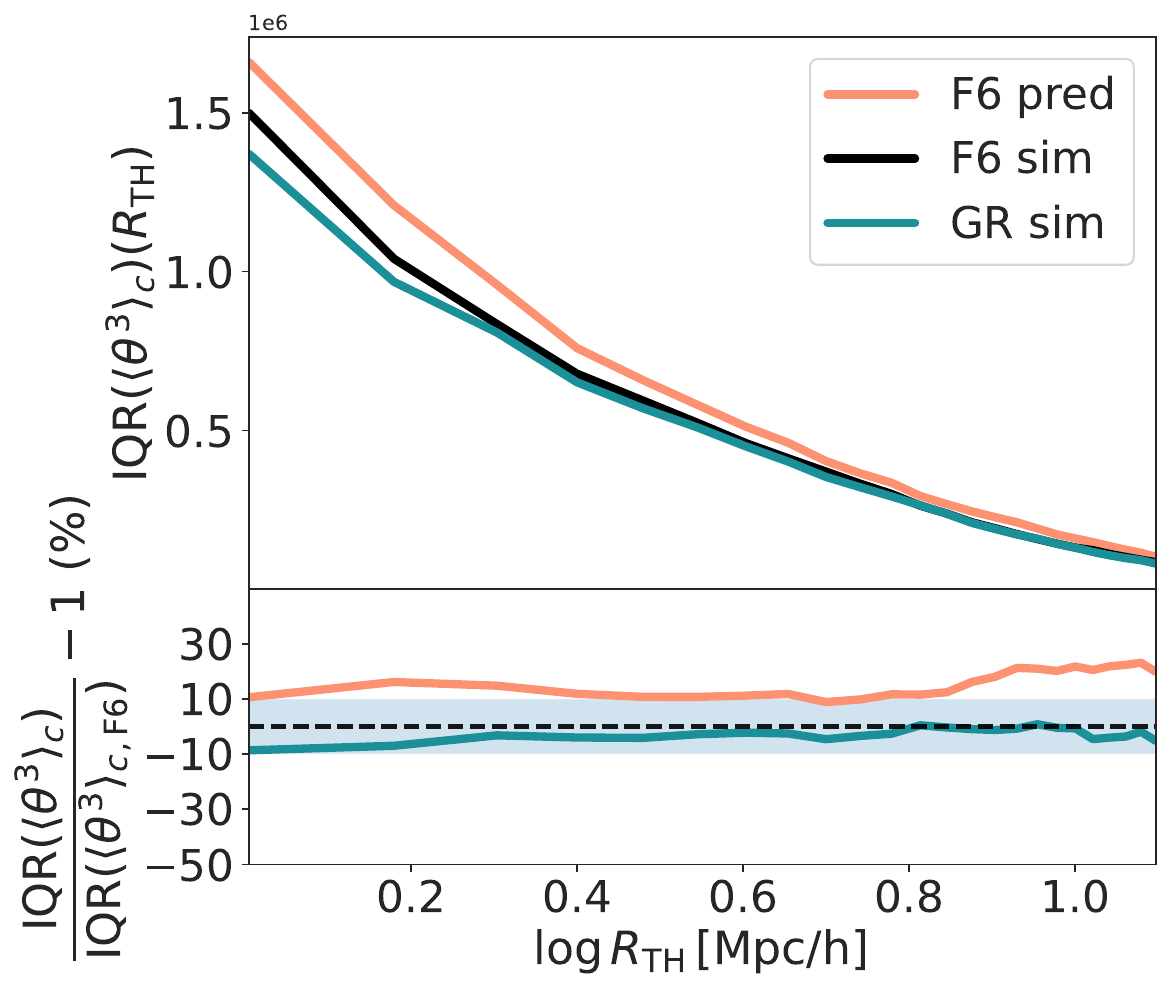}
        \caption{}
        \label{fig:cum-skewnes-f6-veldiv-scatter}
    \end{subfigure}
    \hfill
    \begin{subfigure}{0.33\textwidth}
        \centering
        \includegraphics[keepaspectratio,width=\linewidth]{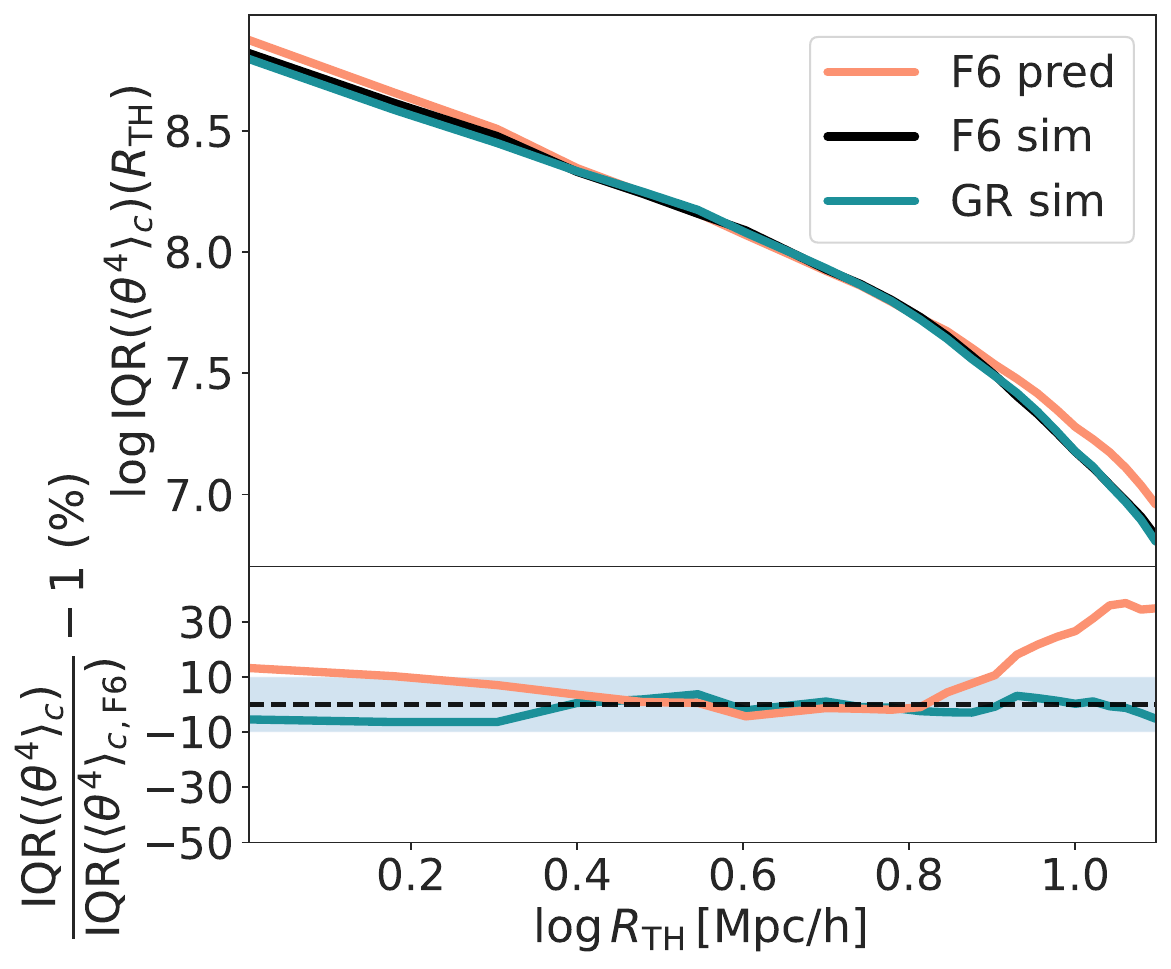}
        \caption{}
        \label{fig:cum-kurtosis-f6-veldiv-scatter}
    \end{subfigure}
    \caption{\emph{F6 velocity divergence (scatter)}: Comparison of the scatter in the evaluation metrics between the simulated GR and F6 velocity divergence fields (GR sim and F6 sim) and the F6 velocity divergence fields predicted by the GAN (F6 pred). See Fig.~\ref{fig:f4-den-scatter} and \ref{fig:f6-veldiv} for a description of the panels.}
    \label{fig:f6-veldiv-scatter}
\end{figure*}

Sect.~\ref{sec:statistical-comparison} compared the mean summary statistics of the simulated GR and $f(R)$ and the predicted $f(R)$ density and velocity divergence fields across the test set. The variation (or scatter) in the summary statistics provides useful supplementary insights, which are detailed in this section. Only results for F4 density and F6 velocity divergence are discussed since their mean agreement was reasonable, as observed in the main text. Figs.~\ref{fig:f4-den-scatter} and \ref{fig:f6-veldiv-scatter} show the comparison.

For F4 density, the scatter in the power spectrum agrees within $10\%$ up to $k \sim 0.7\,h$ $\mathrm{Mpc}^{-1}$. The differences increase to $20-30\%$ at smaller scales, indicating that the variation in the predictions is overestimated at these scales. The scatter in the pixel histogram is well reproduced by the GAN, also at the under-densities where the differences in the scatter of GR and F4 sim are most apparent. The scatter in the cumulants of the prediction does not match perfectly with the F4 sim, although the relative differences are decreased by $\sim$5-10\% compared to GR sim. For F6 velocity divergence, the scatter in the power spectrum agrees within $10\%$ for all scales. The pixel histogram scatter is overestimated for divergences around zero. The scatter in the cumulants agree within $10\%$ for the variance, but the scatter in the skewness and the kurtosis is overestimated.

Since the scatter is not severely underestimated in any of these cases, it hints that the model does not suffer from mode collapse, which is expected due to the choice of the GAN architecture used in this study (Sect.~\ref{sec:bicyclegan}). The figures show that the scatter of the evaluation metrics shows similar qualitative patterns between GR and F4 sim as observed in the mean of the evaluation metrics: for each metric, an enhancement is seen in F4 sim compared to GR sim, which is noteworthy.

%%%%%%%%%%%%%%%%%%%%%%%%%%%%%%%%%%%%%%%%%%%%%%%%%%

% Don't change these lines
\bsp	% typesetting comment
\label{lastpage}
\end{document}